\documentclass[a4paper,fleqn,usenatbib]{mnras}

\usepackage[T1]{fontenc}
\usepackage{ae,aecompl}

\usepackage{graphicx}	% Including figure files
\usepackage{amsmath}	% Advanced maths commands
\usepackage{amssymb}	% Extra maths symbols
\usepackage{subfig}

\title[Fraction of obscured AGN]{X-ray constraints on the fraction of
  obscured AGN at high accretion luminosities} 

\author[Georgakakis et al.]{A. Georgakakis$^{1,2}$\thanks{E-mail: age@mpe.mpg.de},
M. Salvato$^{1}$, Z. Liu$^{3, 4}$, J. Buchner$^{1,5,6}$,
W. N. Brandt$^{7, 8, 9}$, \newauthor T. Tasnim Ananna$^{10}$,
A. Schulze$^{11}$, Yue Shen$^{12, 13}$,
S. LaMassa$^{14}$,  K. Nandra$^{1}$,  \newauthor A. Merloni$^{1}$, I. D. McGreer$^{15}$
\\
$^1$Max Planck Institut f\"{u}r Extraterrestrische  Physik, Giessenbachstra\ss e, 85748 Garching, Germany\\ 
$^2$National Observatory of Athens, V.  Paulou  \& I.  Metaxa, 11532,  Greece\\ 
$^3$National Astronomical Observatories, Chinese Academy of Sciences, Beijing 100012, People's Republic of China\\
$^4$University of Chinese Academy of Sciences, Beijing 100049, People's Republic of China\\
$^{5}$Millenium Institute of Astrophysics, Vicu\~{n}a. MacKenna 4860, 7820436 Macul, Santiago, Chile\\
$^{6}$Pontificia Universidad Cat\'olica de Chile, Instituto de Astrof\'isica, Casilla 306, Santiago 22, Chile\\
$^7$Department of Astronomy and Astrophysics, 525 Davey Lab, The Pennsylvania State University, University Park, PA 16802, USA\\
$^8$Institute for Gravitation and the Cosmos, The Pennsylvania State University, University Park, PA 16802, USA\\
$^9$Department of Physics, 104 Davey Laboratory, The Pennsylvania State University, University Park, PA 16802, USA\\
$^{10}$Department of Physics, Yale University,  New Haven, CT 06520-8120, P.O. Box 208120, USA\\
$^{11}$Kavli IPMU (WPI), UTIAS, The University of Tokyo, Kashiwa, Chiba 277-8583, Japan\\
$^{12}$Department of Astronomy, University of Illinois at Urbana-Champaign, Urbana, IL 61801, USA\\
$^{13}$National Center for Supercomputing Applications, University of Illinois at Urbana-Champaign, Urbana, IL 61801, USA\\
$^{14}$NASA Goddard Space Flight Center, Greenbelt, MD, USA\\
$^{15}$Steward Observatory, The University of Arizona, 933 North Cherry Avenue, Tucson, AZ 85721-0065, USA\\ 
}

\date{Accepted XXX. Received YYY; in original form ZZZ}

\pubyear{2016}

\begin{document}
\label{firstpage}
\pagerange{\pageref{firstpage}--\pageref{lastpage}}
\maketitle

% Abstract of the paper
\begin{abstract} The wide-area XMM-XXL X-ray survey is used to explore
the fraction of obscured  AGN at high accretion luminosities, $L_X(\rm
2-10\,keV)\ga10^{44}\,erg\,s^{-1}$,     and     out    to     redshift
$z\approx1.5$. The  sample covers an area of  about $\rm14\,deg^2$ and
provides constraints on the space  density of powerful AGN over a wide
range  of  neutral  hydrogen  column densities  extending  beyond  the
Compton-thick  limit, $\rm N_H\approx10^{24}\,cm^{-2}$.   The fraction
of obscured  Compton-thin ($\rm N_H=10^{22}-10^{24}\,cm^{-2}$)  AGN is
estimated    to   be    $\approx0.35$   for    luminosities   $L_X(\rm
2-10\,keV)>10^{44}\,erg\,s^{-1}$  independent  of  redshift. For  less
luminous sources  the fraction of obscured  Compton-thin AGN increases
from $0.45\pm0.10$ at $z=0.25$  to $0.75\pm0.05$ at $z=1.25$.  Studies
that  select AGN in  the infrared  via template  fits to  the observed
Spectral Energy  Distribution of extragalactic  sources estimate space
densities at  high accretion luminosities consistent  with the XMM-XXL
constraints.  There is no evidence for a large population of AGN (e.g.
heavily  obscured) identified  in  the infrared  and  missed at  X-ray
wavelengths.  We  further explore the mid-infrared  colours of XMM-XXL
AGN  as  a  function  of  accretion  luminosity,  column  density  and
redshift.   The  fraction  of  XMM-XXL  sources that  lie  within  the
mid-infrared colour wedges defined in  the literature to select AGN is
primarily a function of  redshift.  This fraction increases from about
20-30\% at z=0.25 to about 50-70\% at $z=1.5$.
\end{abstract}

% Select between one and six entries from the list of approved keywords.
% Don't make up new ones.

%%%%%%%%%%%%%%%%%%%%%%%%%%%%%%%%%%%%%%%%%%%%%%%%%%

%%%%%%%%%%%%%%%%% BODY OF PAPER %%%%%%%%%%%%%%%%%%

\section{Introduction}
 
It is now well established that the dominant fraction of the growth of
supermassive black  holes at the  centres of galaxies is  taking place
behind  dense dust  and  gas  obscuring cloud.   The  black hole  mass
density estimated by integrating the luminosity function of unobscured
UV/optically-selected QSOs for example, falls short of the local black
hole    mass    function    \citep[e.g.][]{Soltan1982,    Marconi2004,
Merloni_Heinz2008}.  Under  the assumption  that the dominant  mode of
black hole growth is accretion, this implies large numbers of obscured
Active Galactic Nuclei (AGN) that are missing from UV/optical surveys.
This is  also supported by studies  of the composition  of the diffuse
X-ray  background  radiation  (XRB)  that  represents  the  integrated
emission of AGN  over cosmic time.  A large  fraction of obscured AGN,
including  some  deeply buried  ones,  are  needed  to synthesise  the
observed XRB spectrum in  the energy interval $\rm \approx 2-100\,keV$
\citep[e.g.][]{Comastri1995,          Worsley2005,          Gilli2007,
Draper_Ballantyne2009, Akylas2012}.   Observational constraints on the
distribution of the obscuration  level for AGN are therefore essential
to understand in detail the accretion history of the Universe and also
compile  unbiased AGN samples  to explore  the relation  between black
hole growth and galaxy evolution \citep[e.g.][]{Brandt_Alexander2015}.

X-ray spectroscopy provides a  direct measure of the line-of-sight gas
obscuration of  AGN, parametrised by the equivalent  column density of
neutral hydrogen,  $N_H$.  As a result,  high-energy observations have
been used  extensively to estimate the  fraction of obscured  AGN as a
function of redshift  and accretion luminosity. Follow-up observations
of     local     AGN     samples     selected    in     the     mid-IR
\citep{Brightman_Nandra2011,    Brightman2011_torus},    hard   X-rays
\citep{Burlon2011, Ajello2012}  or via nuclear  optical emission lines
\citep{Risaliti1999, Panessa2006, Akylas_Georgantopoulos2009}, measure
obscured  ($\rm  N_H>10^{22}\,cm^{-2}$)   AGN  fractions  of  $\approx
60\%$. These studies  also find a population of  deeply buried sources
with  hydrogen column  densities  above the  Thomson scattering  limit
($\rm   N_H\ga10^{24}\,cm^{-2}$;   Compton-thick),   which   represent
$\approx30-50\%$ of the overall obscured population.

Outside   the   local  Universe   {\it   XMM-Newton},  {\it   Chandra}
\citep[e.g.][]{Ueda2003,    Barger2005,   DellaCeca2008,   Yencho2009,
Ebrero2009,  Ueda2014, Aird2015, Buchner2015}  and more  recently {\it
NuSTAR} \citep{Alexander2013, Aird2015_NUSTAR}  surveys provide a rich
dataset for  studies of the obscuration distribution  of AGN.  Nuclear
obscuration  has been  shown  to  be less  common  among powerful  AGN
\citep{Akylas2006,  Tozzi2006,  Hasinger2008,  Ebrero2009,  Ueda2014},
possibly indicating  the impact of black hole  related outflows, which
sweep away  gas and dust  clouds in luminous  sources \citep{Silk1998,
Fabian1999, King2003}.   At low  accretion luminosities there  is also
evidence  for a  decrease in  the incidence  of obscuration  among AGN
\cite[e.g.][]{Brightman_Nandra2011, Burlon2011,  Buchner2015}, a trend
which  may  relate to  the  efficiency  of  cloud formation  when  the
radiative output  of accreting supermassive black holes  drops below a
certain limit.   The obscured  AGN fraction integrated  over accretion
luminosity  may   also  increase  with   redshift  at  least   out  to
$z\approx1.5-2$ \citep[e.g.][]{LaFranca2005, Vito2014, Buchner2015}, a
trend which may be linked to  the overall increase of the gas and dust
content  of galaxies at  earlier epochs  \citep{Dunne2011, Magdis2012,
Tacconi2013}.  The number of  Compton-thick sources among AGN is still
debated with different studies  finding discrepant results in terms of
both      space      density      and      cosmological      evolution
\citep[e.g.][]{Brightman_Ueda2012, Ueda2014, Buchner2015, Akylas2016}.

Current  results on the  obscuration distribution  of AGN  outside the
local Universe  are largely  based on relatively  small area  and deep
X-ray surveys. Because  of the form of the  X-ray luminosity function,
these datasets  are dominated by low and  moderate luminosity sources.
Constraints  at high accretion  luminosities, close  to and  above the
knee of the  X-ray luminosity function \citep[$L_X( \rm  2 - 10 \,keV)
\approx 10^{44} \, erg \, s^{-1}$][]{Aird2010}, are therefore affected
by Poisson noise.  There are for example, only 14 AGN with $L_X( \rm 2
-  10 \,keV)>10^{44}  \, erg  \, s^{-1}$  and $z<1$  in the  deep {\it
Chandra} survey source  catalogues compiled by \cite{Georgakakis2015}.
Because bright  sources are rare,  wide-area surveys are  required for
statistically large  samples.  Such luminous sources  are important as
they  dominate the  accretion history  of  the Universe  and may  also
represent an interesting phase of black hole growth, when AGN feedback
processes are sufficiently violent to  have an impact on the evolution
of their host galaxies \citep[e.g.][]{Hopkins2006}.  Despite the large
number  of deep  and relatively  small area  ($\rm \la  2deg^2$) X-ray
surveys   carried   out  by   {\it   Chandra}   or  {\it   XMM-Newton}
\citep{Brandt_Alexander2015}, there are currently only a few wide-area
X-ray samples.  These include the {\it Chandra} survey in the Bo\"otes
field   \citep[XBo\"otes,   $\rm   \approx   9\,deg^2$;][]{Kenter2005,
Brand2006},        the        XMM-XXL       \citep[$\rm        \approx
2\times25\,deg^2$;][]{Pierre2016}  and   the  X-ray  Survey   in  SDSS
Stripe\,82  \citep[Stripe82X, $\rm  31\,deg^2$ of  which  $\rm \approx
16\,deg^2$ is contiguous;][]{LaMassa2016}.

In this  paper we present  results on the obscuration  distribution of
luminous [$L_X(\rm  2-10\,keV) \ga 10^{44}  \, erg \, s^{-1}$]  AGN to
$z\approx1.5$  using one of  the largest  contiguous X-ray  surveys to
date,  the XMM-XXL  \citep{Pierre2016}. This  dataset, like  any X-ray
selected sample, includes sources with a small number of photons. This
poses a  challenge to  studies that attempt  to infer  AGN parameters,
such as level of line-of-sight obscuration or accretion luminosity, by
modeling their  X-ray spectra.  In our  work we address  this issue by
making realistic assumptions on  the adopted X-ray spectral models and
using  robust  statistical methods  for  unbiased  estimates of  X-ray
spectral parameters  and their associated uncertainties.  We show that
the XMM-XXL despite its relatively shallow depth (typically 10\,ks per
{\it XMM} pointing) provides new measurements of the AGN space density
at luminosities $L_X( \rm 2 -  10 \,keV) \ga 10^{44} \, erg \, s^{-1}$
over a wide range of line-of-sight hydrogen column densities extending
above the Compton-thick limit.  Surveys  like the XMM-XXL, in terms of
depth and area, provide an  excellent complement to deep X-ray samples
for  characterising the  statistical  properties of  AGN  over a  wide
luminosity and obscuration baselines.  In the calculations that follow
we adopt  cosmological parameters  $\rm H_0  = 70 \,  km \,  s^{-1} \,
Mpc^{-1}$, $\Omega_M = 0.3$, $\Omega_\Lambda = 0.7$.

\section{Data analysis and products}

\subsection{X-ray source catalogue}

The X-ray  source catalogue used in  this paper is  extracted from the
equatorial  field   ($\alpha  \approx  2^{h}\,16^{m}$,  $\delta\approx
-4\degr  \,52\arcmin$) of  the  XMM-XXL survey  \citep[][]{Pierre2016}
following  the X-ray data  reduction and  analysis steps  described in
\cite{Liu2016}. In the  rest of this paper we  refer to the equatorial
subregion  of  the   XMM-XXL   presented  by \cite{Liu2016}  as
XMM-XXL-N.   The most  important details  of the  X-ray  data analysis
methodology are outlined below.  

For the reduction of the  {\it XMM} observations of the equatorial XXL
field the {\it XMM} Science  Analysis System (SAS) version 12 is used.
Event   files   for    the   EPIC   \citep[European   Photon   Imaging
Camera;][]{Struder2001, Turner2001} PN  and MOS detectors are produced
using  the  {\sc  epchain}  and  {\sc  emchain}  tasks  of  {\sc  sas}
respectively. High particle background intervals associated with solar
flares are filtered out to produce  clean event files as well as X-ray
images and exposure maps in five energy bands ($0.5-8$, full; $0.5-2$,
soft;  $2-8$,  hard;  $5-8$,  very hard;  $7.5-12$\,keV,  ultra-hard).
X-ray Sources are independently detected in each of the spectral bands
above by  applying a Poisson false-detection  probability threshold of
$P<4\times10^{-6}$.  The {\it XMM}  astrometric frame is refined using
the  {\sc  eposcorr}  task  of  {\sc sas}  and  adopting  as  external
astrometric    reference   the    SDSS   \citep[Sloan    Digital   Sky
Survey;][]{Gunn2006} DR8  catalogue \citep{Aihara2011}.  The resulting
positional  accuracy of  X-ray  sources is  about 1\farcs5  ($1\sigma$
rms).   The final  X-ray  source catalogue  consists  of 8,445  unique
sources detected in at least one of the above spectral bands.  In this
paper we focus on the hard band (2-8\,keV) selected sample which has a
total of 2768 unique sources.

\begin{figure}
\begin{center}
\includegraphics[height=0.85\columnwidth]{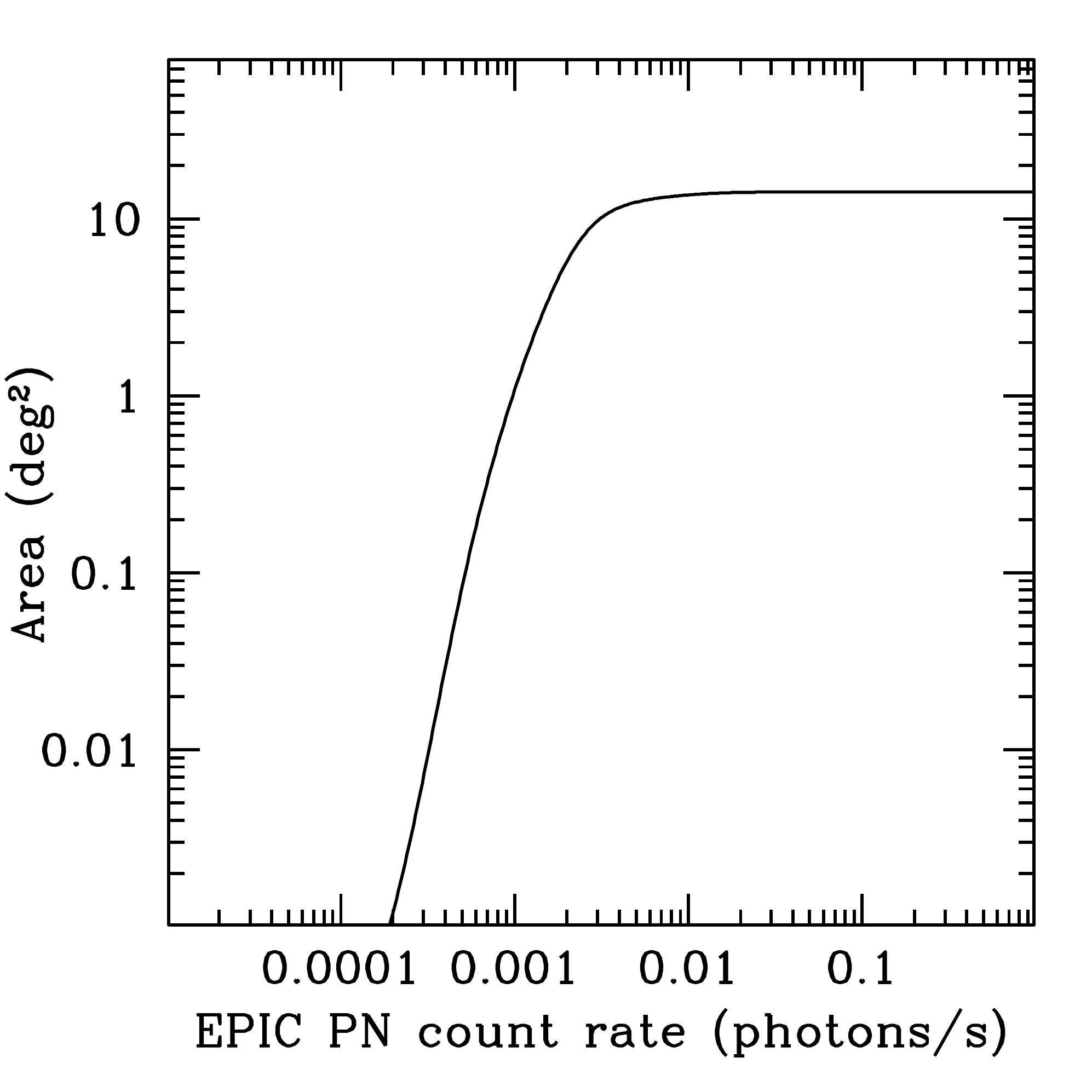}
\end{center}
\caption{XMM-XXL-N survey area (y-axis)  sensitive to sources of a given
EPIC  PN count  rate  in the  2-8\,keV  energy band  (x-axis). In  the
sensitivity map calculation only the part of the XMM-XXL-N that overlaps
with    the   Canada-France    Hawaii    Telescope   Lensing    Survey
\protect\citep[CFHTLenS;][]{Heymans2012, Erben2013}  is used (total of
about $\rm 14\,deg^2$).   The area of the XMM-XXL-N survey that lies
either outside  the CFHTLenS footprint  or within the  CFHTLenS masked
regions associated with optically bright stars is excluded (about $\rm
5\,deg^2$).    This  is  because   the  CFHTLenS   is  used   for  the
identification  of  XMM-XXL-N  sources  with optical  counterparts  (see
Section \ref{sec_optids}). }\label{xmm_area_curve}
\end{figure}

The sensitivity of the survey  is estimated using methods described in
\cite{Georgakakis2008_sense}     and    \cite{Georgakakis_Nandra2011}.
Figure  \ref{xmm_area_curve}  plots  the  XMM-XXL-N area  curve  as  a
function of  EPIC PN count rate.  This curve measures  the total X-ray
survey area that is sensitive to  sources of a given count rate in the
2-8\,keV (hard) band.

\subsection{Optical and near-infrared photometry}\label{sec_optids}

\cite{Liu2016} present optical identifications of the X-ray sources in
the equatorial field  of the XMM-XXL-N survey field  with the SDSS-DR8
photometric   catalogue   \citep{Aihara2011}.    In  this   paper   we
cross-correlate  the X-ray  source positions  with the  deeper optical
photometric catalogue ($ugriz$ bands; AB limiting magnitude $r \approx
24.9$\,mag)  of  the  Canada-France  Hawaii Telescope  Lensing  Survey
\citep[CFHTLenS;][]{Heymans2012, Erben2013}.  The association of X-ray
sources  with counterparts was  based on  the likelihood  ratio method
\citep{Sutherland_and_Saunders1992, Ciliegi2003, Brusa2007, Laird2009,
Luo2010, Xue2010, Georgakakis_Nandra2011}.

Potential counterparts  are searched for  within a radius  of 4\farcs5
from  the X-ray  centroid,  i.e. out  to  a distance  three times  the
$1\sigma$  X-ray  positional   rms  uncertainty.   For  each  possible
identification  we  measure  the  likelihood ratio,  LR,  between  the
probability  that the  source, at  the given  distance from  the X-ray
position and with the given optical magnitude, is the true counterpart
and the probability that the source is a spurious alignment

\begin{equation}\label{eq_lr} {\rm LR}=\frac{q(m)\,f(r)}{n(m)},
\end{equation}

\noindent where  $q(m)$ represents  the magnitude distribution  of the
true  optical counterparts  at  a  given waveband  and  $f(r)$ is  the
probability of finding  the true counterpart at distance  $r$ from the
X-ray  centroid, given  the  typical positional  uncertainties of  the
X-ray  and optical  catalogues. $n(m)$  is the  background  density of
galaxies  with   magnitude  $m$  at   a  given  waveband.    For  each
X-ray/optical  source  pair the  LR  is  estimated  separately in  the
$ugriz$ optical bands  of the CFHTLenS. The pair  is then assigned the
maximum LR of the 5 wavebands.

We  approximate  $f(r)$  by  the  normal  distribution  with  standard
deviation 1\farcs5.  In this calculation we assume that the positional
uncertainty  of the  optical catalogue  is much  smaller than  that of
X-ray  sources.  The  probability $q(m)$  is estimated  by subtracting
from the  magnitude distribution of  all possible counterparts  of all
X-ray  sources  within  a  search  radius  of  4\farcs5  the  expected
magnitude  distribution of  background/foreground galaxies  and stars.
For the latter, we randomly place 4\farcs5 radius apertures within the
survey  area  to  construct  the expected  magnitude  distribution  of
optical sources in random sight-lines.

We assess how secure the optical counterpart of an X-ray source is via
the          reliability         parameter          defined         by
\cite{Sutherland_and_Saunders1992}

\begin{equation}\label{eq_rel} 
\rm Rel =\frac{LR}{\sum_{j} LR_j+(1-Q)},
\end{equation}

\noindent where  the summation is over all  the potential counterparts
of the  X-ray source within the  search radius.  Q is  the fraction of
X-ray  sources with  identifications to  the limiting  magnitude  of a
given  CFHTLenS  waveband.   We  define  as  sample  completeness  the
fraction of  X-ray sources with counterparts above  a given likelihood
threshold, $LR>LR_{th}$.   Sample reliability is then the  mean Rel of
the all counterparts with $LR>LR_{th}$.  We choose the value $LR_{th}$
that  maximises  the  sum   of  the  sample  completeness  and  sample
reliability. We adopt $LR_{th}=0.4$ and  find a total of 4673 CFHTLenS
counterparts out of  5914 XMM-XXL-N X-ray sources that  lie within the
CFHTLenS  footprint  ($\approx   80$\,\%  identification  rate),  i.e.
accounting for the  geometry of the optical survey  and masked regions
associated  with   bright  stars.   In  the  case   of  the  hard-band
($2-8$\,keV) selected sample, which is  the focus of this paper, there
are  1778 CFHTLenS counterparts  out of  2018 XMM-XXL-N  X-ray sources
that   lie   within   the   CFHTLenS   footprint   ($\approx   88$\,\%
identification  rate).  The expected  spurious identification  rate is
about $5\%$.  In  the analysis that follows we  use the 2018 XMM-XXL-N
X-ray sources that overlap with the CFHTLenS footprint.

For   the  estimation   of  photometric   redshifts  (see   below)  we
complemented the CFHTLenS optical data with far-UV, near-infrared, and
mid-infrared  photometry  available   in  the  XMM-XXL-N  field.   The
positions  of  the  CFHTLenS  counterparts  are  cross-matched  within
3\farcs5 to GALEX (Galaxy  Evolution Explorer) sources using the GALEX
All-Sky        Imaging       Survey        Data        Release       5
\cite[GALEX-GR5;][]{Bianchi2011}. The probability  of a spurious GALEX
source  within the search  radius is  about 1.5\%.   Near-infrared
photometry is from public  extragalactic surveys that are been carried
out  by  the  Visible  and  Infrared Survey  Telescope  for  Astronomy
\citep[VISTA;][]{Emerson2004},  the  European  Southern  Observatory's
(ESO) 4-m class telescope dedicated  to imaging the southern sky.  The
VISTA        Deep       Extragalactic        Observations       survey
\citep[VIDEO;][]{Jarvis2013} provides photometry in the $ZYJHKs$ bands
to the depth  $Ks(AB)\approx24$\,mag over a total area  of $\rm 12\,deg^2$
split in three  distinct fields. The VISTA Hemisphere  Survey (VHS; PI
McMahon, Cambridge, UK) will image the entire southern sky in at least
the   $J$   and   $Ks$    filters   to   a   limiting   magnitude   of
$Ks(AB)\approx20$\,mag,  with  certain regions  of  the  survey, like  the
XMM-XXL-N, also observed  in the $H$-band.  The data  we used are from
the  4th Data  Release (DR4)  of  the VIDEO  survey and  the 3rd  Data
Release (DR3) of the VHS  produced by the Vista Science Archive (VSA),
which maintains  data products generated by the  VISTA InfraRed CAMera
(VIRCAM).  The VISTA Data  Flow System pipeline processing and science
archive  are  described  in  \cite{Irwin2004},  \cite{Hambly2008}  and
\cite{Cross2012}.  The VIDEO-DR4  covers about $\rm 3.6\,deg^2$ within
the  XMM-XXL-N  survey field,  while  the  VHS-DR3  covers the  entire
field.  The search radius  for cross-matching CFHTLenS positions with
VIDEO-DR4 and VHS-DR3 counterparts is  fixed to 1\farcs5.  For 99\% of
the  associations   the  radial   offsets  between  the   optical  and
near-infrared positions are  less than 0\farcs5.  At the  depth of the
VIDEO-DR4  survey ($Ks\approx24$\,mag) the  probability of  a spurious
near-infrared  galaxy within 0\farcs5  is 2\%.   The AllWISE  data are
used to compile mid-infrared photometry in the WISE W1, W2, W3, and W4
bands with central  wavelengths of 3.4, 4.6, 12, and  $\rm 22\, \mu m$
respectively.  The X-ray/AllWISE identification follows the likelihood
ratio  method as  described  in \cite{Menzel2016}.   Finally, we  also
cross-match the positions of the optical counterparts of X-ray sources
to  the SWIRE  (Spitzer Wide-area  InfraRed Extragalactic)  DR2 source
catalogue \citep{Surace2005}  using a search radius  of 1\farcs5.  The
Spitzer/SWIRE  survey covers  about  $\rm 9\,deg^2$  in the  XMM-XXL-N
field.  The  Spitzer photometry  is not used  in the  determination of
photometric  redshifts but  only  to explore  the  positions of  X-ray
sources on the Spizter mid-infrared colour-colour diagrams proposed in
the  literature  to select  AGN  (see  section \ref{sec_results}).   A
summary of  the multiwavelength  properties of the  XMM-XXL-N 2-8\,keV
selected sample is presented in Table \ref{table_data}.
 
\begin{table*}
\caption{Hard-band (2-8\,keV) selected sample of the XMM-XXL-N survey}\label{table_data}

\begin{tabular}{c c ccc ccc}

%\hline
%\multicolumn{9}{c}{Full-band selected sample.} \\
\hline
Total number of & X-ray  sources   & Number of      &  VIDEO         &
AllWISE       & GALEX      & {\it Spitzer}    & spectroscopic\\
X-ray sources   & within CFHTLenS  & CFHTLenS IDs   &  photometry    &  photometry  & photometry   & photometry & redshifts    \\

 (1) & (2) & (3) & (4) & (5) & (6) & (7) & (8) \\
\hline

 2768    & 2018  & 1778 &  605 & 1503 & 473 & 944 & 936 \\ 

\hline

\end{tabular}
\begin{list}{}{}
\item  (1) Total  number of  X-ray  sources detected  in the  2-8\,keV
(hard) energy  band in  the XMM-XXL-N sample.  (2) Number  of 2-8\,keV
detected XMM-XXL-N sources that lie within the good photometry area of
the CFHTLenS survey \protect\citep{Heymans2012, Erben2013}. (3) Number
of hard-band  detected X-ray sources with optical  associations in the
CFHTLenS  survey.  (4)  Number  of XMM-XXL-N  2-8\,keV detected  X-ray
sources  with  CFHTLenS counterparts  that  lie  within the  VIDEO-DR4
survey area.  (5) Number  of XMM-XXL-N 2-8\,keV detected X-ray sources
with CFHTLenS identifications and AllWISE counterparts.  (6) Number of
XMM-XXL-N    2-8\,keV   detected    X-ray   sources    with   CFHTLenS
identifications  and GALEX-GR5  \citep{Bianchi2011}  associations. (7)
Number  of XMM-XXL-N  2-8\,keV  detected X-ray  sources with  CFHTLenS
identifications   and   Spitzer/SWIRE-DR2   \protect\citep{Surace2005}
counterparts. (8) Number of  XMM-XXL-N 2-8\,keV detected X-ray sources
with CFHTLenS identifications and spectroscopic redshift measurements.
\end{list}
\end{table*}

\begin{table}
\begin{centering}
\caption{Summary of redshift measurements for the hard-band (2-8\,keV) selected sample of the XMM-XXL-N survey}\label{table_specdata}
\begin{tabular}{c c cc}
\hline
sample         & optically    &  \multicolumn{2}{c}{optically}    \\
               &  unresolved   &  \multicolumn{2}{c}{resolved}     \\
               &             &  CFTHLenS  & VIDEO      \\
   (1)         &    (2)      &    (3)     & (4)      \\
\hline
 spec-zs             &  580  &   264 & 92  \\
 photo-zs            &  331  &   367 & 144 \\
\hline
\end{tabular}
\begin{list}{}{}
\item  (1)  Photometric or  spectroscopic  redshift  subsample of  the
  2-8\,keV (hard) selected XMM-XXL-N  sources.  (2) X-ray sources with
  unresolved  optical  light  profile (point-like).  Sources  in  that
  sample  without  spectroscopy   are  assigned  photometric  redshift
  probability   density  functions   derived  from   the  $dN/dz$   of
  spectroscopically  confirmed sources  (see text  for details).   (3)
  X-ray  sources  with  resolved  optical light  profile  and  without
  available deep  near-infrared photometry from the  VIDEO survey. (4)
  X-ray sources with resolved optical light profile and available deep
  near-infrared photometry from the VIDEO survey.
\end{list}
\end{centering}
\end{table}

\subsection{Optical spectroscopy}

There are  a number of  spectroscopic surveys in the  XMM-XXL-N survey
area  targeting  various classes  of  extragalactic populations,  both
galaxies and  AGN.  A  dedicated follow-up spectroscopic  programme of
X-ray sources in the XMM-XXL-N equatorial field was carried out by the
Sloan  telescope   as  part   of  the  SDSS-III   ancillary  programme
\citep{SDSS_DR12}.   A detailed  description of  these  data including
visual  inspection and  redshift quality  assessment are  presented in
\cite{Menzel2016}.   The XXL  equatorial  field also  lies within  the
footprint  of  the SDSS-III  Baryon  Oscillation Spectroscopic  Survey
(BOSS; Dawson  et al.  2013)  programme, which provides  redshifts for
UV/optically-selected  broad-line  QSOs  and  luminous  red  galaxies.
\cite{Stalin2010} presents redshifts for X-ray sources in the original
XMM-LSS survey  field \citep{Clerc2014}, which  is part of  the larger
XMM-XXL  survey.   The  VIMOS  Public  Extragalactic  Redshift  Survey
\citep[VIPERS,][]{Garilli2014,   Guzzo2014}  targets  galaxies   to  a
magnitude limit  of $i(AB)=22.5$\,mag,  which are pre-selected  on the
basis of their optical colours to maximise the number of spectroscopic
identifications  in the redshift  interval $z  = 0.5-1.2$.   The first
public data release  of the VIPERS includes a  total of 30,523 sources
in the equatorial  XMM-XXL-N field and covers about  $\rm 5\,deg^2$ of
the    X-ray   survey    area.     The   VIMOS    VLT   Deep    Survey
\citep[VVDS;][]{LeFevre2013}  provides spectroscopy for  about 10\,000
optically-selected sources to $i(AB) =  24$\,mag over an area of about
$\rm  0.5\,deg^2$ within  the  XXL equatorial  field.   The ESO  large
programme  180.A-0776 (P.I.  O.   Almaini) followed  with spectroscopy
about  3\,000  galaxies  to  $K(AB)=23$\,mag selected  in  the  UKIDSS
Ultradeep Survey Field (UDS).   A description of these observations is
presented in \cite{Bradshaw2013} and \cite{McLure2013}.

The  spectroscopic  catalogues are  matched  to  the CFHTLenS  optical
positions  within a  radius of  1\,arcsec.  We  select  redshifts with
quality  flags  in  the  published  catalogues from  which  they  were
retrieved  that indicate  a probability  better than  $\approx90$\% of
being correct.  There are 2311 redshift measurements out of 4673 X-ray
sources with  CFHTLenS identifications.  The majority  (2012) are from
the SDSS spectroscopic programmes.   There are a further 129 redshifts
from  \cite{Stalin2010}, 91  from the  UDS follow-up  spectroscopy, 62
from  VIPERS   and  17  from   VVDS.   The  number   of  spectroscopic
identifications of the XMM-XXL-N 2-8\,keV selected sample is listed in
Table \ref{table_data}  and their breakdown to  optically resolved and
unresolved sources is shown in Table \ref{table_specdata}.

\subsection{Photometric redshifts}

For X-ray sources without optical spectroscopy we estimate photometric
redshifts using the {\sc LePhare} code \citep[Photometric Analysis for
Redshift  Estimate;][]{Arnouts1999, Ilbert2006}.   We  apply different
cuts to the X-ray selected sample to optimise the template fits to the
observed  Spectral Energy  Distributions (SEDs)  of certain  groups of
sources, and  improve the overall quality of  the photometric redshift
estimates.   We  present results  separately  for  X-ray sources  that
overlap with the VIDEO survey area and therefore benefit from the deep
near-infrared  photometry  of   that  programme.   Different  template
libraries  are  applied to  X-ray  sources  associated with  optically
resolved (extended optical  light profile) and unresolved (point-like)
counterparts.  This is motivated by the work of \cite{Salvato2009} who
found that  the morphological information of  the optical counterparts
of X-ray  sources can  be used as  a prior when  computing photometric
redshifts.  X-ray  AGN associated with optically  extended sources are
typically at  low and  moderate redshifts and  have SEDs with  a large
contribution  from  the  host  galaxy. Point-like  sources  have  SEDs
dominated by emission  from the central engine and  are likely to lie,
on  average,  at  higher  redshifts.   In addition  to  the  different
template   library   for   each   morphological  class,    we   follow
\cite{Salvato2009}  and  impose a  $B$-band  absolute magnitude  prior
$-30<M_B<-20$\,mag  when estimating the  photometric redshifts  of the
optically unresolved  sample. X-ray sources  with stellarity parameter
in the publicly available  CFHTLenS catalogues {\sc class\_star > 0.8}
are considered optically unresolved, and are likely to be dominated by
emission from  the AGN in  the optical bands.   It is also  found that
using  different  templates  for  samples  split by  X-ray  flux  also
improves the photometric redshift results. Previous studies have shown
that the  template libraries  used to determine  photometric redshifts
for    X-ray   selected    AGN   in    bright    surveys   \citep[e.g.
XMM-COSMOS][]{Salvato2009}  cannot be applied  to sources  detected at
fainter flux limits in deeper  samples, such as the 4\,Ms Chandra Deep
Field South \citep{Hsu2014}.

The  model  template  libraries  that  are  applied  independently  to
different  sub-samples of  X-ray sources  are (i)  the galaxy  SEDs of
\cite{Ilbert2009}, (ii)  the hybrid QSO/galaxy templates  presented by
\cite{Salvato2011} and (iii) the  templates developed by Tasnim Ananna
et al.  (in prep.)  for the determination of photometric redshifts for
X-ray  sources  in  the Stripe82X  survey  field  \citep{LaMassa2016}.
Extinction is  added to the  templates as  a free parameter  using the
\cite{Calzetti2000} and  the \cite{Prevot1984} attenuation  laws.  The
photometry used  for X-ray sources in  different wavebands corresponds
to the ``total'' source flux estimates in the corresponding catalogue.
For the VIDEO and the VHS surveys  in particular we use either Kron or
Petrosian-type  magnitudes  (appropriate  for  resolved  sources),  or
fluxes  integrated within  fixed-size apertures  (corrected for  loses
because of the size of the Point Spread Function), which are listed in
the VSA databases.  Before estimating photometric redshifts we explore
potential systematic offsets among  photometric bands, which may arise
because  of e.g.  variations in  seeing  conditions as  a function  of
observing  time  and  wavelength,  differences in  the  definition  of
`total' magnitude  in different  catalogues.  Such  zero-point offsets
among wavebands are  estimated by {\sc LePhare} using a  total of 2463
spectroscopically    confirmed    galaxies   from    the    VIPERS-DR1
\citep{Garilli2014,  Guzzo2014}.   The  redshift   is  fixed  and  the
best-fit template for  each source is found.   Photometric offsets are
then estimated for  each waveband to minimize  the differences between
the model  templates and the  observed magnitudes.  These  offsets are
then applied to the observed  photometry prior to the determination of
photometric redshifts.  Spectroscopically confirmed X-ray selected AGN
are not the  optimal sample for this calculation  because of potential
variability issues  affecting different photometric bands  observed at
separate epochs.

\begin{figure}
\begin{center}
\includegraphics[height=0.85\columnwidth]{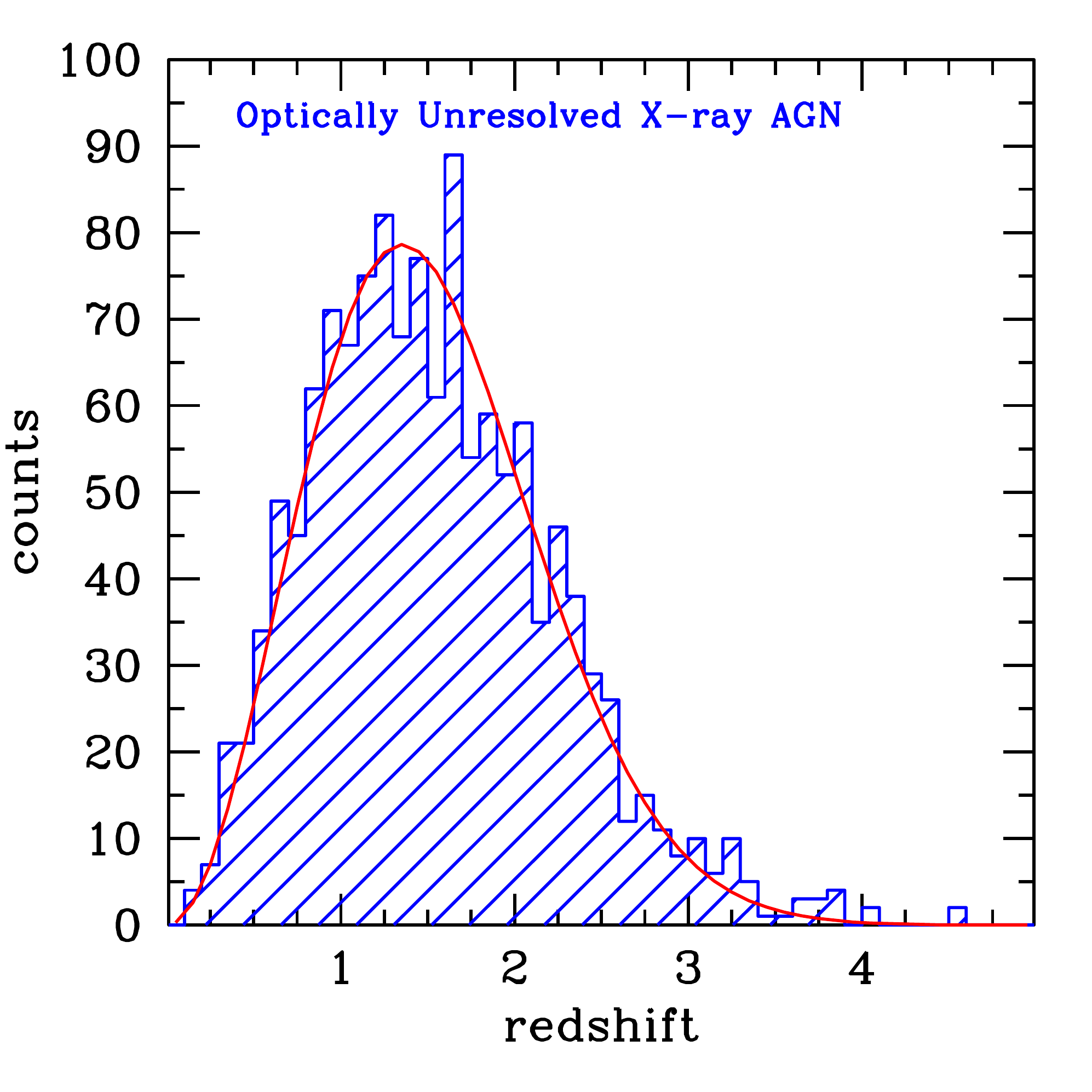}
\end{center}
\caption{Spectroscopic redshift distribution  of X-ray selected AGN in
the   XMM-XXL-N   field  with   unresolved   optical  light   profiles
(point-like).   The red  curve is  the best-fit  function of  the form
$dN(z)/dz \propto \exp[-(z-\bar{z})^2/2\,\sigma_z^2]$, where $\bar{z}$
and $\sigma_z$ are free parameters.  }\label{fig_nz_point}
\end{figure}

In the case of optically-unresolved  X-ray sources ({\sc class\_star >
  0.8})   we   use  the   hybrid   QSO/galaxy   template  library   of
\cite{Salvato2011}  for X-ray  fluxes  $f_X \rm  >10^{-13}  \, erg  \,
s^{-1} \,  cm^{-2}$, and the  templates of  Tasnim Ananna et  al.  (in
prep.) for fainter X-ray sources.  The quality\footnote{The quality of
  photometric   redshifts   is   quantified  by   the   rms   scatter,
  $\sigma_{NMAD}$,  of  the  quantity   $|z_{spec}  -  z_{phot}|/(1  +
  z_{spec})$  and the  outlier fraction,  $\eta$, defined  as $\eta  =
  |z_{spec} - z_{phot}|/(1  + z_{spec}) > 0.15$.}   of the photometric
redshift  measurements is  rather poor  with $\sigma_{NMAD}=0.07$  and
outlier  fraction  $\eta=27\%$.   These  numbers do  not  improve  for
sources with near-infrared information from  the VIDEO or VHS surveys.
\cite{Salvato2011} showed that for broad-line X-ray selected QSOs with
blue  featureless continua,  narrow-band  photometry as  well as  some
handle on the source variability  are essential to reduce $\eta$ below
the  10\% level.   Without  this additional  information  and for  the
spectral  bands  used in  this  work,  \cite{Salvato2011} estimate  an
expected catastrophic  failure redshift rate  of $\eta\approx20-40\%$,
similar to what is found here.  Another potential issue relates to the
separation  between optically  extended and  point-like sources  using
morphological  information  from seeing-limited  ground-based  images.
\cite{Hsu2014} for example, underlined  the importance of high spatial
resolution imaging from  e.g.  the Hubble Space  Telescope to identify
optically  unresolved  X-ray sources  and  apply  the correct  set  of
templates to the different mophological  classes.  The poor quality of
the  photometric redshift  estimates  for  optically unresolved  X-ray
sources  is   mirrored  by  the  corresponding   photometric  redshift
Probability  Distribution   Functions  (PDZs),  which   are  typically
broad. It  is found for example,  that for the majority  (89\%) of the
optically unresolved  X-ray AGN with available  spectroscopy, the true
redshift  (i.e.    spectroscopic)  lies   within  the  5th   and  95th
percentiles of the  estimated PDZ distribution.  We  wish to propagate
the uncertainties in the  photometric redshift estimates of point-like
X-ray AGN  in the  analysis. There  are at least  two ways  to achieve
that.  The  first is to use  directly the PDZs of  individual sources.
The second  exploits the  fact that most  of the  optically unresolved
XMM-XXL-N X-ray sources have spectroscopic redshift measurements (e.g.
580  out  911   in  the  hard-band  selected   sub-sample;  see  Table
\ref{table_specdata}), and therefore the  redshift distribution of the
population is well constrained (see Fig.  \ref{fig_nz_point}).  We may
therefore  assume  that  optically unresolved  X-ray  sources  without
spectroscopic redshifts  follow the  same redshift  distribution, $\rm
dN(z)/dz$, as  the spectroscopically confirmed part  of the population
in  the XMM-XXL-N  field.   This  latter approach  is  adopted in  the
analysis that follows.   We note however, that our results  on the AGN
space density (Section \ref{sec_results}) do  not change if instead we
adopt the PDZs of individual sources.   This is because the broad PDZs
produced by {\sc  LePhare} are representative of the  overall level of
uncertainty in  the determination  of photometric redshifts  for X-ray
sources with point-like optical-light profiles.

Figure \ref{fig_nz_point}  plots the spectroscopic  redshift histogram
of  X-ray  AGN  that  are  unresolved  in  the  optical  bands.   This
distribution   is   fit    with   the   function   $dN(z)/dz   \propto
\exp[-(z-\bar{z})^2/2\,\sigma_z^2]$,  where  $\bar{z}$ and  $\sigma_z$
are  free   parameters  (red  curve   in  Figure  \ref{fig_nz_point}).
Spectroscopically unidentified and  optically unresolved X-ray sources
are assigned a redshift PDZ that follows the
relation  above.   We  caution  that the  optically  unresolved  X-ray
sources  without  spectroscopic redshifts  extend  to fainter  optical
magnitudes   (median   $r    \approx   22$\,mag)   compared   to   the
spectroscopically  confirmed part of  the population  (median $\approx
20.5$\,mag).  Nevertheless, we find  that the redshift distribution of
the optically unresolved X-ray sources  with secure redshifts is not a
strong function of magnitude.  The  median and 68th percentiles of the
distribution  increase  from  $z=1.4^{+0.8}_{-0.6}$ for  sources  with
$r<21.5$\,mag to $z=1.7^{+0.6}_{-0.7}$ for $r>22$\,mag.  These numbers
are  broadly confirmed  using  the recent  COSMOS-Legacy X-ray  source
catalogue \citep{Marchesi2016}.  We select  a total of about 100 X-ray
sources from that sample with relatively bright X-ray fluxes $f_X( \rm
2-10  \, keV  ) >  10^{-14} \,  erg \,  s^{-1} \,  cm^{-2}$ similar  to the
XMM-XXL-N  limit,  optically   faint  counterparts  $r>22.0$\,mag  and
unresolved  optical  light  profiles.  The  redshift  (photometric  or
spectroscopic)  distribution of  this  subset has  a  median and  68th
percentiles $z=1.5^{+0.4}_{-0.3}$.

\begin{figure*}
\begin{center}
\includegraphics[height=0.85\columnwidth]{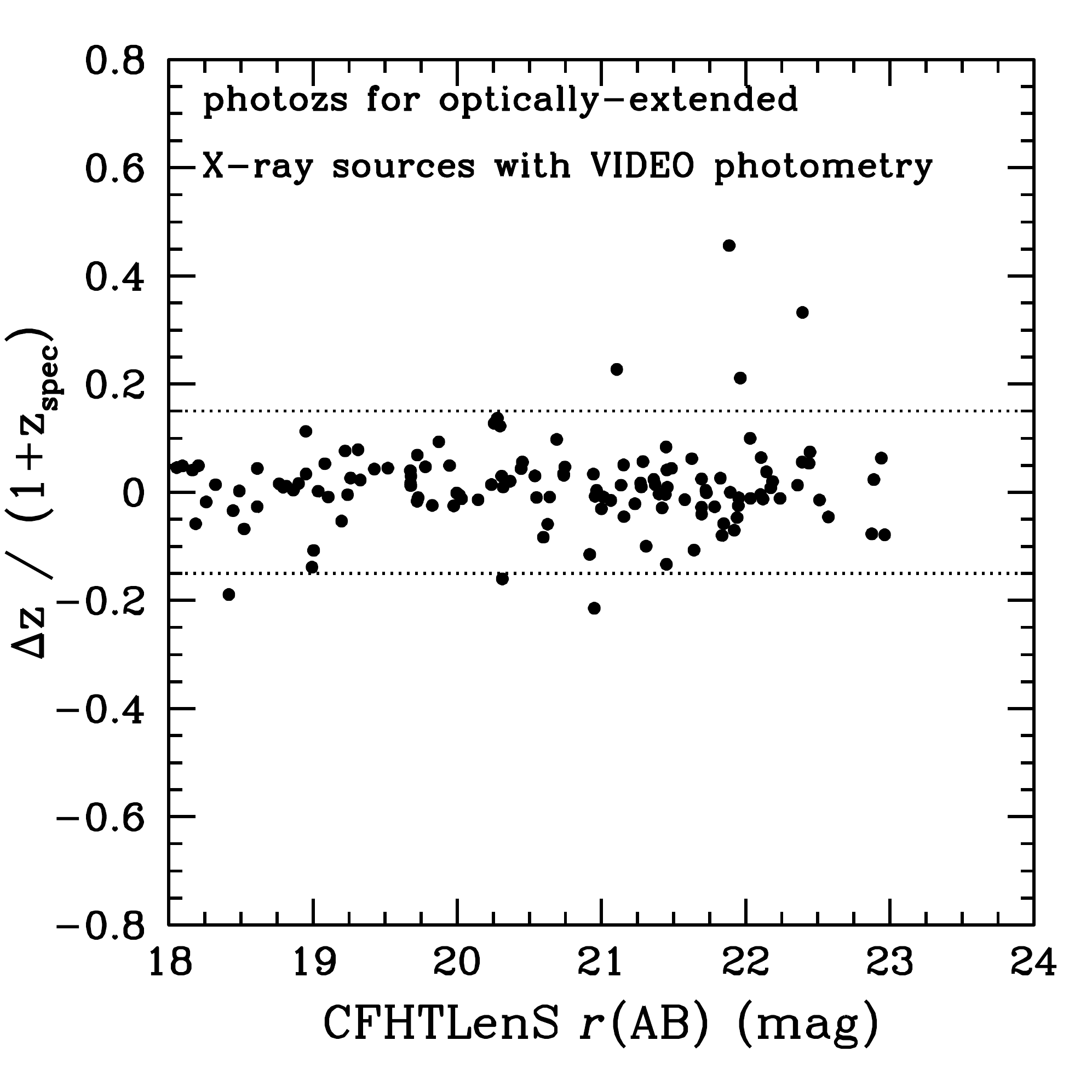}
\includegraphics[height=0.85\columnwidth]{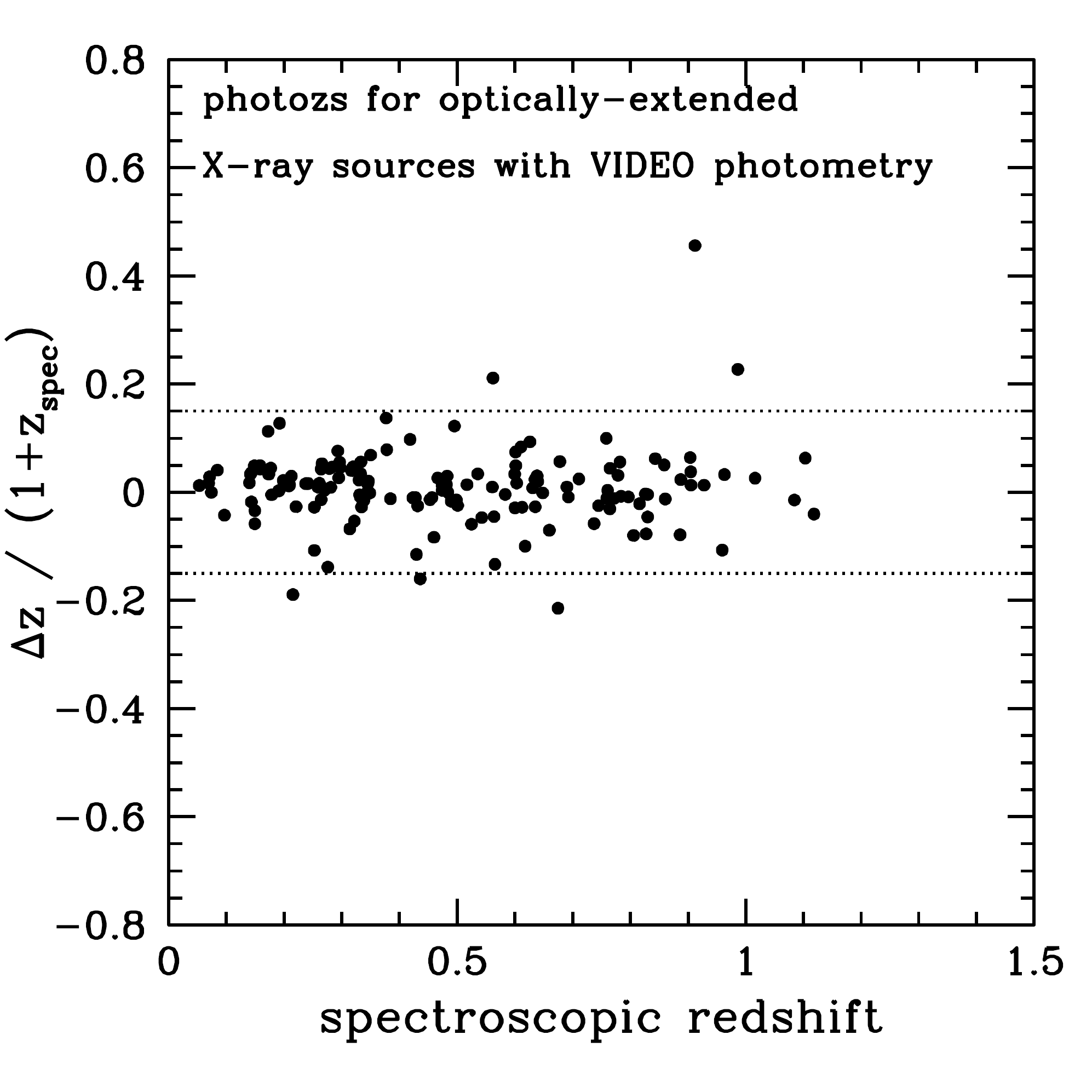}
\end{center}
\caption{The  quantity $(z_{phot}  - z_{spec})  / (1  +  z_{spec})$ is
plotted as a function of  CFHTLenS $r$-band magnitude (left panel) and
spectroscopic redshift  (right panel) for  spectroscopically confirmed
XMM-XXL-N AGN with extended optical light  profiles that  lie within
the VIDEO-DR4 area. The horizontal dotted lines mark   the
catastrophic       failure        limit,       $\eta>0.15$. 
}\label{fig_qphotoz_video}
\end{figure*}

X-ray sources with extended  optical profiles (CFHTLenS parameter {\sc
class\_star    <    0.8})    are    likely    to    have    broad-band
optical/near-infrared emission dominated  by stellar light with little
contamination  from  the central  engine.   For  these sources  normal
galaxy  templates  may  provide  an  adequate  representation  of  the
broad-band  SEDs.  Experimentation  showed that  the  best photometric
redshift results  for this class  of sources are obtained  by adopting
the \cite{Ilbert2006} normal galaxy  template library for sources with
X-ray fluxes  $f_X<\rm 10^{-14} \, erg  \, s^{-1} \,  cm^{-2}$ and the
hybrid  QSO/galaxy templates developed  by Tasnim  Ananna et  al.  (in
prep.)  for  brighter sources.   For optically extended  X-ray sources
with  additional near-infrared  photometry  from the  VIDEO survey  we
estimate photometric redshift  quality measures $\sigma_{NMAD} = 0.06$
and $\eta=4.5\%$.  Figure \ref{fig_qphotoz_video} explores further the
quality of  the photometric  redshifts for X-ray  AGN by  plotting the
quantity $(z_{phot}  - z_{spec})  / (1 +  z_{spec})$ as a  function of
spectroscopic   redshift   and   $r$-band  optical   magnitude.    The
catastrophic  redshift failure  rate does  not appear  to  change with
optical  magnitude, at  least  to the  limit $r\approx23$\,mag,  where
sufficient   numbers  of   spectroscopically   confirmed  sources   is
available.   The  fraction  of  catastrophic  redshift  failures  also
appears to  be stable with redshift  at least out  out to $z\approx1$.
In  the analysis that  follows optically  extended X-ray  sources with
VIDEO  near-infrared  photometry  are  assigned  the  PDZs  determined
directly by the {\sc LePhare} code.

\begin{figure}
\begin{center}
\includegraphics[height=0.85\columnwidth]{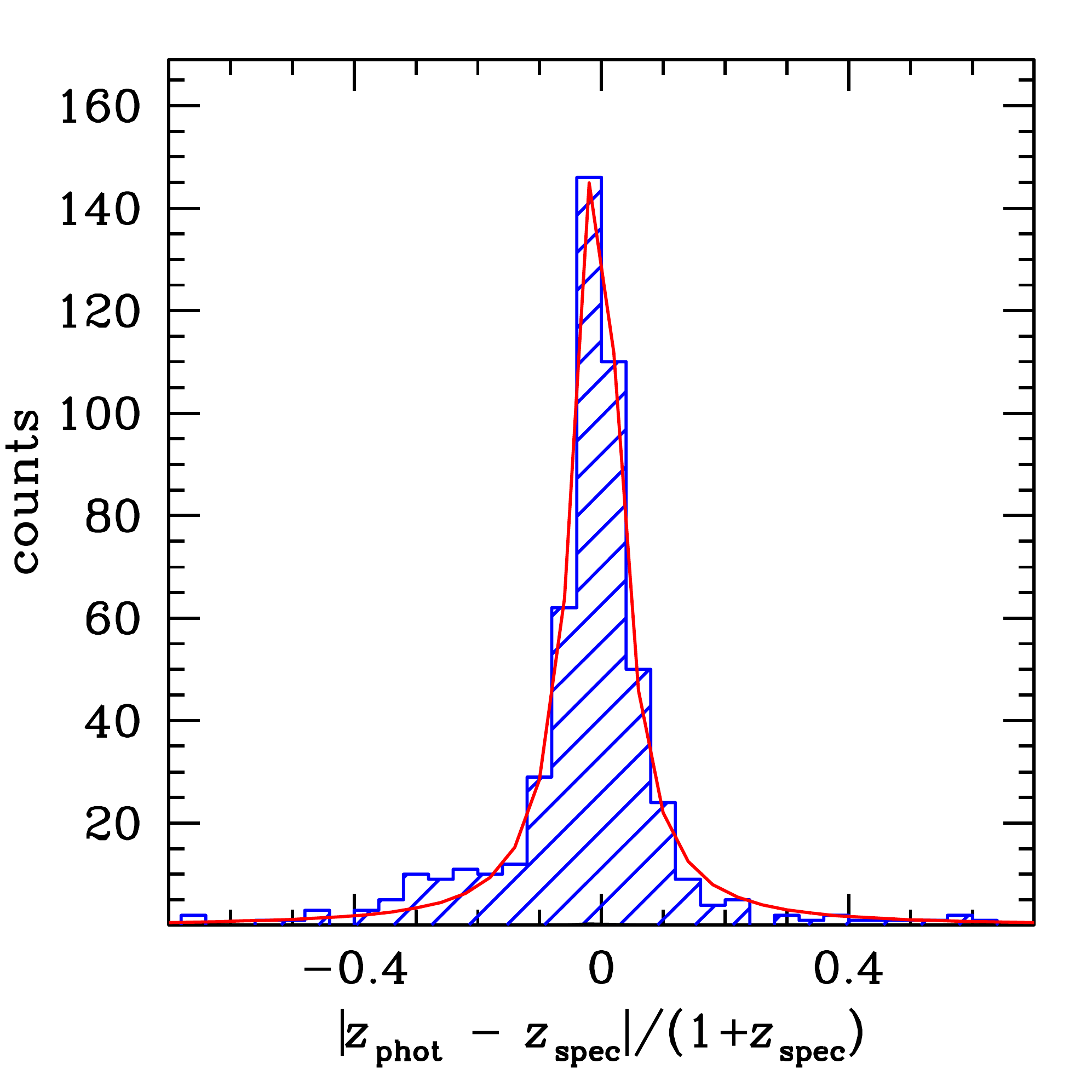}
\end{center}
\caption{Histogram  of the  quantity  $(z_{phot} -  z_{spec}) / (1  +
z_{spec})$ for XMM-XXL-N X-ray  sources that (i) lie outside the footprint
of  the VIDEO-DR4 deep near-infrared survey,  (ii) have counterparts
with  extended  optical  light  profiles and  (iii) have spectroscopic
redshift  measurements.  The red  curve  is  the  best-fit  Lorentzian
distribution  to the  observed  histogram.  The  broad  wings of  that
distribution  provide  a  reasonable  representation  of  the  overall
uncertainty  of  the  photometric  redshifts,  including  catastrophic
failures.}\label{fig_nz_ext_lens}
\end{figure}

The availability  of deep infrared  photometry in addition  to optical
data is key for good  quality photometric redshift measurements in the
case  of optically  extended X-ray  sources.  For  the subset  of this
population that  lies outside the  VIDEO near-infrared survey  area we
estimate photometric redshift quality  measures $\sigma_{NMAD} = 0.06$
and  $\eta=16\%$. The  VHS survey  near-infrared photometry,  which is
available   over   the   entire    XMM-XXL-N   field,   is   shallower
($Ks\approx20$\,mag) than  the VIDEO survey  ($Ks\approx24$\,mag), and
does   not  substantially   improve  the   photometric  redshifts   of
optically-extended   X-ray  sources.    Also,   the  relatively   high
catastrophic redshift fraction is not  represented in the broadness of
the  PDZs estimated  by the  {\sc LePhare}.   Only about  half of  the
optically extended X-ray sources outside the VIDEO area with available
optical  spectroscopy have  PDZs with  5th and  95th percentiles  that
bracket the true  spectroscopic redshift. Therefore for  this class of
sources we choose a different approach for propagating the photometric
redshift  uncertainties   in  the   analysis  that   follows.   Figure
\ref{fig_nz_ext_lens}   shows  the   distribution   of  the   quantity
$|z_{phot}  -  z_{spec}|   /  (1  +  z_{spec})$  for   the  sample  of
spectroscopically  confirmed and  optically  extended XMM-XXL-N  X-ray
sources  outside  the VIDEO  area,  where  the parameters  $z_{phot}$,
$z_{spec}$  in  the  relation  above   are  the  photometric  and  the
spectroscopic-redshift  measurements, respectively.   The distribution
of this quantity  can be represented with a Lorentzian  that has wings
which  are  sufficiently  broad  and  consistent  with  the  estimated
catastrophic  redshift fraction,  $\eta=16\%$.  In  the analysis  that
follows uncertainties and  potential systematics affecting photometric
redshift  measurements  for  optically  extended  X-ray  selected  AGN
without VIDEO  photometry are compensated  for, at least to  the first
approximation, by  using PDZs derived  from the Lorentzian  plotted in
Figure \ref{fig_nz_ext_lens}.

\begin{figure*}
\begin{center}
\includegraphics[height=0.85\columnwidth]{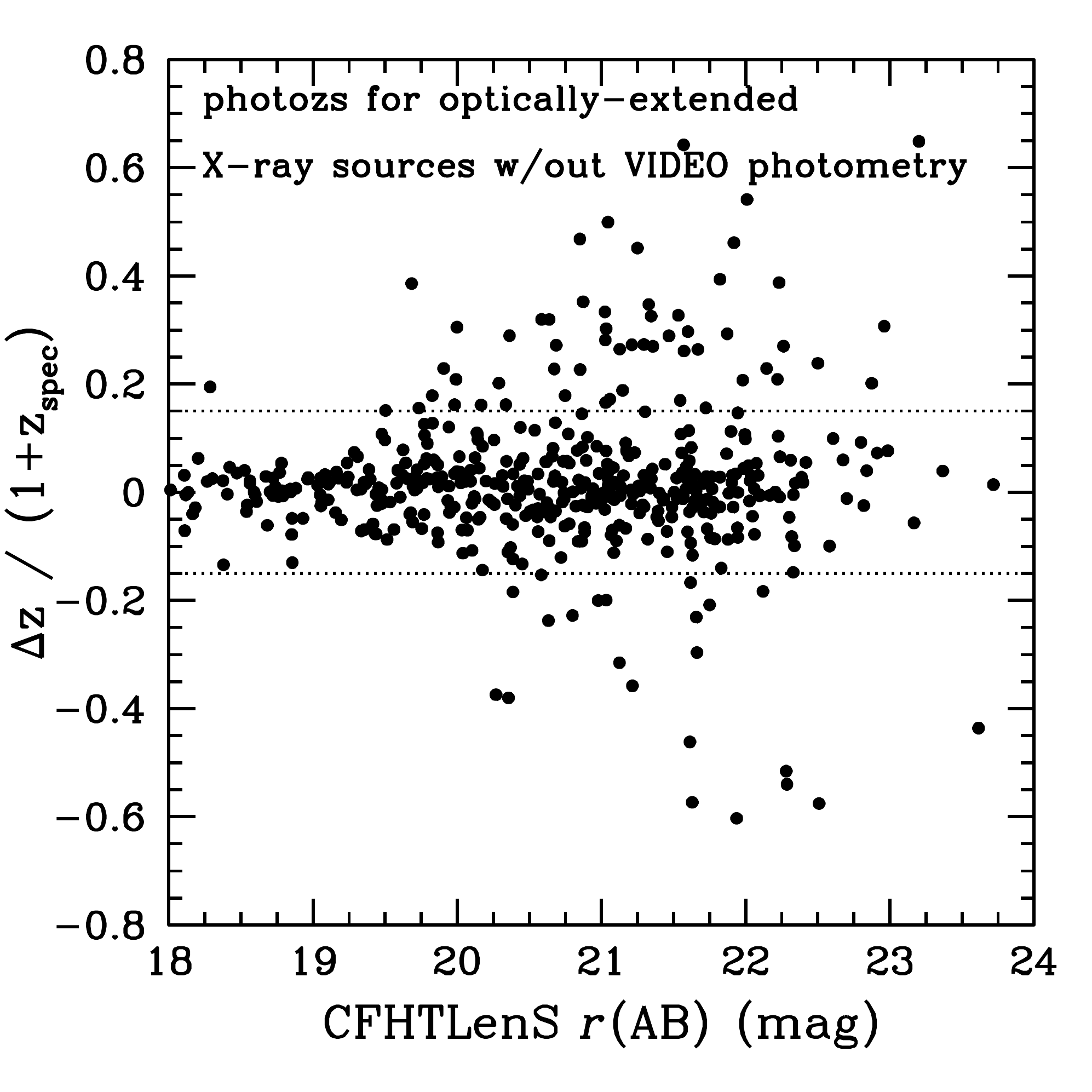}
\includegraphics[height=0.85\columnwidth]{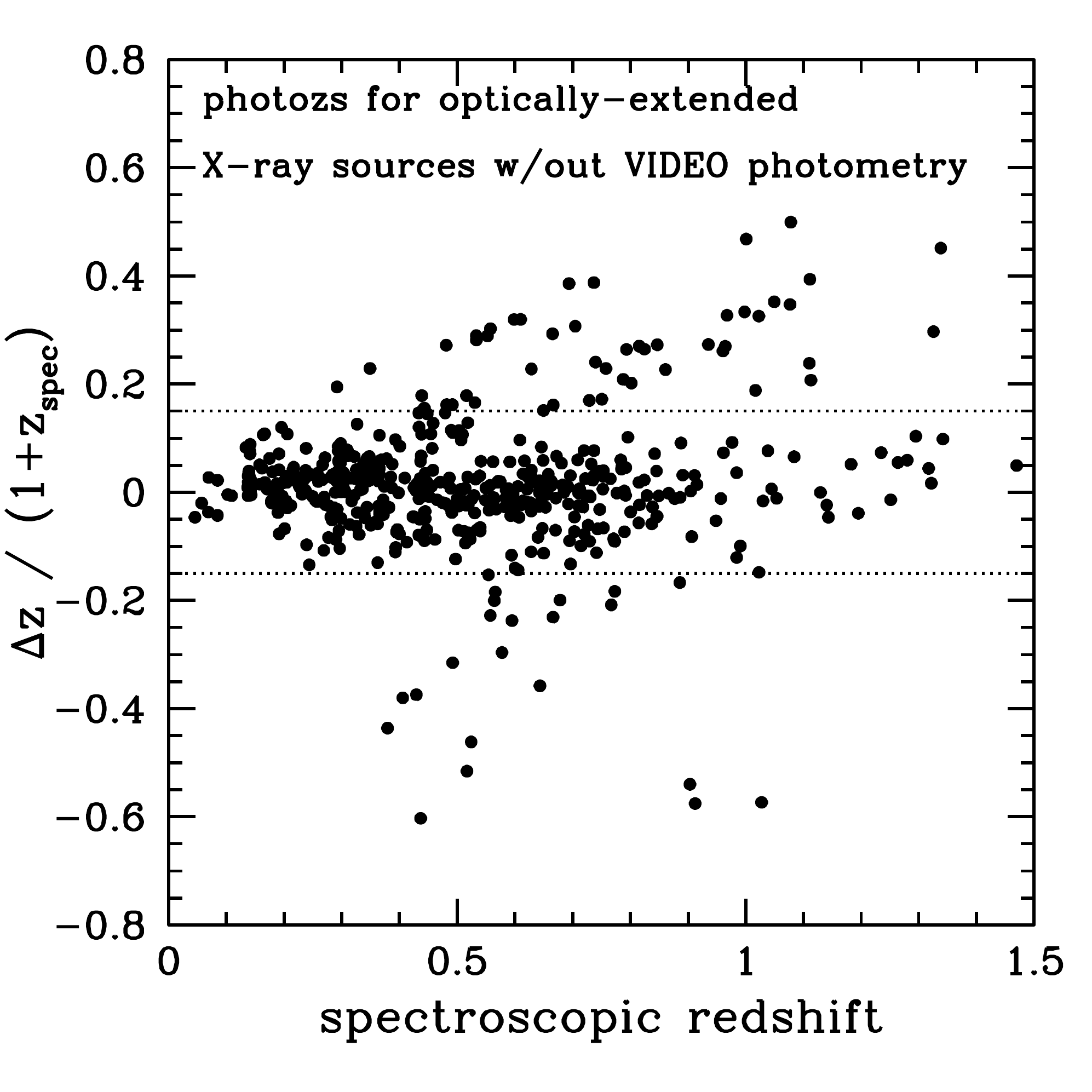}
\end{center}
\caption{The  quantity $(z_{phot}  - z_{spec})  / (1  +  z_{spec})$ is
plotted as a function of  CFHTLenS $r$-band magnitude (left panel) and
spectroscopic redshift  (right panel) for  spectroscopically confirmed
XMM-XXL-N AGN with extended optical light  profiles that lie outside
the VIDEO-DR4 footprint. The horizontal dotted lines mark
the        catastrophic       failure        limit,       $\eta>0.15$.
}\label{fig_qphotoz_lens}
\end{figure*}

In Figure \ref{fig_qphotoz_lens} we explore further the quality of the
photometric redshift  estimates for  optically extended  light profile
and without VIDEO  photometry. There is no strong  trend of increasing
$\eta$   at  faint   optical  magnitudes,   although  the   number  of
spectroscopically confirmed X-ray AGN beyond $r\approx22.5$\,mag drops
substantially  and  therefore  small   number  statistics  affect  any
conclusions.  The fraction of catastrophic redshift failures in Figure
\ref{fig_qphotoz_lens} appears  to increase at redshifts  $z\ga1$.  We
note however  the small number of  spectroscopic redshift measurements
of X-ray  sources with  extended optical  light profile  beyond $z=1$,
which has an impact on any conclusions.

\begin{figure}
\begin{center}
\includegraphics[height=0.85\columnwidth]{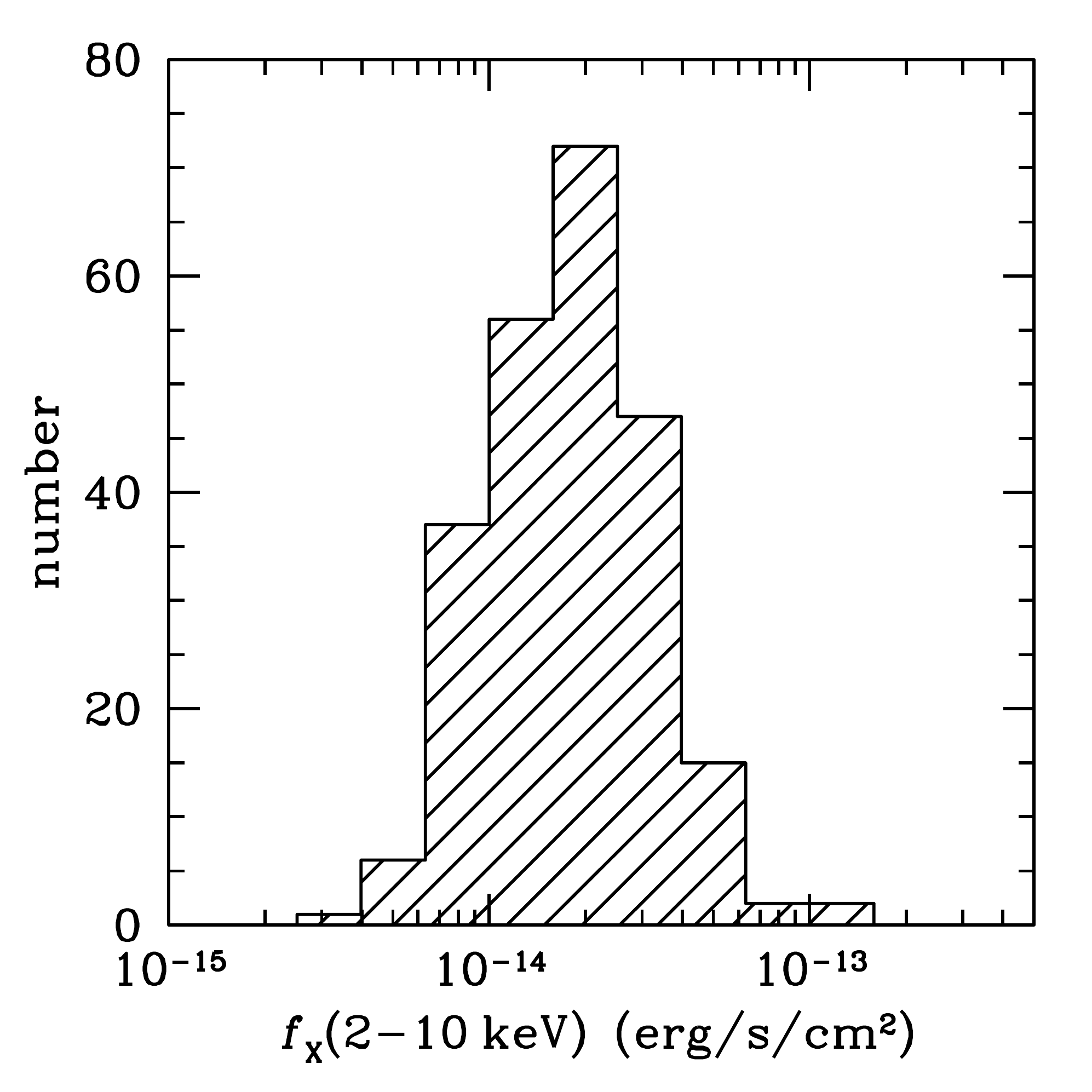}
\end{center}
\caption{2-10\,keV  X-ray  flux   distribution  of  XMM-XXL-N  sources
detected  in  the  2-8\,keV   spectral  band  without  secure  optical
counterparts in the CFHTLenS.  }\label{fig_xflux}
\end{figure}

\begin{figure}
\begin{center}
\includegraphics[height=0.85\columnwidth]{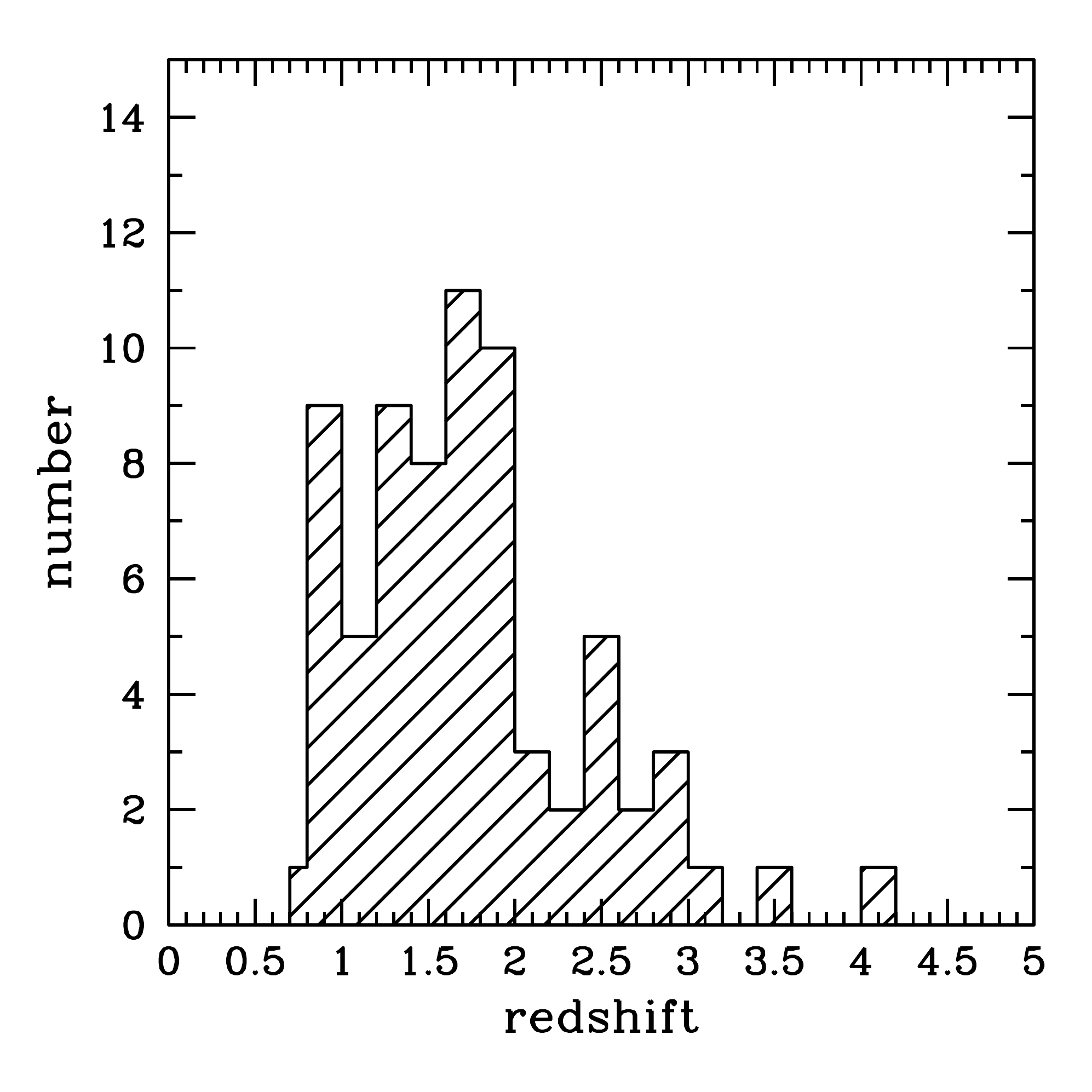}
\end{center}
\caption{Redshift distribution of X-ray bright, $f_X(\rm 2 - 10 \, keV
)  > 10^{-14}  \,  erg  \, s^{-1}  \,  cm^{-2}$, and  optically-faint,
$r>24.5$\,mag,   X-ray   sources   in   the   COSMOS   Legacy   survey
\protect\citep{Civano2016,  Marchesi2016}.  The  redshifts  are either
spectroscopic              or             photometric             from
\protect\cite{Marchesi2016}. }\label{fig_cosmos_dndz_faint}
\end{figure}

In addition to optically identified  X-ray sources in the sample there
are  also 240  2-8\,keV  detections in  the  XMM-XXL-N without  secure
counterparts  in  the   CFHTLenS  (12\%  of  the   sample;  see  Table
\ref{table_data}).   Figure \ref{fig_xflux}  presents  the X-ray  flux
distribution of  these sources. The majority  have $f_X(\rm 2 -  10 \,
keV  ) >  10^{-14}  \, erg  \,  s^{-1} \,  cm^{-2}$.   We explore  the
expected  redshift  distribution of  this  population  using the  $\rm
2\deg^2$-wide  COSMOS-Legacy survey  \citep{Civano2016, Marchesi2016}.
We select COSMOS-Legacy  survey X-ray sources with $f_X(\rm 2  - 10 \,
keV  ) >  10^{-14} \,  erg  \, s^{-1}  \, cm^{-2}$  and faint  optical
counterparts, $r>24.5$\,mag,  i.e. close to limiting  magnitude of the
CFHTLenS  survey data  used in  this paper.   We use  spectroscopic or
photometric redshifts for these sources listed in \cite{Marchesi2016}.
The   resulting    redshift   distribution   is   shown    in   Figure
\ref{fig_cosmos_dndz_faint}.   Most sources  are  in  the interval  at
$z=1-3$.   We  choose  to  assign XMM-XXL-N  sources  without  optical
counterparts  a weak  prior  that  is flat  in  the redshift  interval
$z=1-6$.  The  breakdown  of  redshift  measurements,  photometric  or
spectroscopic,  for   the  2-8\,keV  selected  XMM-XXL-N   sample  are
presented in Table \ref{table_specdata}.

\subsection{X-ray spectral analysis}\label{sec:xspectral}

The  X-ray  spectra  of  individual sources  are  extracted  following
methods described in \cite{Liu2016}.  Events with patterns up to 4 and
12  for the  PN and  MOS  detectors, respectively,  are included.  The
relevant ARF  and RMF  calibration files are  generated using  the SAS
tasks {\sc arfgen} and {\sc  rmfgen}. Because of the similar responses
of the  MOS1 and MOS2 CCDs,  the spectra from those  two detectors are
coadded into a single spectrum.

The  Bayesian   X-ray  Analysis   package  \citep[BXA;][]{Buchner2014,
  Buchner2015} is used to fit the X-ray spectra of individual sources.
The   adopted  model   includes  (i)   components  that   account  for
photoelectric  absorption   and  Compton  scattering   from  obscuring
material along the  line-of-sight that is distributed  with a toroidal
geometry  close  to  the  central supermassive  black  hole,  (ii)  an
independent soft component, which is  often observed in the spectra of
obscured  AGN \citep[e.g][]{Brightman_Nandra2012colours,  LaMassa2012,
  LaMassa2014, Buchner2015,  Lanzuisi2015} and could be  attributed to
Thomson  scattering  of  the  intrinsic  X-ray  emission  off  ionised
material       within       the        torus       opening       angle
\citep[e.g.][]{Guainazzi_Bianchi2007},   (iii)    Compton   scattering
(reflection)   of   radiation   on    dense   material   outside   the
light-of-sight. We adopt the torus model of \cite{Brightman2011_torus}
to approximate  the transmitted spectrum  of an AGN  through obscuring
material.  The  intrinsic spectrum  is assumed  to follow  a power-law
parametrised     by    the     spectral    index     $\Gamma$.     The
\cite{Brightman2011_torus} model assumes a  sphere of constant density
with two symmetric  conical wedges with vertices at the  centre of the
sphere removed. The opening angle of the cones is fixed to 45\degr and
the viewing angle of the observer is set to the maximum allowed in the
model,  87\degr,  i.e.   nearly  edge   on.   The  soft  component  is
approximated  by  a power-law  model  with  the same  spectral  index,
$\Gamma$,  as  the  intrinsic   power-law  spectrum.   The  reflection
component  is  modelled  with  the {\sc  pexrav}  model  presented  by
\cite{pexrav}. The  spectral index  of that component  is the  same as
that of the intrinsic power-law spectrum.

The adopted  model ({\sc  torus +  zpowerlw +  pexrav} in  {\sc xspec}
terminology)  has five  free parameters,  the slope  of the  intrinsic
power-law spectrum, $\Gamma$, the  line-of-sight column density of the
obscurer, $\rm  N_H$, the  normalisations of the  intrinsic power-law,
soft scattering and reflection components.  The adopted spectral model
is a  trade-off between a  reasonable representation of the  basic AGN
X-ray  spectral components  found in  local samples  and a  relatively
small  number of  free parameters  to  be constrained  from data  with
typically low  number of photons. \cite{Buchner2014}  also showed that
the adopted  spectral components are  the minimum needed  to represent
the X-ray properties of AGN in the {\it Chandra} Deep Field South.  We
impose that  the normalisations of  the soft power-law  and reflection
components  cannot exceed  10\% of  the intrinsic  power-law spectrum.
The redshift of the soft power-law and reflection component is tied to
that of  the torus model.   We use a  Gaussian prior for  the spectral
index   $\Gamma$  with   mean   1.95  and   standard  deviation   0.15
\citep{Nandra1994}. The  hydrogen column density of  the line-of-sight
obscuration  is  assigned  a  flat  prior  in  the  logarithmic  range
$\log_{10} (N_H/\rm cm^{-2})=20  - 25$.  The source  redshift is fixed
to the spectroscopic  value if available.  In the  case of photometric
redshift estimates the corresponding PDZs are used as priors. The {\it
  XMM}  PN  and  MOS  background  spectrum  is  fit  with  the  models
constructed by \cite{Maggi2014} that include an empirical instrumental
component   \citep{Sturm2012_PHD}  and   an  astrophysical   component
\citep{Kuntz_Snowden2010}.   The energy  range  used  in the  spectral
analysis is $0.5-8$\,keV.   The output of the  spectral analysis using
the  BXA  are  posterior  probability distribution  functions  in  the
multidimensional space of the spectral fit free parameters.  These are
then converted to posterior chains in the parameter space of intrinsic
AGN  luminosity [$\log  L_X( \rm  2 -  10 \,  keV)$], hydrogen  column
density ($\log N_H$) and redshift ($z$).

\begin{figure}
\begin{center}
\includegraphics[height=0.85\columnwidth]{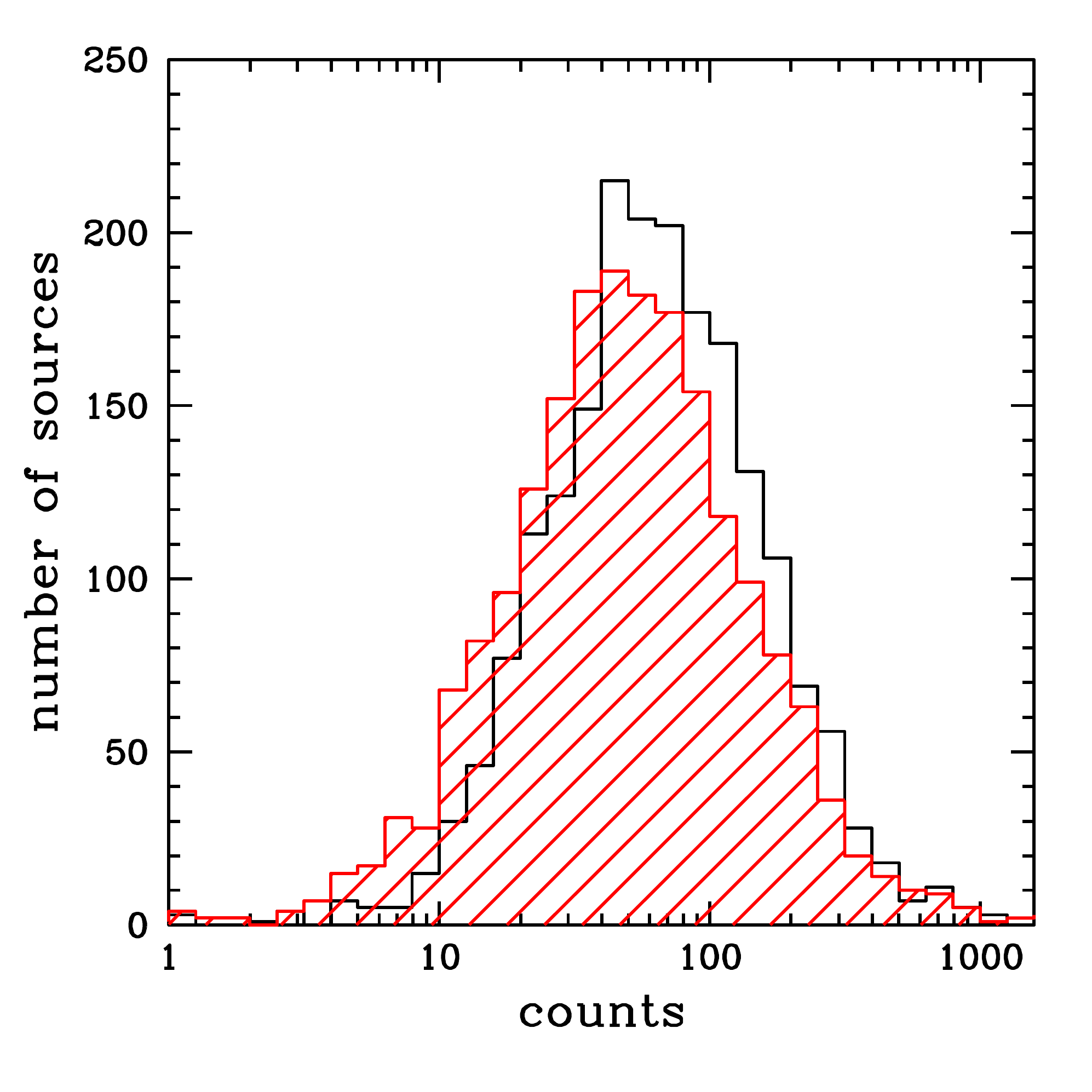}
\end{center}
\caption{ Distribution of spectral photon counts in the 0.5-8\,keV
energy range of XMM-XXL-N sources.  The open black histogram shows the
total  spectral counts, i.e.  source and  background, the  hatched red
histogram   corresponds  to   the  source-only   counts,   i.e.  after
subtracting   for each source the     mean    expected    background    contribution.
}\label{fig:hist_cnts}
\end{figure}

Figure \ref{fig:hist_cnts} plots the distribution of total (source and
background) spectral  counts in the  0.5-8\,keV band of  the XMM-XXL-N
X-ray sources.   The median  of this distribution  is 63  photons. For
completeness also  shown in  Figure \ref{fig:hist_cnts} is  the source
only (i.e. after subtracting  the expected background counts) spectral
count  histogram.  We  note  however, that  when  analysing the  X-ray
spectra  of individual  sources the  background is  not  subtracted but
fitted to the  data as a separate model  component.  The uncertainties
of  the derived  parameters (e.g.   $\log N_H$)  depend on  the energy
distribution of the photons (i.e.   overall spectral shape) and on the
accuracy of the adopted  redshift measurements (i.e.  spectroscopic vs
photometric).  Examples of confidence intervals for the $\log N_H$ and
$\log   L_X$  of  individual   sources  with   spectroscopic  redshift
measurements   are   presented   in  Figures   \ref{fig_example}   and
\ref{fig_example_type1} of Section \ref{sec_results}.

\begin{figure}
\begin{center}
\includegraphics[height=0.85\columnwidth]{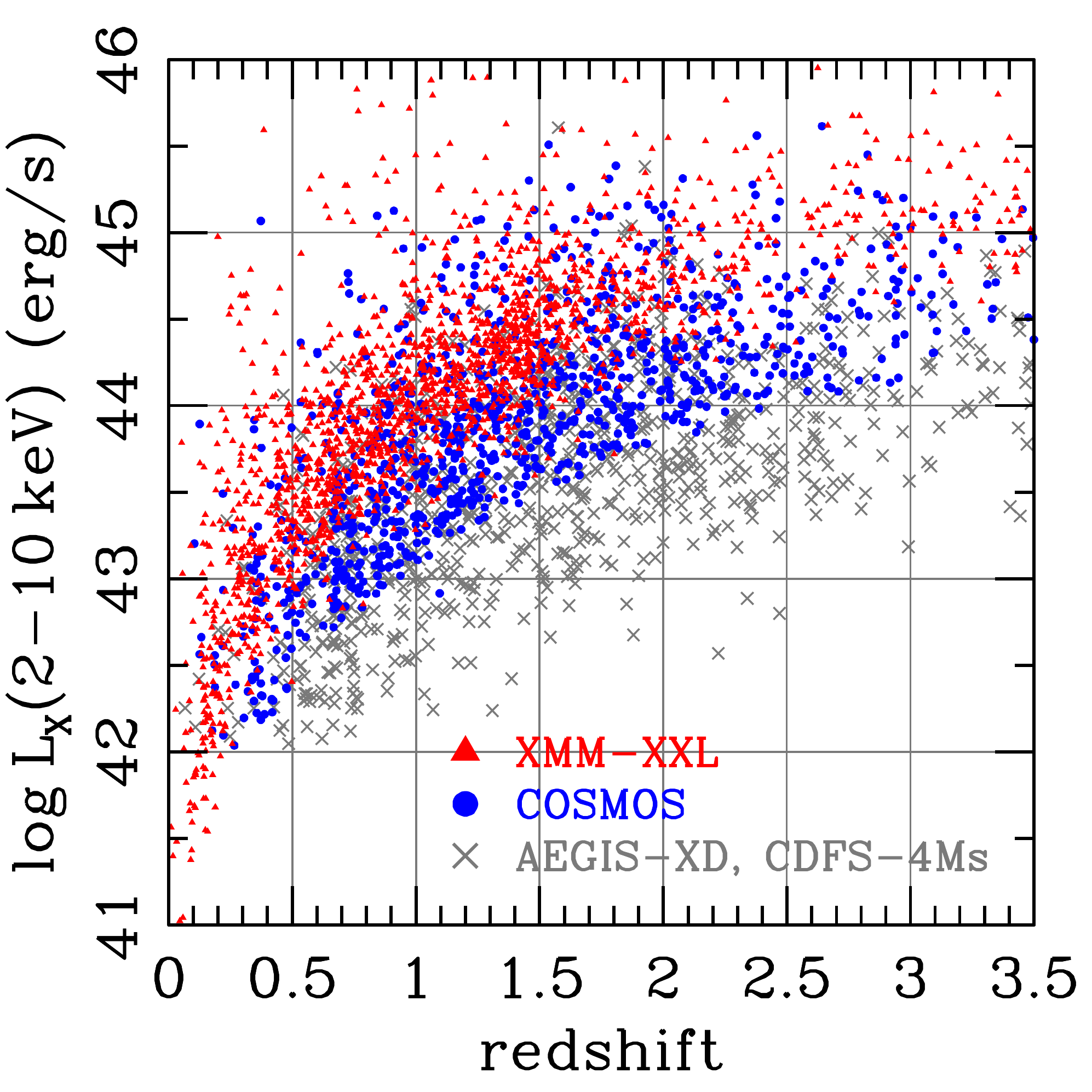}
\end{center}
\caption{   Intrinsic  X-ray   luminosity   (i.e.  corrected   for
line-of-sight  obscuration)  against   redshift  for  AGN  detected  in
different  X-ray  surveys.   The   red  triangles  correspond  to  the
XMM-XXL-N sample used in this work. The blue circles are X-ray sources
detected  in the  Chandra COSMOS  survey \protect\citep[][]{Elvis2009}
and  the grey  crosses  are X-ray  selected  AGN in  the deep  Chandra
surveys        of        the        Extended       Growth        Strip
\protect\citep[AEGIS-XD;][]{Nandra2015}  and  the  4\,Ms Chandra  Deep
Field  South   \protect\citep[CDFS-4\,Ms;][]{Rangel2013}  fields.   The
X-ray  luminosities  for  the   Chandra  survey  datapoints  are  from
\protect\cite{Buchner2015}  and  are   estimated  from  X-ray  spectra
analysis.  Each  data point is  randomly drawn from  the 3-dimensional
X-ray luminosity, column density and redshift probability distribution
function of individual sources inferred from the X-ray spectral fits.
}\label{fig:lxz}
\end{figure}

Figure \ref{fig:lxz} presents the  distribution on the intrinsic X-ray
luminosity (i.e. corrected  for line-of-sight obscuration) vs redshift
space of  the XMM-XXL-N X-ray  sources in comparison with  the Chandra
surveys  in  the   COSMOS  \citep{Elvis2009},  Extended  Growth  Strip
\citep[AEGIS-XD][]{Nandra2015}  and  4\,Ms  Chandra Deep  Field  South
\citep[][]{Xue2011,  Rangel2013}. For  the Chandra  survey  fields the
data-points   are  based   on   the  spectral   analysis  results   of
\cite{Buchner2015}.  This figure demonstrates the complementarity
of the XMM-XXL-N sample to Chandra survey fields.

\section{X-ray luminosity function determination}\label{sec_method}

This  section  describes our  methodology  for  determining the  X-ray
luminosity function  of AGN. This is  defined as the  space density of
the population  at a given redshift ($z$),  X-ray accretion luminosity
in the  2-10\,keV energy band [$L_X(\rm  2-10\,keV)$], and obscuration
along the line-of-sight parametrised  by the column density of neutral
hydrogen ($N_H$).  The likelihood  of a particular luminosity function
parametrisation  given  a set  of  observations  is  described by  the
product of the Poisson probabilities of individual sources.

\begin{equation}\label{eq_likeli}
\begin{split}
 \mathcal{L}( d_i \;|\; \theta) & =  e^{-\lambda} \times \\
                                & \prod_{i=1}^{N} \int \mathrm{d}\log L_\mathrm{X} \,\, \frac{\mathrm{d}
V}{\mathrm{d}   z}  \mathrm{d}  z   \,\, \mathrm{d} \log N_\mathrm{H} \times \\
                                 & p(d_i   |\;  L_\mathrm{X},z, N_\mathrm{H})  \;
\phi(L_\mathrm{X},z, N_\mathrm{H} \;|\; \mathbf{\theta}),
\end{split}
\end{equation}

\noindent where ${\rm  d}V/{\rm d}z$ is the comoving  volume per solid
angle at redshift  $z$, $d_i$ signifies the data  available for source
$i$ and $\theta$ represents  the parameters of the luminosity function
model,  $\phi(L_\mathrm{X}, z,  N_\mathrm{H}  \;|\; \mathbf{\theta})$,
that are  to be determined.   The multiplication is over  all sources,
$N$,  in  the sample  and  the  integration  is over  redshift,  X-ray
luminosity  and hydrogen  column  density. The  quantity  $p( d_i  |\;
L_\mathrm{X},z, N_H)$ is the probability of a particular source having
redshift $z$, X-ray luminosity  $L_{\rm X}$ and line-of-sight hydrogen
column   density   $N_H$.    This   captures  uncertainties   in   the
determination  of all  three parameters  because of  e.g.  photometric
redshift  errors or  uncertainties  in the  X-ray  spectra, under  the
assumption  that  the adopted  spectral  model  provides a  reasonable
representation of the basic  AGN X-ray spectral properties \citep[e.g.
see][]{Buchner2014}.   In equation~\ref{eq_likeli},  $\lambda$  is the
expected number of  detected sources in a survey  for a particular set
of model parameters $\theta$

\begin{equation}\label{eq_lambda}
\begin{split}  \lambda  = &  \int  \mathrm{d}  \log L_\mathrm{X}  \,\,
\frac{\mathrm{d} V}{  \mathrm{d} z} \mathrm{d} z  \,\, \mathrm{d} \log
N_\mathrm{H}    \times   \\   &    A(L_\mathrm{X},z,N_\mathrm{H})   \;
\phi(L_\mathrm{X},z,N_\mathrm{H}\;|\; \mathbf{\theta}).
\end{split}
\end{equation}

\noindent  where, $A(L_\mathrm{X},z,N_\mathrm{H})$ is  the sensitivity
curve, which quantifies the survey area over which a source with X-ray
luminosity  $L_X$,  redshift  $z$  and  column density  $N_H$  can  be
detected. A  non-parametric approach for the determination  of the AGN
space  density  is adopted.   A  three-dimensional  grid in  redshift,
luminosity and  column density  is defined and  $\phi(L_\mathrm{X}, z,
N_\mathrm{H})$ is assumed  to be constant within each  cube pixel with
dimensions ($\log L_\mathrm{X} \pm \mathrm{d}\log L_\mathrm{X}$, $z\pm
\mathrm{d}z$,  $\log N_\mathrm{H}\pm  \mathrm{d}  \log N_\mathrm{H}$).
The value  of the AGN space  density in each grid  pixel is determined
via equation~\ref{eq_likeli}. The edges of  the grid pixels in each of
the three dimensions  are $\log L_X = (41.0$,  $42.0$, $42.5$, $43.0$,
$43.5$, $44.0$,  $45.0$, $46.0$,  $47.0)$, $z=$( $0.0$,  $0.5$, $1.0$,
$1.5$, $2.0$,  $6.0$), $\log N_H  =$ ($20.0$, $22.0$,  $23.0$, $24.0$,
$25.0)$. The total number of free model parameters is 160.

Importance sampling  \citep{NR1992} is used to  evaluate the integrals
in  equation~\ref{eq_likeli}.  For  each source  we draw  ($\log L_X$,
$z$, $\log N_H$) points from  the equal probability chains produced by
the  X-ray spectral  analysis step.  The luminosity  function  is then
evaluated for each  sample point, ($L_X, z$, $N_H$).   The integral of
equation~\ref{eq_likeli} is simply  the average luminosity function of
the   sample.   The   Hamiltonian  Markov   Chain  Monte   Carlo  code
Stan\footnote{\url{http://mc-stan.org/shop/}} \citep{Carpenter2016} is
used for Bayesian statistical inference.

Finally  we define  the quantity  $N_{obs}$, the  observed  number of
sources  within in  each bin  of the  3-dimensional grid  in redshift,
luminosity and column density \citep[see also ][]{Miyaji2001, Aird2015}

%\begin{equation}\label{eq_Nobs}
%\begin{split}  
%N_{mdl}  = &  \int  \mathrm{d}  \log L_\mathrm{X}  \,\,
%\frac{\mathrm{d} V}{  \mathrm{d} z} \mathrm{d} z  \,\, \mathrm{d} \log
%N_\mathrm{H}    \times   \\   &    A(L_\mathrm{X},z,N_\mathrm{H})   \;
%\phi(L_\mathrm{X},z,N_\mathrm{H}\;|\; \mathbf{\theta}),
%\end{split}
%\end{equation}

\begin{equation}\label{eq_Nobs}
\begin{split}  
N_{obs} = & \sum_{i=1}^{N} \,\frac{1}{\aleph_{i}}\, \int\mathrm{d}\log L_\mathrm{X} \,\, \frac{\mathrm{d}
V}{\mathrm{d}   z}  \mathrm{d}  z   \,\, \mathrm{d} \log N_\mathrm{H} \times \\
                                 & p(d_i   |\;  L_\mathrm{X},z, N_\mathrm{H})  \;
\phi(L_\mathrm{X},z, N_\mathrm{H} \;|\; \mathbf{\theta}),
\end{split}
\end{equation}

\noindent where the integration limits correspond to the edges of each
grid pixel  defined above.  The summation  is over all sources  in the
sample.  The  normalisation factor $\aleph_i$  is the integral  of the
quantity     $p(d_i    |\;     L_\mathrm{X},z,    N_\mathrm{H})     \;
\phi(L_\mathrm{X},z,  N_\mathrm{H}  \;|\; \mathbf{\theta})$  over  the
full range of luminosities, redshifts and column densities.  $N_{obs}$
is  used  for  the  visualisation  of the  results.   Only  bins  with
$N_{obs}>1$ are shown  in the relevant figures and  tables.  Bins with
$N_{obs}<1$   typically   have   large  errors   and   therefore   the
observational constraints on the space density of AGN are loose.

\begin{figure*}
\begin{center}
\includegraphics[height=0.7\columnwidth]{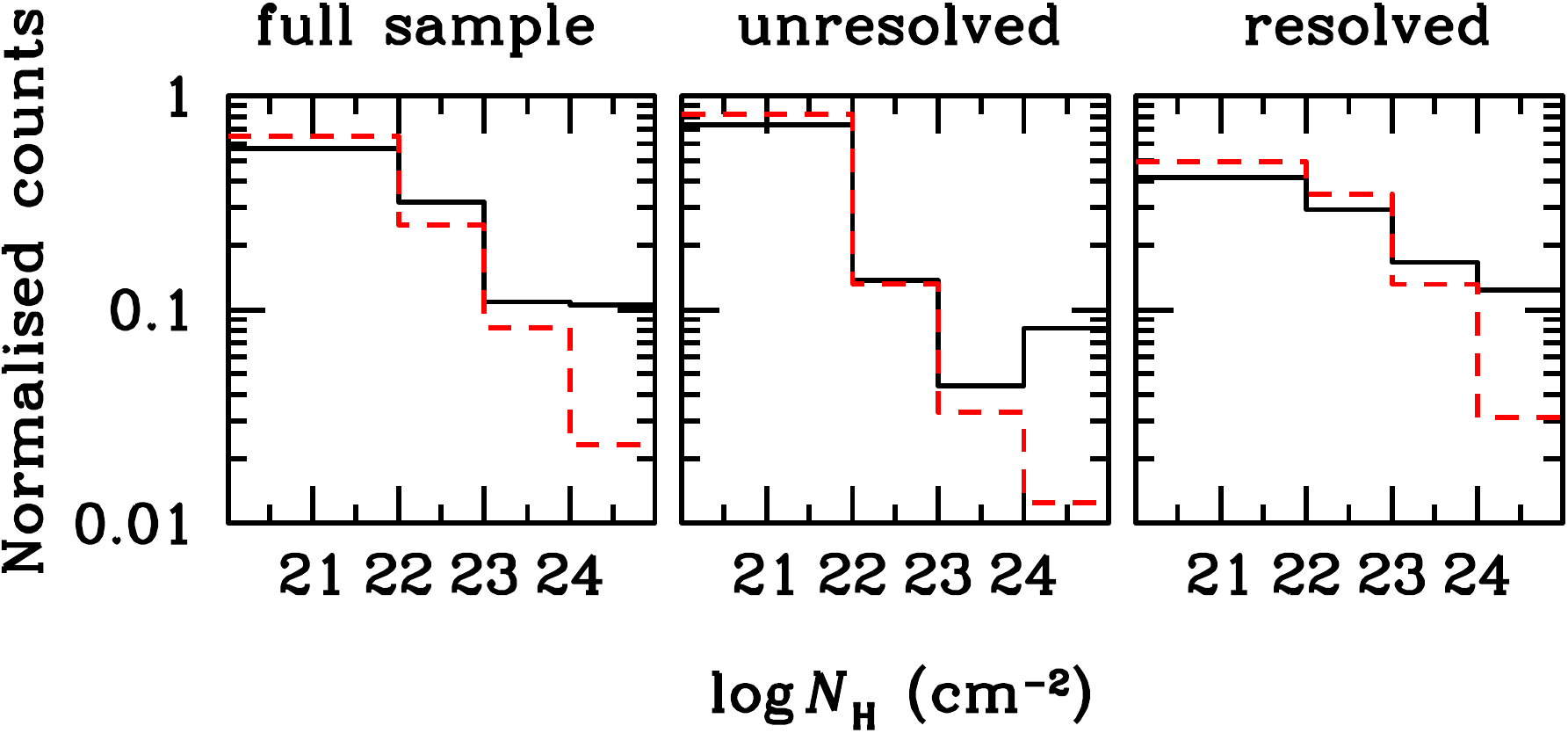}
\end{center}
\caption{Observed  hydrogen   column  density  distribution   for  the
  XMM-XXL-N hard X-ray selected sample.  This black solid histogram is
  estimated   by  summing   the  $\log   N_H$  posterior   probability
  distribution  functions  of  individual  sources.   The  red  dashed
  histogram weights  the posterior distribution of  individual sources
  with  the  corresponding  value  of the  X-ray  luminosity  function
  (equation   \ref{eq_Nobs}),   and    therefore   provides   a   more
  representative view of the detected  AGN population in the XMM-XXL-N
  survey.  The left  panel presents the full sample,  the middle panel
  corresponds to X-ray sources  with optically unresolved (point-like)
  counterparts and the right panel is for X-ray sources with optically
  resolved      (extended       light      profile)      counterparts.
}\label{fig_hist_nh}
\end{figure*}

\begin{figure*}
\begin{center}
\includegraphics[height=6cm]{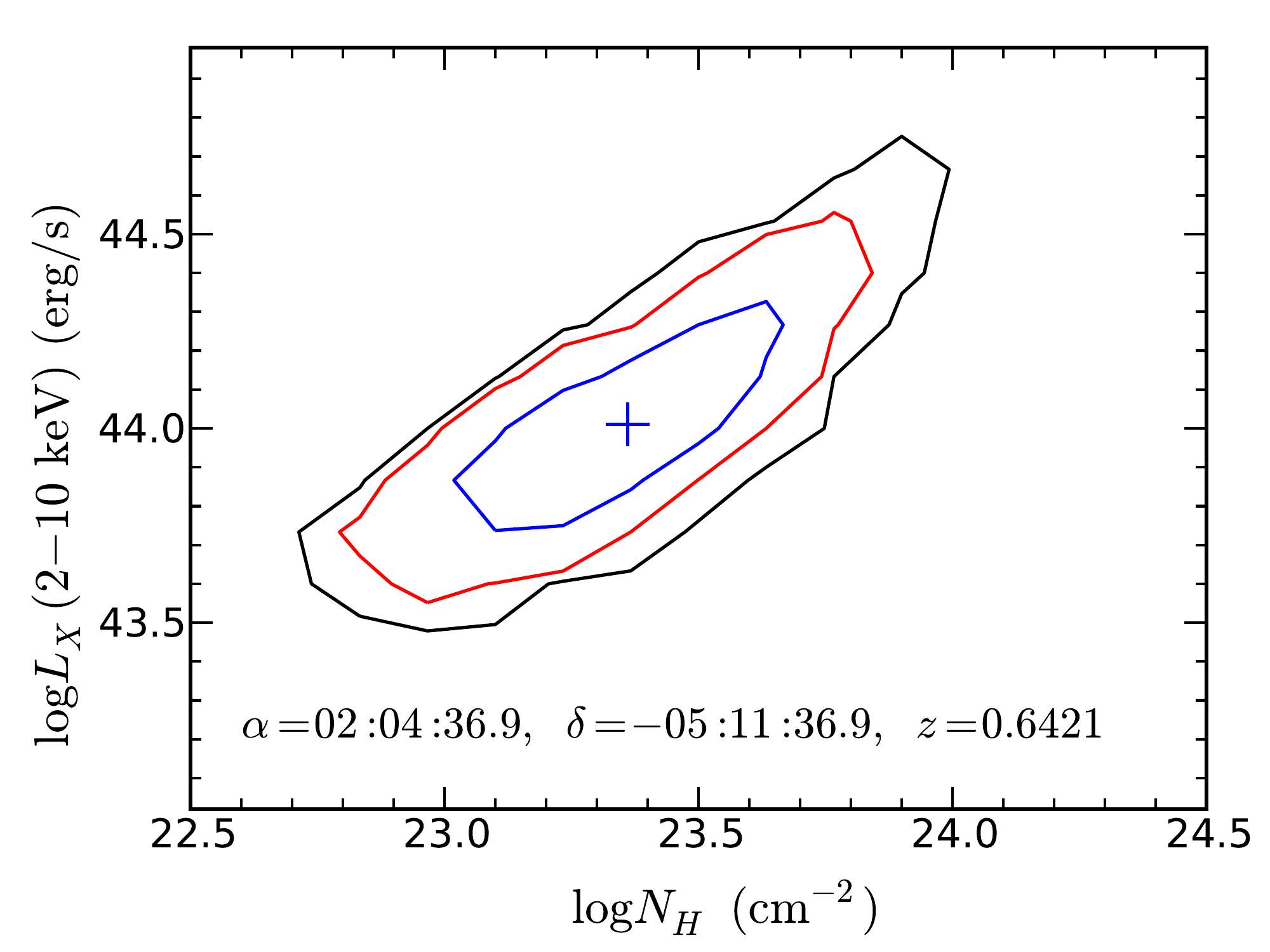}
\includegraphics[height=6cm]{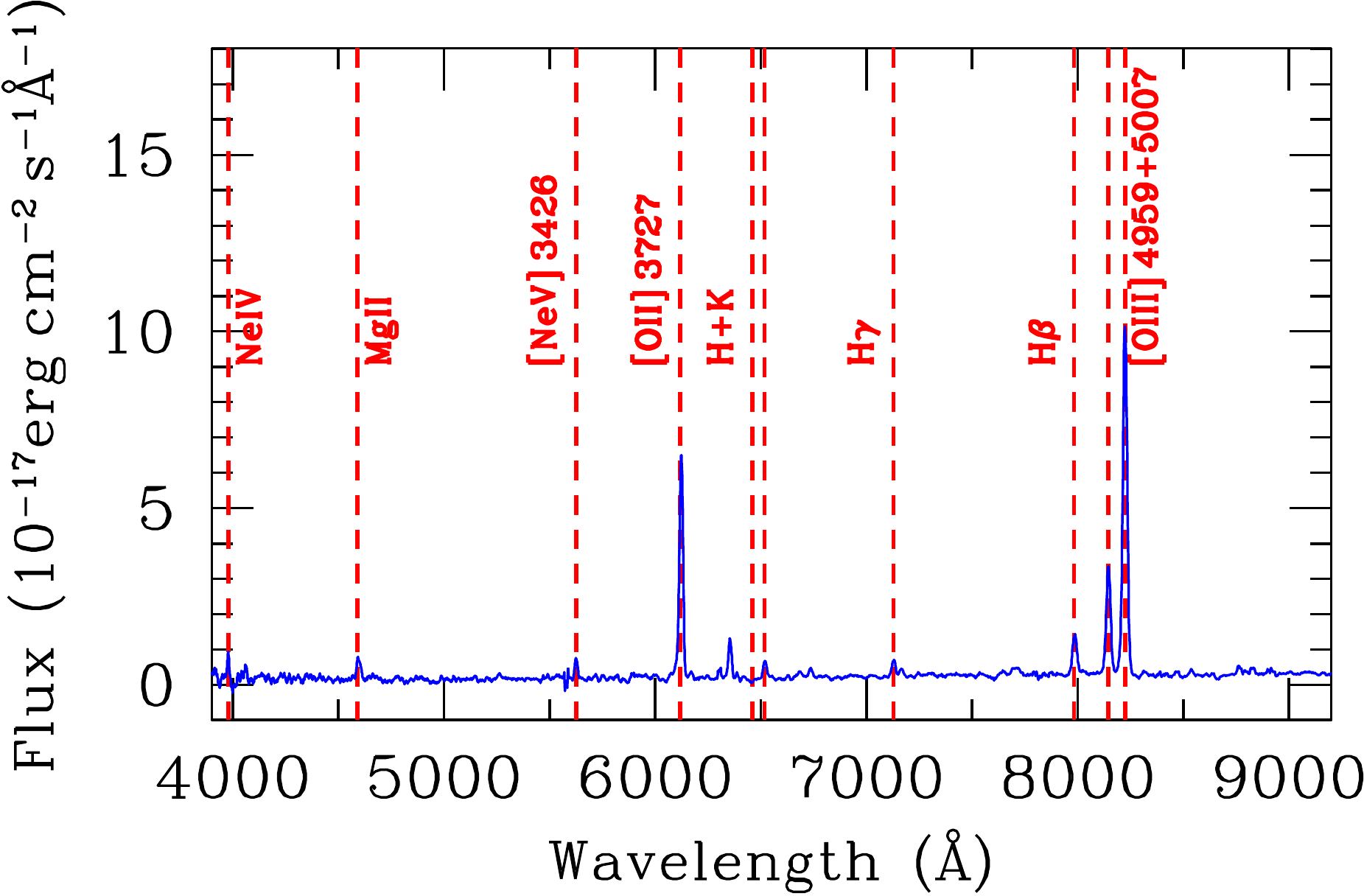}
\includegraphics[height=6cm]{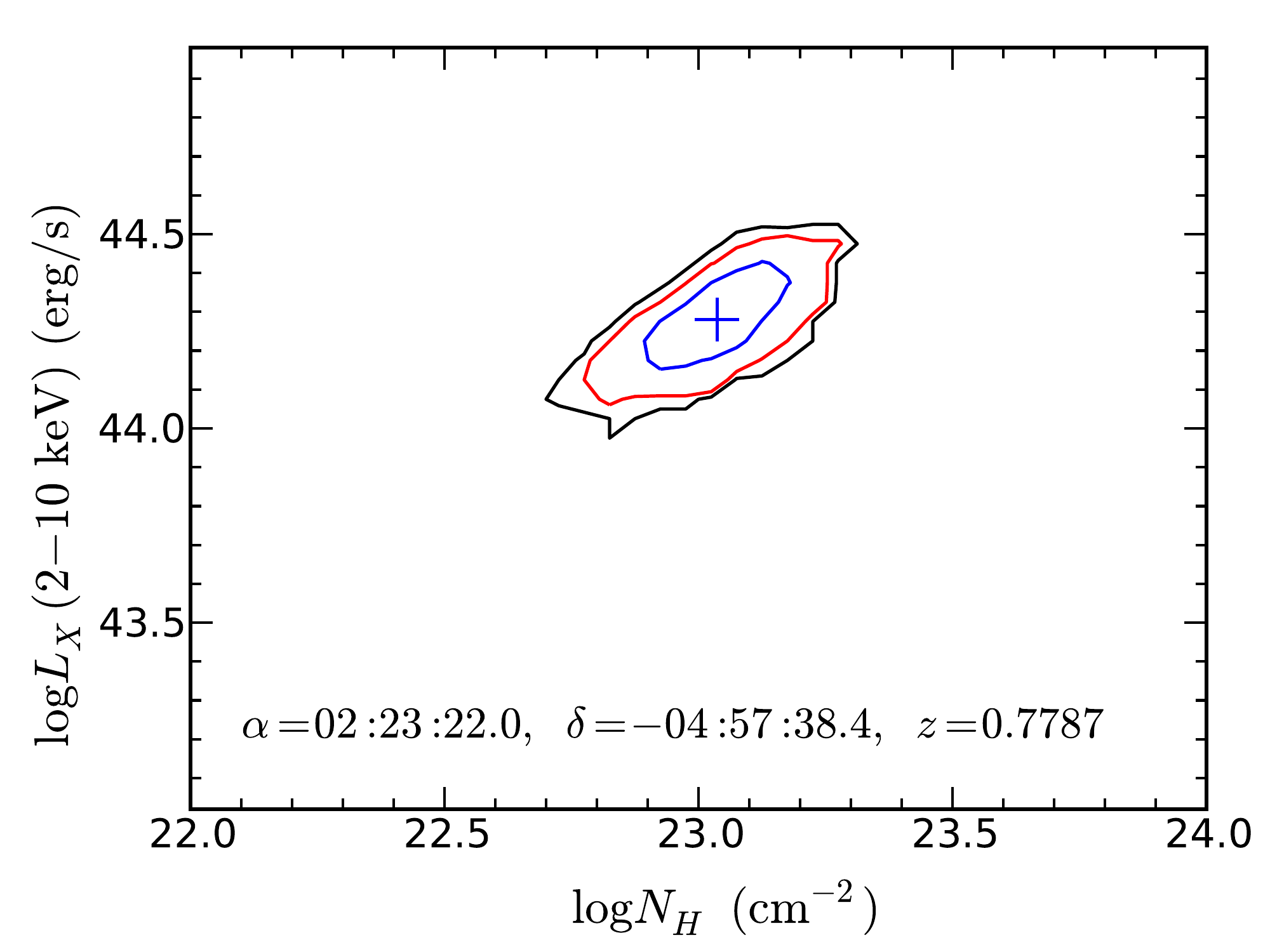}
\includegraphics[height=6cm]{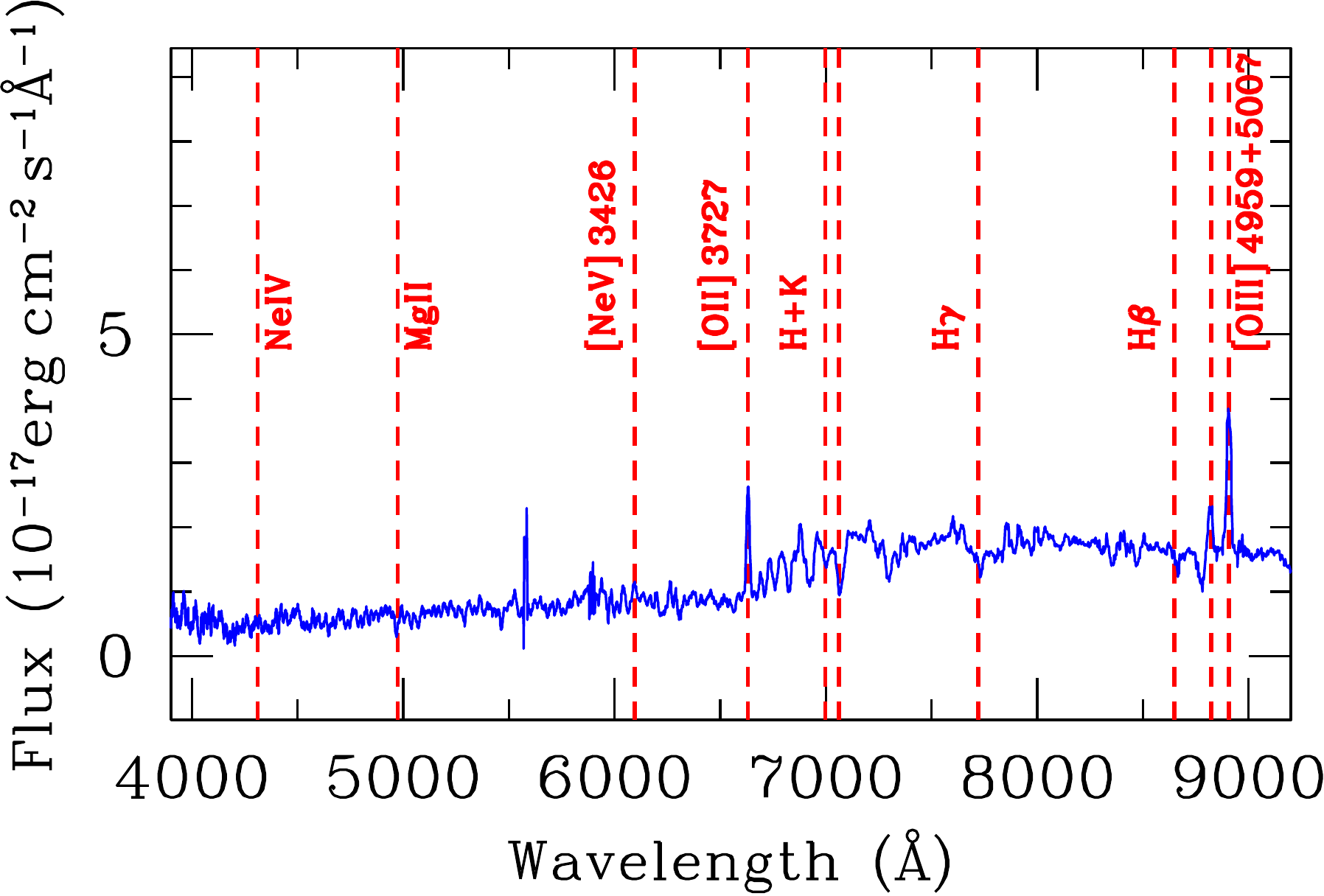}
\includegraphics[height=6cm]{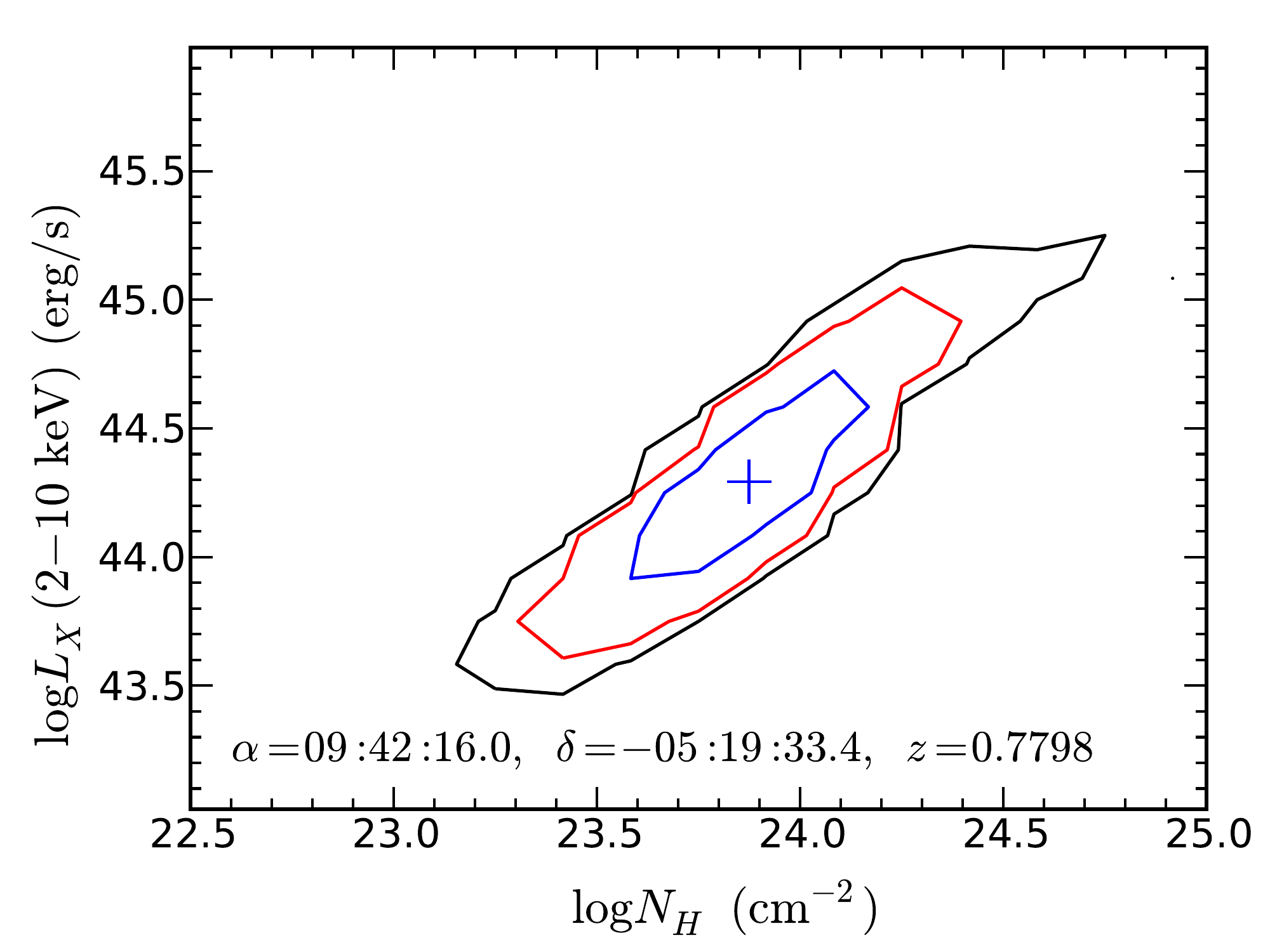}
\includegraphics[height=6cm]{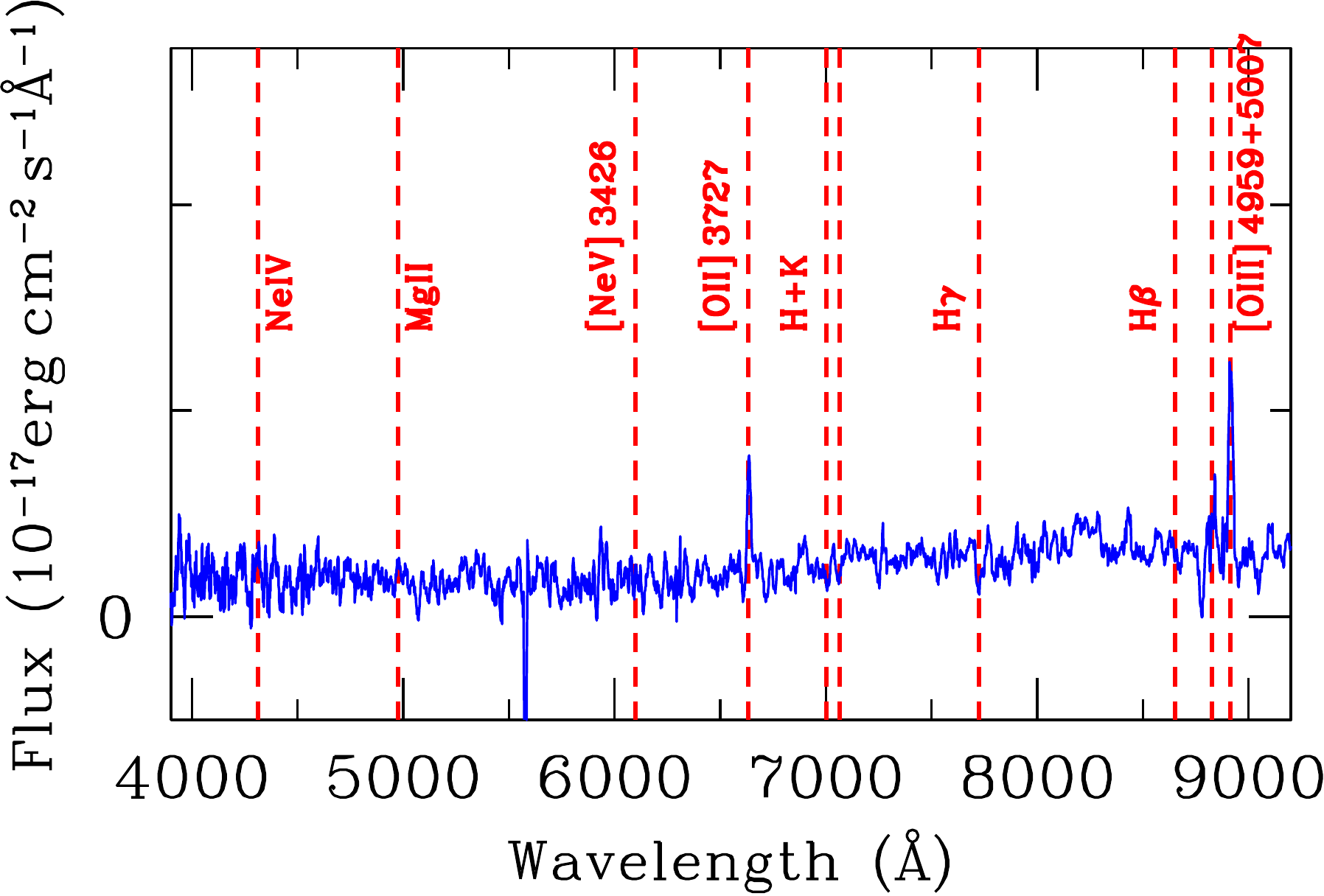}
\end{center}
\caption{Examples  of  X-ray  spectral  constraints and  SDSS  optical
spectra of  three relatively luminous, obscured X-ray  selected AGN in
the XMM-XXL-N  field with available SDSS spectroscopy.   The panels on
the left  show the posterior  probability distribution function in the
two-dimensional space  of intrinsic X-ray luminosity  in the 2-10\,keV
band and  the line-of-sight hydrogen column density  inferred from the
X-ray spectral analysis.  The median luminosity and column density are
marked by  the cross.   The contours enclose  68 (blue), 95  (red) and
99\% (black)of  the posterior probability distribution.  The right panels
show the SDSS optical spectra  of each source.  The redshift ($z$) and
sky coordinates  of the optical  counterpart of each X-ray  source are
also shown.  }\label{fig_example}
\end{figure*}

\begin{figure*}
\begin{center}
\includegraphics[height=6cm]{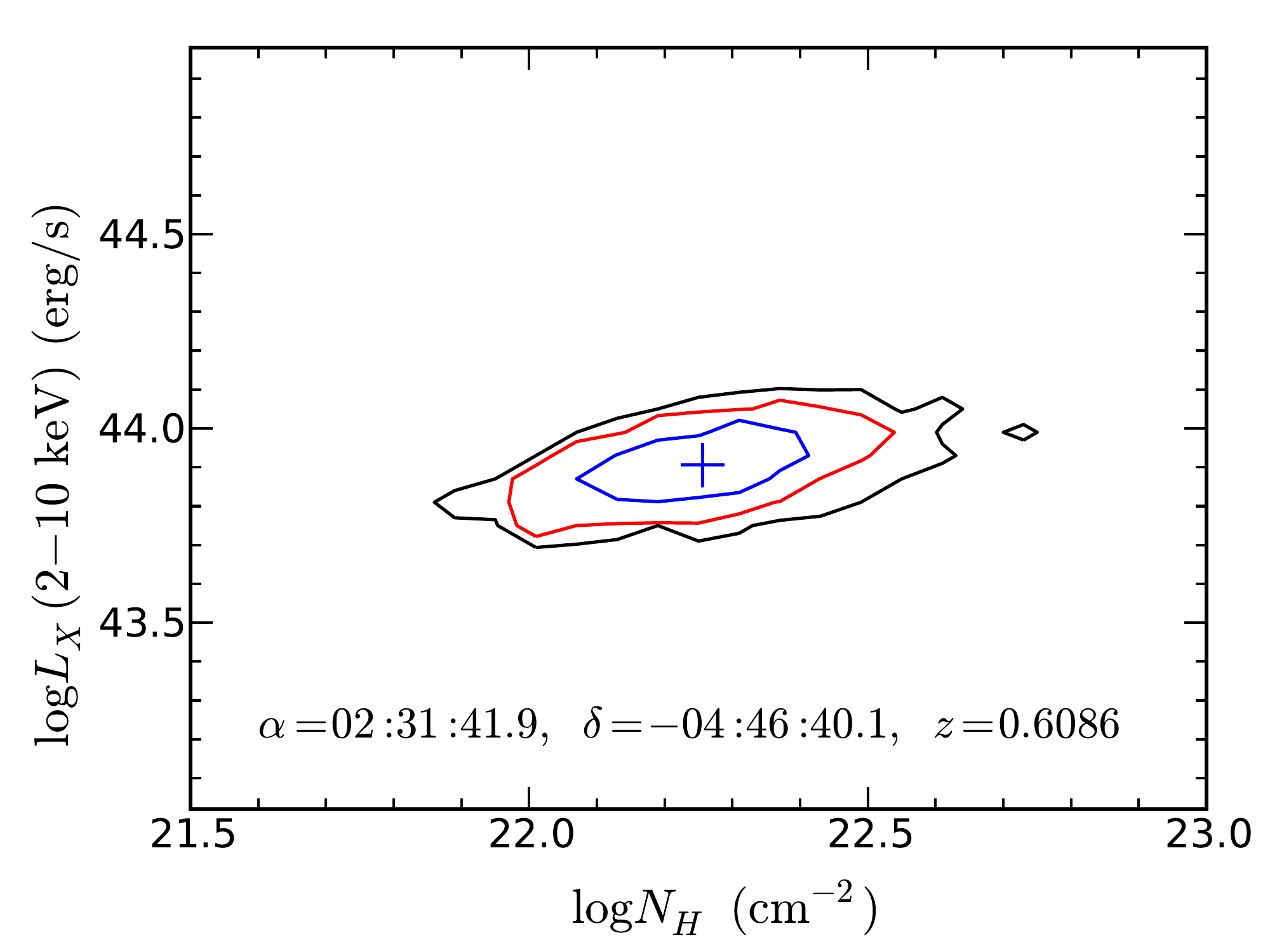}
\includegraphics[height=6cm]{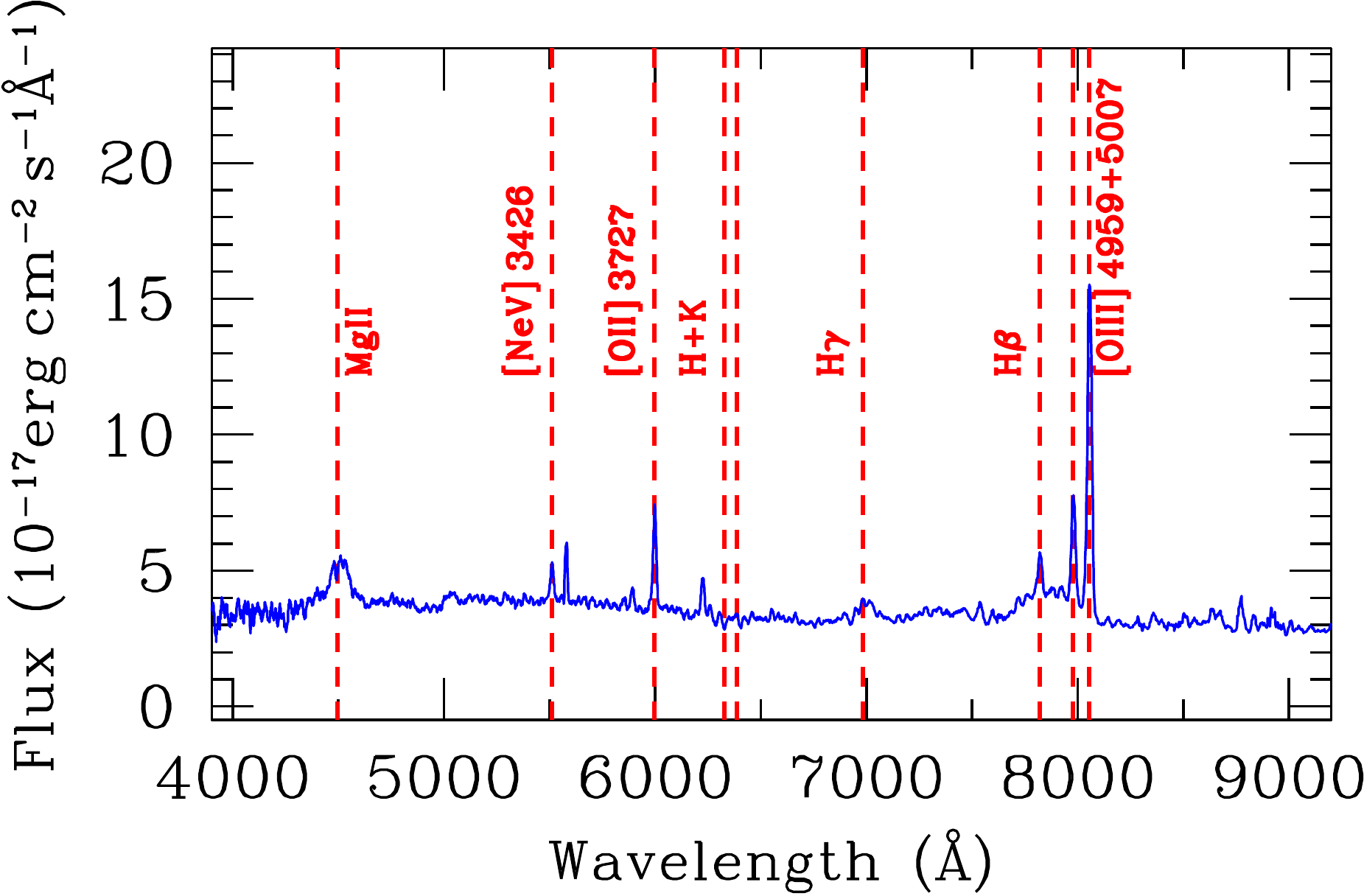}
\includegraphics[height=6cm]{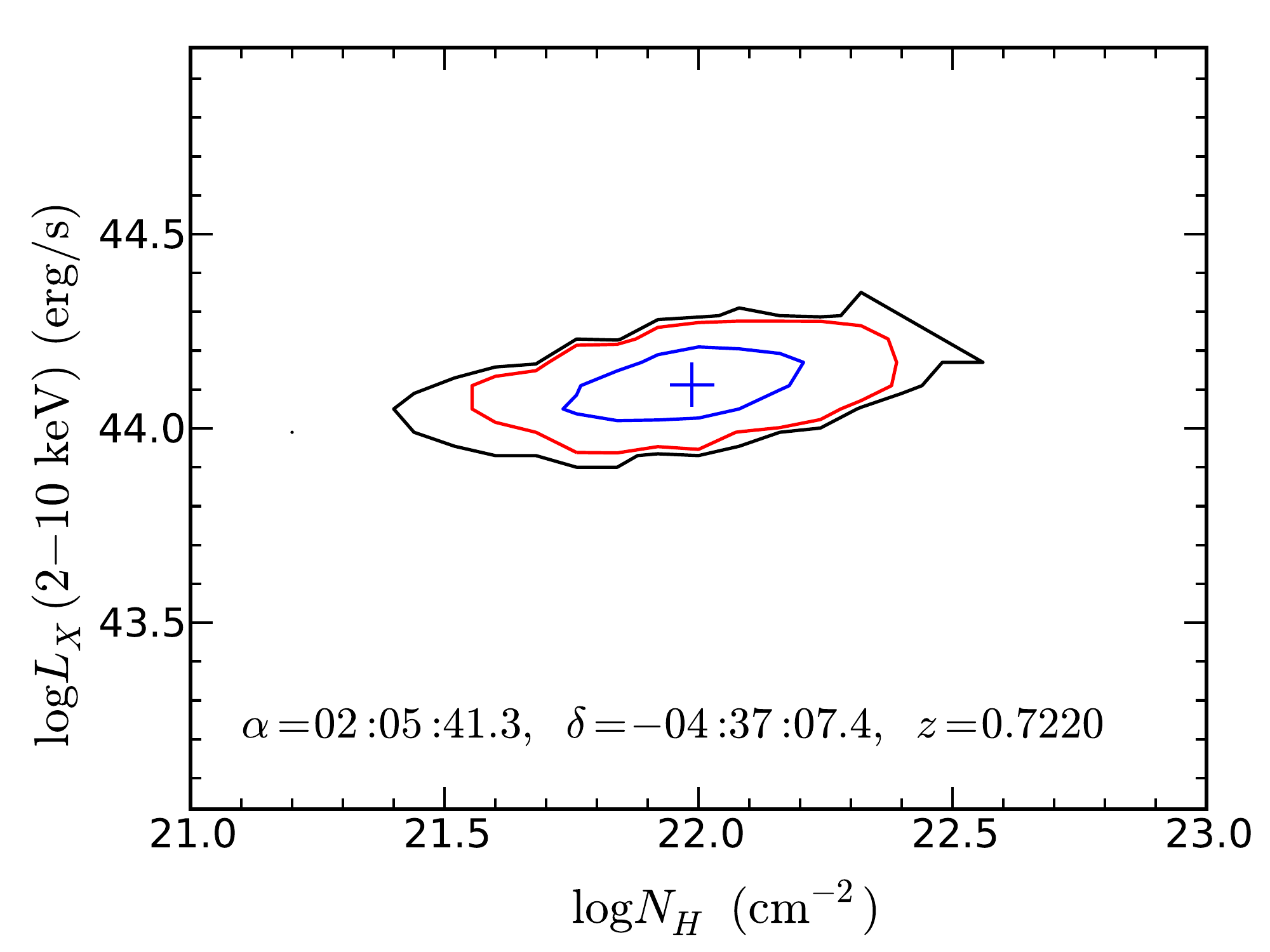}
\includegraphics[height=6cm]{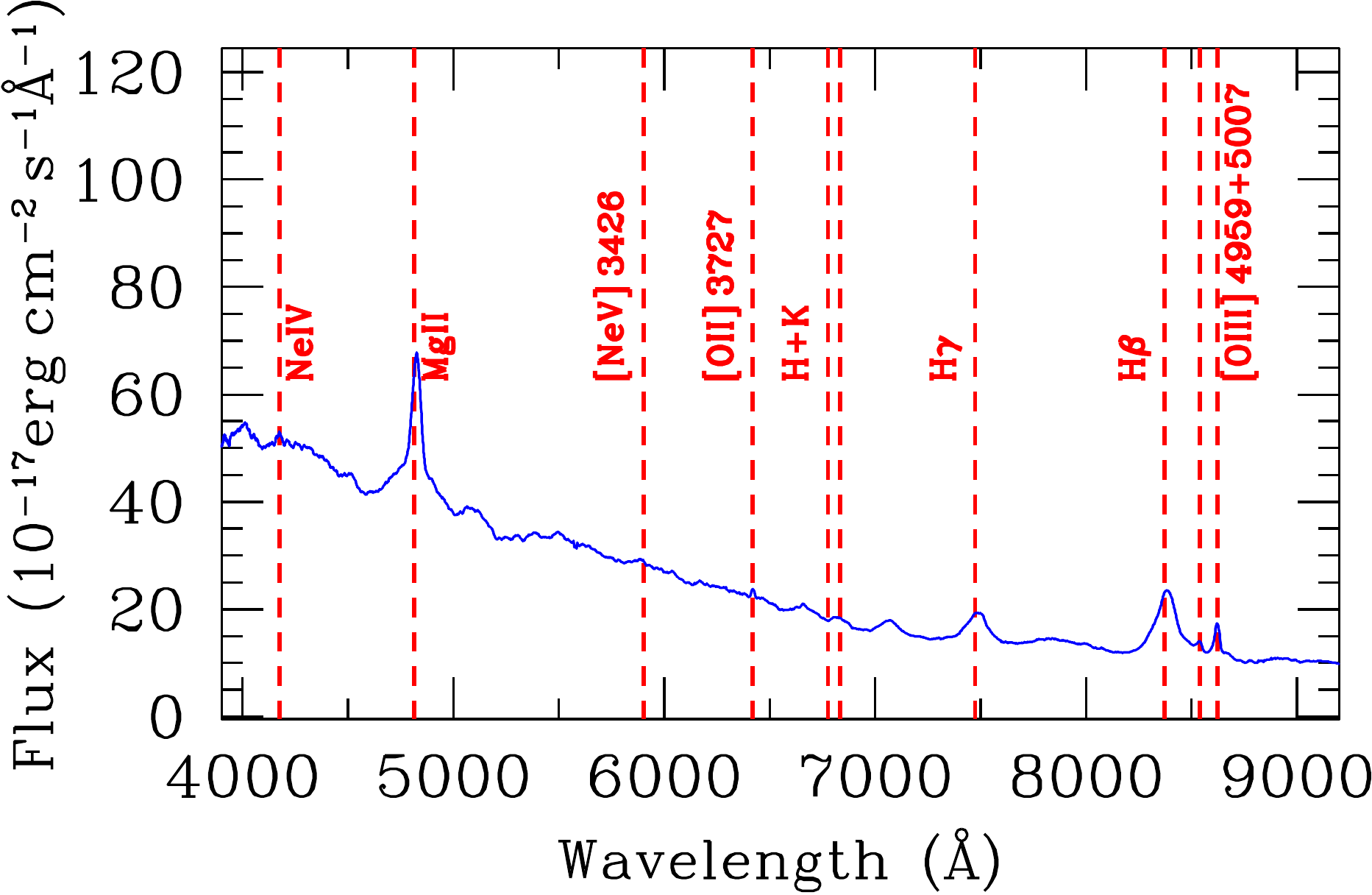}
\includegraphics[height=6cm]{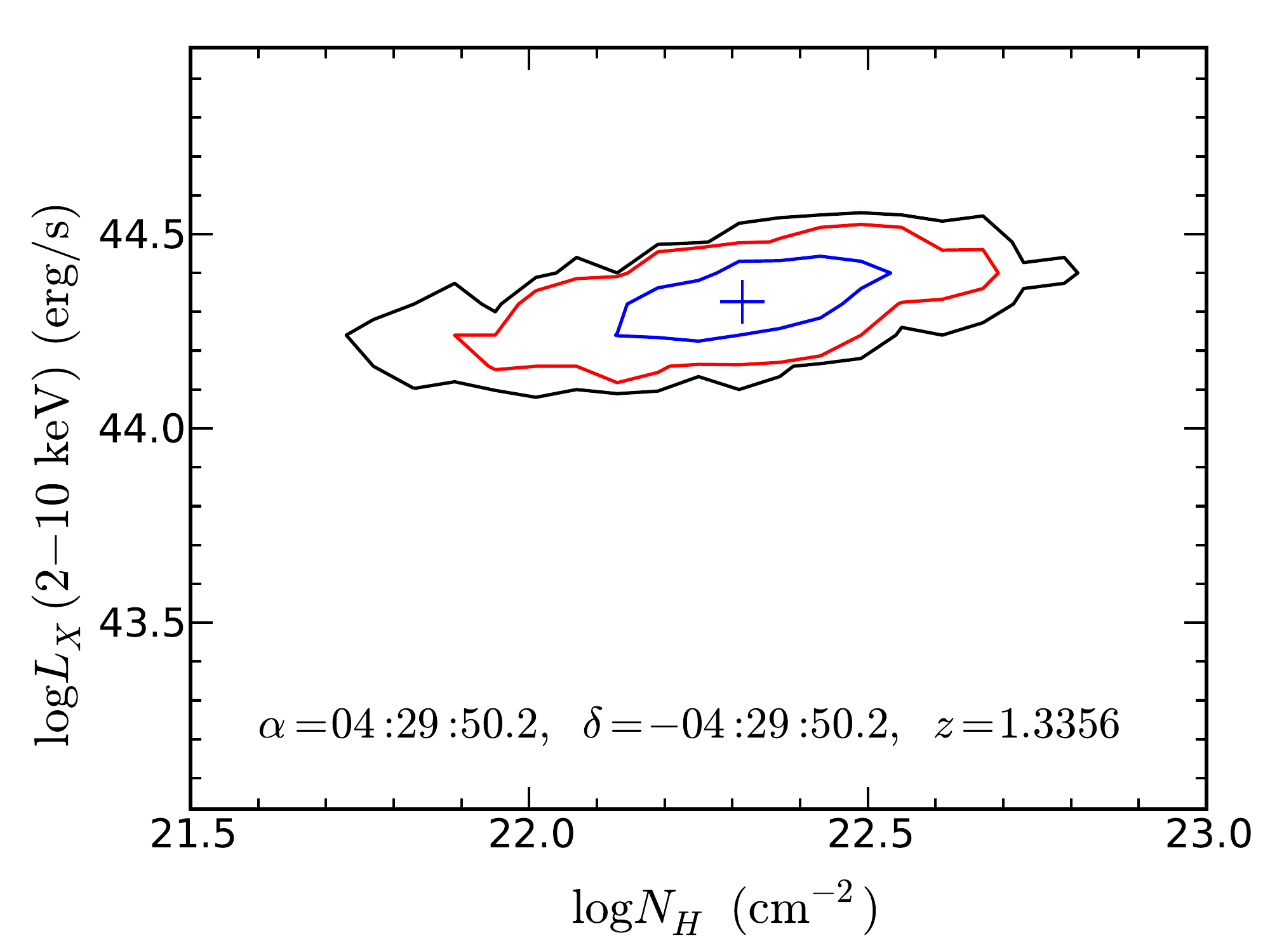}
\includegraphics[height=6cm]{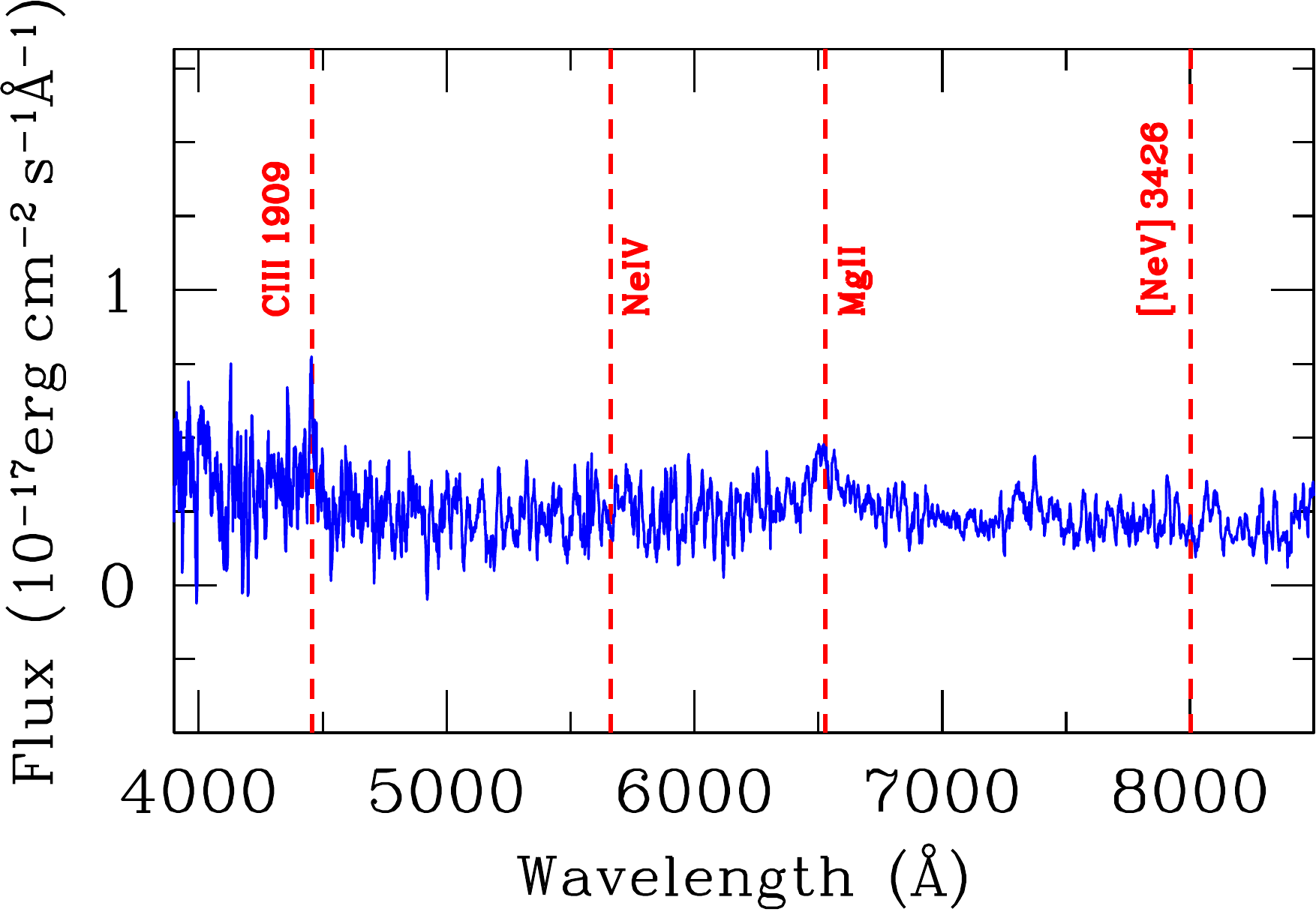}
\end{center}
\caption{Examples  of XMM-XXL-N AGN  with broad  optical emission-line
spectra   and   evidence  for   X-ray   obscuration,   $N_H  \ga   \rm
10^{22}\,cm^{-2}$.   The  panels  on   the  left  show  the  posterior
probability  distribution  in  the two-dimensional  space  of
intrinsic X-ray luminosity in the 2-10\,keV band and the line-of-sight
hydrogen  column density  inferred from  the X-ray  spectral analysis.
The median luminosity and column density are marked by the cross.  The
contours enclose 68 (blue), 95  (red) and 99\% (black)of the posterior
distribution function. The right  panels show the SDSS optical spectra
of each source.  The redshift ($z$) and sky coordinates of the optical
counterpart    of    each     X-ray    source    are    also    shown.
}\label{fig_example_type1}
\end{figure*}

\begin{figure*}
\begin{center}
\includegraphics[height=1.5\columnwidth]{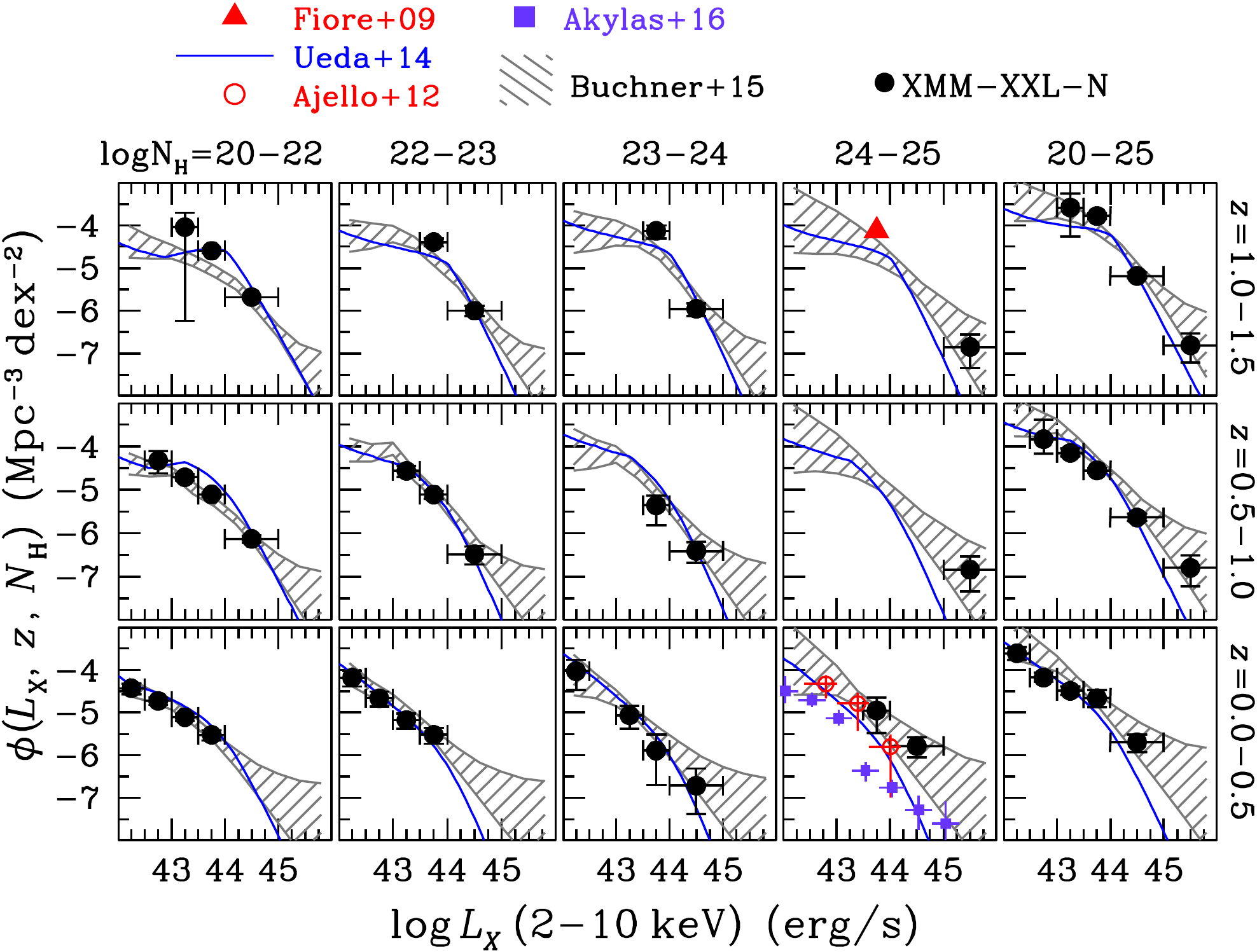}
\end{center}
\caption{Space  density  of  AGN  per luminosity  and  column  density
decades (i.e. in units of $\rm Mpc^{-3}\,dex^{-2}$).  Different panels
correspond to  different levels of line-of-sight  obscuration (left to
right) and different  redshift intervals (bottom: $z=0.0-0.5$; middle:
$z=0.5-1.0$; top:  $z=1.0-1.5$). The last  column of panels  plots the
total luminosity function, i.e.  integrated across the hydrogen column
density  interval $N_H=\rm 10^{20}  - 10^{-25}\,cm^{-2}$.   Black data
points (filled circles) are the non-parametric space density estimates
using the XMM-XXL-N data.  The uncertainties in the vertical direction
correspond  to the  90\% percentiles  around the  median of  the space
density probability distribution function.  The horizontal error-bars show
the luminosity bin size. The XXL data points are compared with the XLF
estimates  of   \protect\citet[][grey  hatched  regions]{Buchner2015},
\protect\citet[][blue  curves]{Ueda2014}  and   in  the  case  of  low
redshift  Compton-thick AGN  with the  local Swift/BAT  constraints of
\protect\citet[][red circles]{Ajello2012} and \protect\citet[][purple
  squares]{Akylas2016}. The red triangle in the $z=1.0-1.5$ Compton  
thick panel is the space density estimated by \protect\cite{Fiore2009}
for  infrared-selected Compton  thick  AGN candidates  in the  redshift
interval $z=0.7-1.2$}\label{fig_xlf_nh_z}
\end{figure*}

\section{Results}\label{sec_results}

\subsection{X-ray AGN obscuration distribution}

A key aim of this work is  to estimate the fraction of obscured AGN at
the relatively bright X-ray  luminosities sampled by the wide-area and
shallow  XMM-XXL-N  survey.   This  sample includes  a  non-negligible
fraction of AGN with hydrogen column densities in excess of $\rm N_H =
10^{22} \, cm^{-2}$.  This  is shown in Figure \ref{fig_hist_nh} (left
panel), which plots the observed  $\log \rm N_H$ histogram of the full
XMM-XXL-N hard-band  selected sample.  We  construct this distribution
using two different approaches. The  first is summing up the posterior
probability distribution  functions of individual  sources produced by
the X-ray spectral analysis.  This  method however, does not take into
account the form of the  AGN X-ray luminosity function, i.e.  the fact
that  luminous sources are  rarer than  moderate luminosity  ones.  We
therefore use  equation \ref{eq_Nobs} to estimate  the observed number
of sources, $N_{obs}$,  as a function of the  hydrogen column density.
This  calculation  weighs  the  $\rm N_H$  posterior  distribution  of
individual  sources with  the  value of  the corresponding  luminosity
function.   The histogram of  the quantity  $N_{obs}$ provides  a more
representative  view of  the observed  $N_H$ distribution  of  the AGN
population  detected  in  the  XMM-XXL-N.  The  fraction  of  observed
sources with column densities $\rm  N_H > 10^{22} \, cm^{-2}$ is about
$35\%$.   Compton-thick  AGN  with   $N_H  \rm  >  10^{24}\,  cm^{-2}$
represent a  small fraction, $\approx2\%$, of  the observed population
in the  XMM-XXL field.  Figure \ref{fig_hist_nh} also  shows the $\log
\rm N_H$ distribution  of of the sample split  into AGN with optically
unresolved  (middle   panel)  and  optically-extended   (right  panel)
counterparts.  As expected  X-ray sources with unresolved (point-like)
counterparts are mostly associated with low hydrogen column densities,
$N_H \rm > 10^{22}\, cm^{-2}$.  AGN identified with optically-resolved
galaxies  have  a  flatter  $\log  \rm  N_H$  distribution.   Selected
examples of  individual XMM-XXL-N obscured  AGN are also  presented in
Figure  \ref{fig_example}.  The  SDSS optical  spectra in  this figure
show narrow optical  emission lines (e.g.  [OII]\,3727\,\AA, H$\beta$,
[OIII]\,5007\,\AA) and/or absorption  features (e.g. H+K break, Balmer
lines)  associated  with  the  AGN  host  galaxy.   The  corresponding
posterior  probability  distributions   in  accretion  luminosity  and
hydrogen column density inferred  from the X-ray spectral analysis are
also shown in Figure \ref{fig_example}.

 Within   the  XMM-XXL-N   sample   there  are   also  broad   optical
emission-line AGN  (type-1) that show evidence  for X-ray obscuration,
$N_H \ga \rm 10^{22}\,cm^{-2}$.   Such sources represent about 10\% of
the type-1 AGN  sample in the XMM-XXL-N survey  \citep[see Figure 9 of
][]{Liu2016}.   Examples of  this class  of  AGN are  shown in  Figure
\ref{fig_example_type1}. The X-ray obscuration in these systems may be
associated with dust-free material close to the central black hole and
within the broad-line region \citep{Liu2016}.

 Figure \ref{fig_xlf_nh_z} and Table \ref{table:phi} present the space
 density of  AGN as a function  of 2-10\,keV X-ray luminosity  for the
 different redshift  and hydrogen column density  intervals defined in
 section \ref{sec_method}.  Results are shown  only up to the redshift
 bin  $z=1.0-1.5$.  At  higher redshifts  the XMM-XXL-N  space density
 constraints  suffer  large  uncertainties  because  of  small  number
 statistics,  increasing photometric  redshift  uncertainties and  the
 fraction of optically  unidentified sources in the  sample, which are
 expected  to   be  associated   preferentially  with   moderate-  and
 high-redshift AGN.  Further observations are needed to mitigate these
 issues.  Deep  near-infrared photometry  (e.g.  VIDEO  survey depths)
 over  the  full   XMM-XXL-N  field  for  example,   will  allow  both
 identifications  and  improved  photometric  redshift  estimates  for
 obscured  and high  redshift AGN  at $z\ga1.5$.   Dedicated follow-up
 spectroscopic surveys of optically faint (e.g. $r\ga22.5$\,mag) X-ray
 sources in  the XMM-XXL-N are  also needed to improve  constraints on
 the AGN space density, particularly for obscured sources, at moderate
 and high  redshift.  In Figure \ref{fig_xlf_nh_z}  the non-parametric
 constraints from the XMM-XXL-N field  are also compared with previous
 estimates    of    the    X-ray   luminosity    function    of    AGN
 \citep[][]{Ueda2014, Buchner2015}.   Overall there is  good agreement
 with  previous  studies.  For  Compton-thin  AGN  in particular,  the
 shallow XMM-XXL-N survey provides  competitive constraints on the AGN
 space density  to $L_X(\rm 2-10\,keV)\approx10^{45}\,erg  \, s^{-1}$,
 i.e.  close to  and above the knee of the  X-ray luminosity function.
 For  Compton-thick  levels  of  obscuration  ($\rm  N_H  >  10^{24}\,
 cm^{-2}$) we find AGN space densities consistent with the recent work
 of  \cite{Buchner2015}.    This  is  also  broadly   consistent  with
 \cite{Ajello2012}  in  the case  of  low  redshift Compton-thick  AGN
 selected     in    the     60-month     BAT    \citep[Burst     Alert
   Telescope][]{Barthelmy2005}  all sky  survey.  Also  shown are  the
 recent constraints from \cite{Akylas2016}  on the luminosity function
 of Compton-thick AGN using the 70-month Swift-BAT all-sky survey hard
 X-ray  catalogue.  The  \cite{Akylas2016} space  densities are  lower
 than previous  studies, including those from  \cite{Ajello2012}. This
 discrepancy  is  surprising given  that  the  two studies  find  very
 similar observed fractions  of Compton thick sources  among the total
 AGN sample.  The source of  the difference  is likely related  to the
 X-ray spectral models adopted  to determine the intrinsic luminosity
 of heavily obscured  sources and the method used  to extrapolate from
 the  observed   number  of   Compton-thick  sources  to   the  parent
 population. Our constraints are also systematically higher than those
 of  \cite{Ueda2014}.   Current  estimates  of the  space  density  of
 Compton thick  AGN will improve  by adding high  energy observations,
 $>10$\,keV (e.g.   {\it NuSTAR}), to the  existing X-ray spectroscopy
 at observed  energies below  about 10\,keV provided  by {\it  XMM} or
 {\it Chandra}.

\begin{figure*}
\begin{center}
\includegraphics[height=0.85\columnwidth]{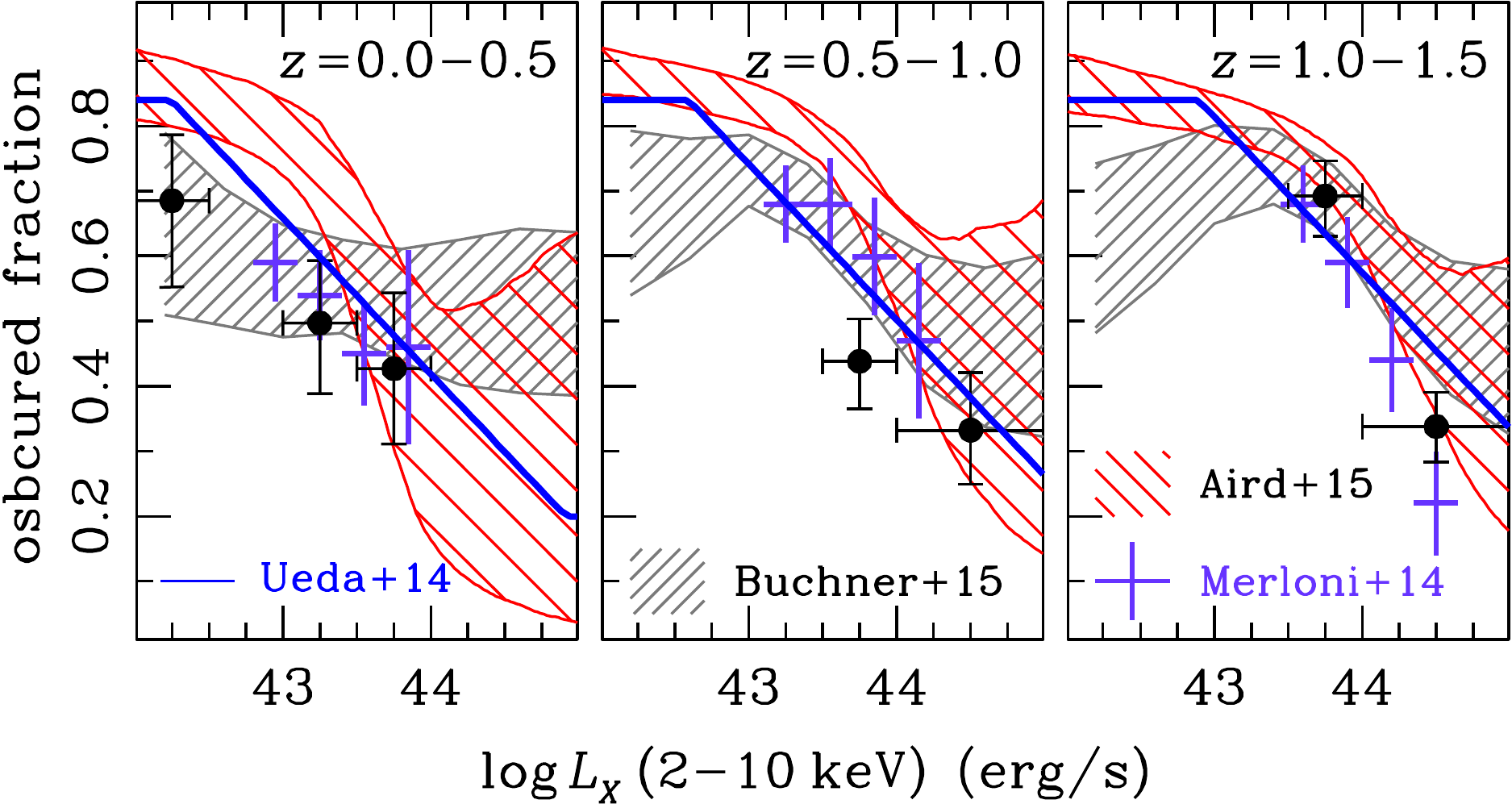}
\end{center}
\caption{Fraction  of obscured  (Compton-thin)  AGN as  a function  of
luminosity.   The obscured  fraction is  the  ratio of  the AGN  space
density in the line-of-sight column density intervals $\log N_H=22-24$
and $\log  N_H=20-24$ (in units  of $\rm cm^{-2}$).   Different panels
correspond to different  redshift intervals (left: $z=0.0-0.5$; middle
$z=0.5-1.0$; right  $z=1.0-1.5$). The  black data points  are inferred
from the  non-parametric space  density estimates using  the XMM-XXL-N
data.  The  uncertainties in the vertical direction  correspond to the
90\%  percentiles around  the median  of the  probability distribution
function. The horizontal error-bars show the luminosity bin size.  The
XMM-XXL-N  constraints  are   compared  with  the  obscured  fractions
determined by Buchner et al. (2015; grey hatched regions), Aird et al.
(2015;  red hatched  regions), Ueda  et al.   (2014; blue  curves) and
Merloni et al. (2014; purple crosses).}\label{fig_fobs}
\end{figure*}

We  also  explore the  fraction  of  obscured  Compton-thin AGN  as  a
function of luminosity  and redshift.  We define this  fraction as the
ratio of  the space densities  of AGN in the  column-density intervals
$\rm  N_H  =   10^{22}-10^{24}\,cm^{-2}$  and  $\rm  10^{20}-10^{24}\,
cm^{-2}$.  The  results are  presented in  Table \ref{table:fobscured}
and are plotted  in Figure \ref{fig_fobs}.  We find  that the obscured
Compton-thin AGN  fraction in the  luminosity interval $\log  L_X (\rm
2-10    \,   keV)    /    erg\,s^{-1}=    43.5-44$   increases    from
$46\pm_{-12}^{+11}\%$    at    a    mean    redshift    $z=0.25$    to
$75\pm_{-5}^{+5}\%$  at   a  mean  redshift  $z=1.25$.    At  brighter
luminosities, $L_X (\rm  2 - 10 \,keV) >10^{44}\,erg  \, s^{-1}$, this
fraction is  found to  be about 35\%  independent of  redshift. Figure
\ref{fig_fobs} also  compares our  results with recent  constraints in
the   literature    \citep[][]{Merloni2014,   Ueda2014,   Buchner2015,
  Aird2015}. Within  the uncertainties  there is  reasonable agreement
with previous  estimates.  It  is also  worth emphasising  the smaller
uncertainties of the XMM-XXL-N data points at $L_X ( \rm 2 - 10 \,keV)
\ga  10^{44} \,  erg \,  s^{-1}$ compared  to previous  studies.  This
demonstrates   the  complementarity   of   deep   {\it  Chandra}   and
wide/shallow {\it XMM}  surveys in studies of the  distribution of AGN
in obscuration over a wide luminosity baseline.

\subsection{Comparison with infrared-selected AGN}

\begin{figure*}
\begin{center}
\includegraphics[height=0.85\columnwidth]{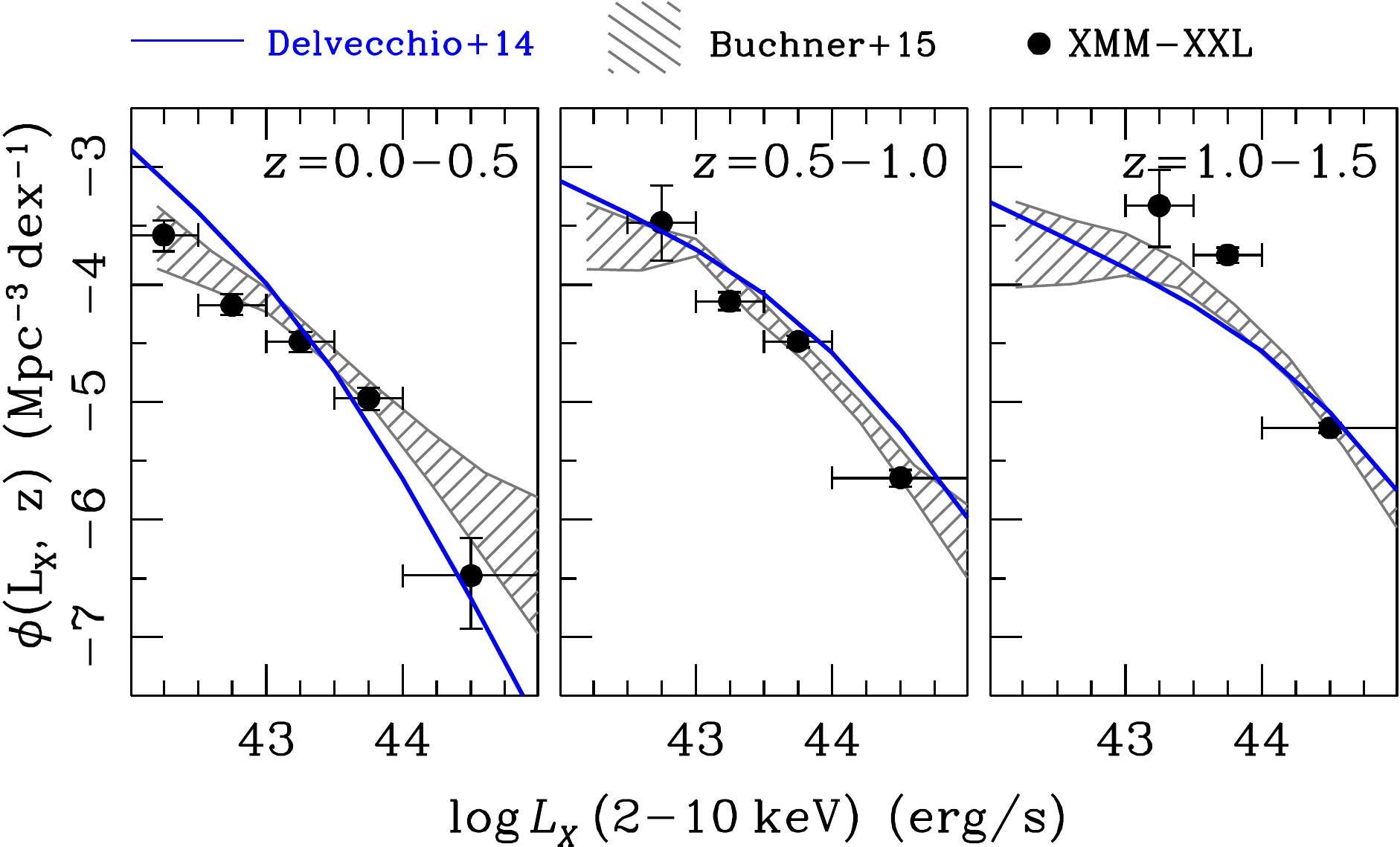}
\end{center}
\caption{Total X-ray luminosity function of Compton-thin AGN, $\rm N_H
= 10^{20}-10^{24}\,cm^{-2}$ (space density per luminosity decade). The
black datapoints  are the constraints  from the XMM-XXL-N,  the shaded
area represents the  recent results of \protect\cite{Buchner2015}. The
blue  curve  is  the  luminosity  function  of  infrared-selected  AGN
estimated by \protect\cite{Delvecchio2014}.  }\label{fig_txlf}
\end{figure*}

 Next  it is investigated  if the  total space  density of  X-ray AGN,
including  obscured  sources,  is  consistent  with  constraints  from
infrared-selected samples.   Like in the X-ray, AGN  selected at these
longer  wavelengths are  less  prone to  the  effects of  obscuration,
albeit with non-negligible levels of contamination by non-AGN sources.
For this exercise  the X-ray luminosity function is  integrated in the
range $\rm N_H =  10^{20} - 10^{24}\,cm^{-2}$ (i.e.  only Compton-thin
sources) to  avoid current uncertainties  in the determination  of the
space density  of Compton  thick AGN.  We  also adopt the  most recent
results  on the space  density of  infrared-selected AGN  presented by
\cite{Delvecchio2014}.   They  fit galaxy  and  AGN  templates to  the
multi-waveband   spectral  energy   distribution  of   {\it  Herschel}
infrared-selected    galaxies   in   the    GOODS-South   \citep[Great
Observatories Origins Deep  Survey-South,][]{Elbaz2011} and the COSMOS
fields  to  identify  those   with  a  statistically  significant  AGN
component.  These are then used to constrain the space density of AGN,
including heavily  obscured systems, as  a function of  bolometric AGN
luminosity  inferred from  the SED  fits.  We  use the  parametric AGN
luminosity    functions   at    different   redshifts    provided   by
\cite{Delvecchio2014} to interpolate to the mean of the three redshift
intervals adopted in this paper. Bolometric luminosities are converted
to   X-ray   luminosities   using   the  bolometric   corrections   of
\cite{Marconi2004}.    These    results   are   plotted    in   Figure
\ref{fig_txlf}.

There  is broad  agreement between  the X-ray  luminosity  function of
Compton-thin X-ray  selected AGN  and that of  infrared-selected ones,
given the current  level of uncertainties in the  determination of AGN
space densities, and systematics  associated with e.g.  the conversion
from  bolometric to  X-ray  luminosities. Additionally,  the level  of
agreement between  the X-ray and infrared  luminosity functions argues
against  a  large population  of  heavily  obscured  sources that  are
identified  at the infrared  and are  missing from  X-ray wavelengths.
The addition  of Compton  thick AGN to  the X-ray  luminosity function
constraints  in Figure  \ref{fig_txlf} would  further  strengthen this
argument. Assuming for  example a flat 30\% fraction  of Compton thick
AGN independent  of redshift and  luminosity \citep{Buchner2015} would
shift the X-ray curves and data-points in Figure \ref{fig_txlf} upward
by about 0.15\,dex.

Next we explore the  completeness of infrared colour-selection methods
proposed  in the  literature to  identify obscured  and  luminous AGN,
using   as  starting   point   the  XMM-XXL-N   AGN  sample.    Figure
\ref{fig_wise}  plots  the  distribution   of  XMM-XXL-N  AGN  in  the
2-dimensional plane  that consists  of WISE $W_1-W_2$  infrared colour
and  X-ray  luminosity.   Different  panels  correspond  to  different
redshift and hydrogen  column-density intervals.  The density contours
in  that figure  are constructed  using the  redshift,  luminosity and
column  density probability  distribution functions  derived  from the
X-ray spectral  analysis of individual  sources.  A single  source may
therefore   spread   out   in   different  luminosity   and   redshift
bins. Iso-density curves correspond to the 68th and 95th percentile of
the distribution.

   We adopt the colour limit $W_1-W_2>0.8$ defined by \cite{Stern2012}
for selecting AGN.  Formally their selection also includes a magnitude
cut in the  WISE W2 filter, $W2<15.05$\,mag. This  is primarily driven
by the depth  of the WISE observations in the  COSMOS field, which was
used  to calibrate  the WISE  AGN  selection, and  the requirement  to
minimise contamination.   Our work uses the  deeper AllWISE catalogue.
Additionally,  contamination  is not  an  issue  since  all the  X-ray
sources in the  sample are AGN.  Figure \ref{fig_wise}  shows that the
fraction of sources that  pass the cut $W_1-W_2>0.8$ changes primarily
with redshift.  The impacts of obscuration and X-ray luminosity appear
to  be  second-order  effects.   We caution,  nevertheless,  that  the
luminosity baseline of our sample is not large.  The fraction of X-ray
selected AGN  above the WISE  colour cut proposed  by \cite{Stern2012}
increases from about 20-30\% at  $z<0.5$ to $>50\%$ at $z>1$ with some
variation  among  column  density  bins at  fixed  redshift.   Similar
conclusions  apply to the  AGN mid-IR  selection criteria  proposed by
\cite{Donley2012}.   Figure \ref{fig_swire}  shows  how the  XMM-XXL-N
X-ray  selected AGN  are distributed  on Spitzer  mid-IR colour-colour
plane relative  to the selection wedge  proposed by \cite{Donley2012}.
The fraction of  XMM-XXL-N AGN within that wedge  increases from about
20-30\% at $z\approx0.5$ to 60-70\% at $z\approx1.5$.  These fractions
are not particularly sensitive to intrinsic AGN X-ray luminosity cuts.

\subsection{Comparison with UV-selected AGN}

 In this section the space density of unobscured X-ray selected AGN is
compared  to  that  of  UV-selected  QSOs.  In  the  X-ray  literature
unobscured  AGN  are often  defined  as  those  with column  densities
$N_H<\rm  10^{22}\,cm^{-2}$.    Recent  evidence  suggests   that  the
consistency   between  optically-classified   type-1  AGN   and  X-ray
unobscured ones  is maximised  at a lower  threshold, $\log N_H  = \rm
10^{21.5}  \, cm^{-2}$  \citep{Merloni2014, Liu2016}.   This  limit is
adopted  here  to  compare  the  luminosity  functions  of  X-ray  and
UV/optically-selected  AGN  samples.    The  space  density  of  X-ray
unobscured  AGN (i.e.   $\rm  \log  N_H/ cm^{2}  =  20-21.5$) is  then
approximated by  simply scaling the  estimated space densities  in the
logarithmic interval $\rm  \log N_H/ cm^{2} = 20 - 22$  by a factor of
1.5.  The underlying  assumption is a flat distribution  of AGN in the
logarithmic column density interval $\rm  \log N_H/ cm^{2} = 20 - 22$.
For  the luminosity  of  UV/optically-selected QSOs  we  use the  LEDE
(Luminosity     and    Density    Evolution)     parametrisation    of
\cite{Croom2009}, which is constrained by observations in the redshift
interval $0.4-1.5$. The conversion  of QSO absolute optical magnitudes
to X-ray luminosities follows the steps described in \cite{Croom2009}.
At  lower redshift,  $z<0.4$, we  also  compare our  results with  the
broad-line  AGN luminosity  function presented  by \cite{Schulze2009}.
From the latter  study we use the double  power-law parametrisation of
the redshift $z=0$ bolometric  luminosity function.  This is converted
to  X-rays   using  the  \cite{Marconi2004}   bolometric  corrections.
Following  \cite{Schulze2009}  pure  density  evolution  of  the  form
$(1+z)^{5}$  is   also  included  to  determine   the  broad-line  AGN
luminosity function at redshifts  $z>0$.  The comparison between X-ray
and  optical type-1 AGN  luminosity functions  is presented  in Figure
\ref{fig_uxlf}.   Overall there  is fair  agreement between  the space
densities  of  unobcured ($\rm  \log  N_H/  cm^{2}  = 20-21.5$)  X-ray
selected  AGN and  UV/optically selected  QSOs.  At  low  redshift the
\cite{Schulze2009}  luminosity function  appears to  exceed  the X-ray
space densities of unobscured AGN below  about $L_X(\rm 2 - 10 \, keV)
\approx10^{43} \, erg  \, s^{-1}$.  A similar trend  has been reported
by  \cite{Schulze2009} when comparing  their luminosity  function with
the    soft-band   (0.5-2\,keV)    X-ray   luminosity    function   of
\cite{Hasinger2005}.  A possible explanation  is that the faint-end of
the  \cite{Schulze2009}  luminosity function  includes  a fraction  of
Seyfert-1.8s  and 1.9s, i.e.   not purely  type-1 AGN.   These sources
correspond to a higher X-ray column density threshold, e.g.  $\rm \log
N_H/   cm^{2}\approx22.3$   \citep{Burtscher2016}.    Differences   in
e.g. the  adopted bolometric corrections  or the definition  of type-1
AGN may also contribute to the difference between the faint-end of the
\cite{Schulze2009} and X-ray luminosity functions.

\subsection{eROSITA predictions}

Finally we explore expectations from the eROSITA All Sky Survey on the
determination  of the  AGN space  density as  a function  of redshift,
accretion  luminosity and line-of-sight  hydrogen column  density. For
this exercise we assume (i) that the AGN population follows the median
space density  constraints of \cite{Buchner2015},  (i) the uncertainty
of  the luminosity  function scales  as  $\delta \phi(L_X,  z, N_H)  /
\phi(L_X,  z, N_H)  \propto  1/\sqrt{N}$, where  $N$  is the  expected
number of AGN in a given $(L_X, z, N_H)$ bin, (ii) the 4-year depth of
the eROSITA All Sky Survey in the  soft-band, $f_X( \rm 0.5 - 2 \, keV
)=    1.5   \times    10^{-14}   \,erg    \,   s^{-1}    \,   cm^{-2}$
\citep{Merloni2012}, over a total area of $\rm 5000\,deg^2$, (iii) the
X-ray  spectral model  described in  section  \ref{sec:xspectral} with
$\Gamma=1.9$  and  the  normalisations   of  the  soft  power-law  and
reflection  components fixed  to  10\% and  1\%  respectively, of  the
normalisation of the intrinsic power-law spectrum.

The   resulting   eROSITA    predictions   are   plotted   in   Figure
\ref{fig_xlfero} along with the  non-parametric constraints of the AGN
space density from \cite{Buchner2015}  and the XMM-XXL-N.  This figure
shows  that the  eROSITA  will provide  excellent  statistics for  AGN
population  studies at redshifts  $z \la  0.5$ over  a range  of X-ray
luminosities, $L_X( \rm 2-10\,keV) \approx 10^{43} - 10^{45} \, erg \,
s^{-1}$, and for column densities approaching the Compton thick limit,
$\rm  N_H \approx  10^{24}  \,  cm^{-2}$.  We  choose  not to  provide
eROSITA predictions for Compton-thick AGN because current measurements
remain  somewhat  uncertain  and  the detectability  of  such  heavily
obscured sources  by the  eROSITA All Sky  Survey is sensitive  to the
adopted AGN X-ray  spectral model, and in particular  the level of the
soft   scattering   component   \citep[see  also][]{Akylas2012}.    At
redshifts  $z>0.5$ the  eROSITA All  Sky Survey  will be  sensitive to
moderate  obscured AGN,  $\rm  N_H  \la 10^{23}  \,  cm^{-2}$, at  the
bright-end  of the luminosity  function $L_X  (\rm 2  - 10  \,keV) \ga
10^{44}\, erg  \, s^{-1}$. At  higher levels of obscurations  only the
most  extreme  sources in  terms  of  luminosity  are expected  to  be
detected, $L_X (\rm 2 - 10 \,keV) \ga 10^{45}\, erg \, s^{-1}$.

We   caution  that   the   eROSITA  predictions   plotted  in   Figure
\ref{fig_xlfero}  only  depend  on  the  expected  number  of  AGN  in
different $(L_X, z, N_H)$ bins.  They do not account for uncertainties
in the determination of  redshifts (e.g. via photometric methods), the
measurement errors  of the line-of-sight column density  from the X-ray
spectra or uncertainties in the determination of the AGN space density
at the bright end of the luminosity function from current surveys.

\section{Discussion}

This paper presents  constraints on the space density  of obscured AGN
at relatively  high accretion luminosities, $L_X  (\rm 2 -  10 \, keV)
\approx 10^{44} \, erg \, s^{-1}$, using one of the largest contiguous
X-ray surveys  currently available,  in the equatorial  XMM-XXL field.
We show  that despite the relatively shallow  X-ray depth \citep[$f_X(
\rm   2-10\,keV)   \approx   3\times   10^{-15}\,  erg   \,   s^{-1}\,
cm^{-2}$;][]{Liu2016},  this sample provides  robust estimates  of the
space  density  of Compton-thin  ($\rm  N_H<10^{24}\,cm^{-2}$) AGN  at
luminosities close to  and above the break of  the luminosity function
\citep[$L_X  (\rm  2   -  10  \,  keV)  \approx   10^{44}  \,  erg  \,
s^{-1}$;][]{Aird2010,  Ueda2014, Aird2015},  where smaller  area X-ray
surveys  are  affected by  small  number  statistics.   This point  is
demonstrated in  Figure \ref{fig:lxz}, from  which it can  be inferred
that the XMM-XXL  improves by a factor of four the  number of AGN with
$L_X  (\rm 2  - 10  \, keV)  >10^{44} \,  erg \,  s^{-1}$  and $z<1.5$
compared to the combined Chandra COSMOS, AEGIS-XD and CDFS-4Ms samples
presented by  \cite{Buchner2015}. For heavily  obscured, Compton-thick
AGN,  the XMM-XXL-N  provides constraints  only  in the  case of  very
luminous sources, i.e.  $L_X  (\rm 2 - 10 \, keV) >  10^{45} \, erg \,
s^{-1}$  at $z>0.5$.  Additional  observations at  rest-frame energies
$>10$\,keV are  also needed to complement  existing X-ray spectroscopy
below  about 10\,keV  and  confirm the  high  levels of  line-of-sight
obscuration of  these systems. Therefore  the Compton thick  AGN space
density constraints  from the XMM-XXL-N  sample should be  viewed with
caution.

We estimate a Compton-thin  obscured AGN fraction of $\approx0.35$ for
luminosities $\log L_X (\rm 2 - 10 \, keV) = 44 - 45 \, erg \, s^{-1}$
independent  of  redshift   to  $z\approx1.5$.   At  somewhat  fainter
luminosities [$\log L_X (\rm 2 - 10 \, keV) = 43.75 \, erg \, s^{-1}$]
there  is  evidence that  the  obscured  AGN  fraction increases  with
redshift from $0.45\pm0.10$ at  $z=0.25$ to $0.75\pm0.05$ at $z=1.25$.
Similar results  are claimed by  \cite{Buchner2015} who show  that the
Compton-thin  AGN   fraction  is  a  complex   function  of  accretion
luminosity and  redshift.  At  high accretion luminosities  [$\log L_X
(\rm  2  - 10  \,  keV)  > 44  \,  erg  \,  s^{-1}$] feedback  process
associated with AGN winds may be responsible for clearing dust and gas
clouds in the vicinity of supermassive black holes and hence, lowering
the obscured AGN fraction.  At lower accretion luminosities [$\log L_X
(\rm 2  - 10 \, keV)  < 44 \, erg  \, s^{-1}$] AGN  feedback is likely
subdominant.  The increase of  the obscured AGN fraction with redshift
at lower luminosities  may be associated with the  overall increase in
the dust content of galaxies with redshift, which in turn is linked to
the higher specific star-formation rates of galaxies at earlier epochs
\citep[e.g.][]{Noeske2007, Magdis2012, Santini2014}.

We also show that infrared surveys that identify AGN via template fits
to   the   observed   multi-waveband  Spectral   Energy   Distribution
\citep{Delvecchio2014}  estimate  AGN  luminosity functions  that  are
broadly  consistent  with  the  space densities  of  Compton-thin  AGN
determined in the XMM-XXL-N  at high accretion luminosities [$\log L_X
(\rm 2  - 10 \, keV)  \ga 43 \,  erg \, s^{-1}$].  This  comparison is
limited by the current level  of random uncertainties in the AGN space
densities  at both  X-ray and  infrared wavelengths  and the  level of
contribution   of  Compton   thick   AGN  to   the  X-ray   luminosity
function. Despite  these points the  level of agreement  between X-ray
and  infrared AGN  luminosity functions  is evidence  against  a large
population  of  obscured  sources   that  is  identified  by  infrared
selection methods  but is missing at X-ray  wavelengths.  For example,
in Figure \ref{fig_xlf_nh_z} there is reasonable agreement between the
space density  of Compton thick AGN candidates  estimated using either
X-ray   or   mid-infrared   \citep{Fiore2009}  selected   samples   at
$z\approx1$.     Similar   conclusions    are   also    presented   by
\cite{Buchner2015}  by  comparing   X-ray  constraints  with  infrared
selected  samples of  Compton thick  AGN  \citep{Fiore2008, Fiore2009,
Alexander2011} to redshift $z\approx2$.   Despite the current level of
agreement between  the space densities of X-ray  and infrared selected
AGN, a  population of deeply buried sources  that remains unidentified
at both wavelengths cannot be excluded.

\begin{figure*}
\begin{center}
\includegraphics[height=1.1\columnwidth]{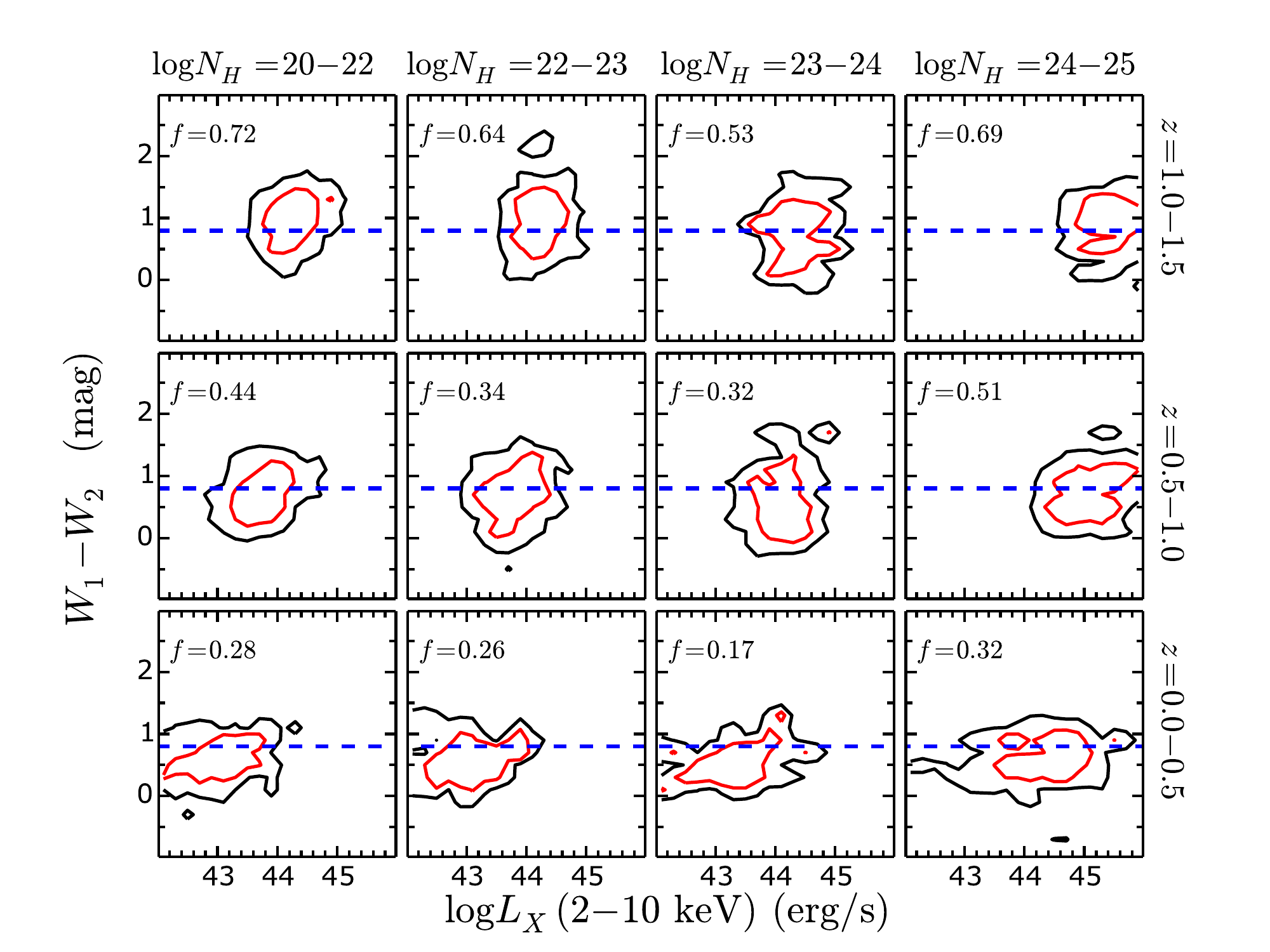}
\end{center}
\caption{WISE  $W_{1}-W_{2}$ infrared colour  as a  function 2-10\,keV
X-ray luminosity for XMM-XXL-N  X-ray selected AGN.  The contours show
density  of  AGN and  the  different  panels  correspond to  different
redshifts and column  density intervals as indicated by  labels at the
top and  the right of the  panels. The contours  are constructed using
the probability distribution  function of individual XMM-XXL-N sources
in redshift, luminosity and  column density.  The iso-density contours
correspond to the  68th and 95th percentile of  the distribution.  The
vertical dashed  line indicates the  \protect\cite{Stern2012} WISE AGN
colour  selection. The numbers  in each  panel indicate  the fraction,
$f$,  of X-ray  sources  above the  \protect\cite{Stern2012} WISE  AGN
colour cut.  }\label{fig_wise}
\end{figure*}

\begin{figure*}
\begin{center}
\includegraphics[height=1.1\columnwidth]{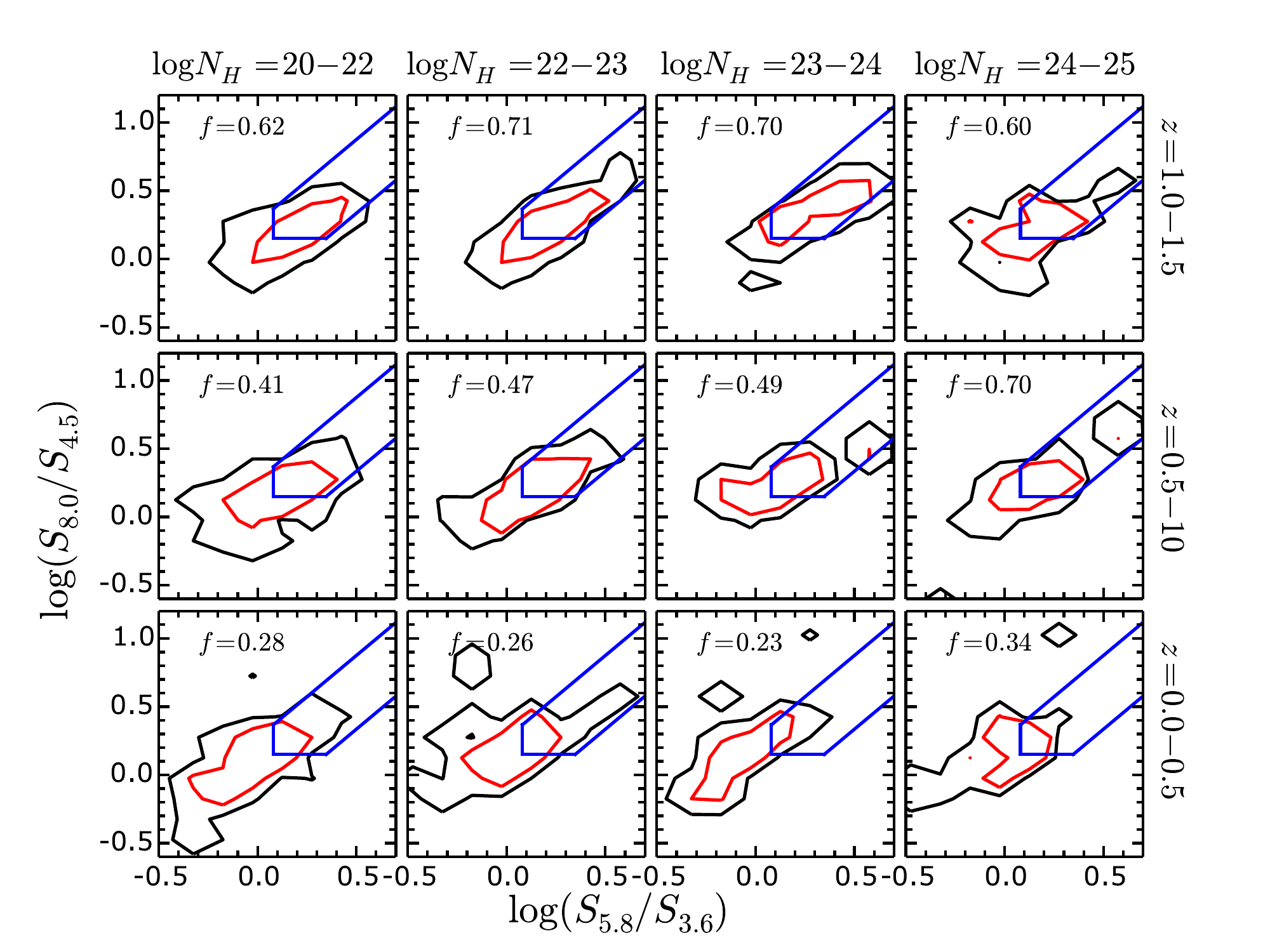}
\end{center}
\caption{The Spitzer mid-infrared  colour AGN selection plane proposed
by \protect\cite{Donley2012} .  The contours show density of XMM-XXL-N
AGN  and the different  panels correspond  to different  redshifts and
column density  intervals as  indicated by labels  at the top  and the
right  of  the  panels.    The  contours  are  constructed  using  the
probability  density  function  of  individual  XMM-XXL-N  sources  in
redshift,  luminosity and  column density.   The  iso-density contours
correspond to the  68th and 95th percentile of  the distribution.  The
blued   solid  lines   indicates  the   \protect\cite{Donley2012}  AGN
selection wedge. The numbers in each panel indicate the fraction, $f$,
of X-ray sources within the wedge.  }\label{fig_swire}
\end{figure*}

\begin{figure*}
\begin{center}
\includegraphics[height=1.1\columnwidth]{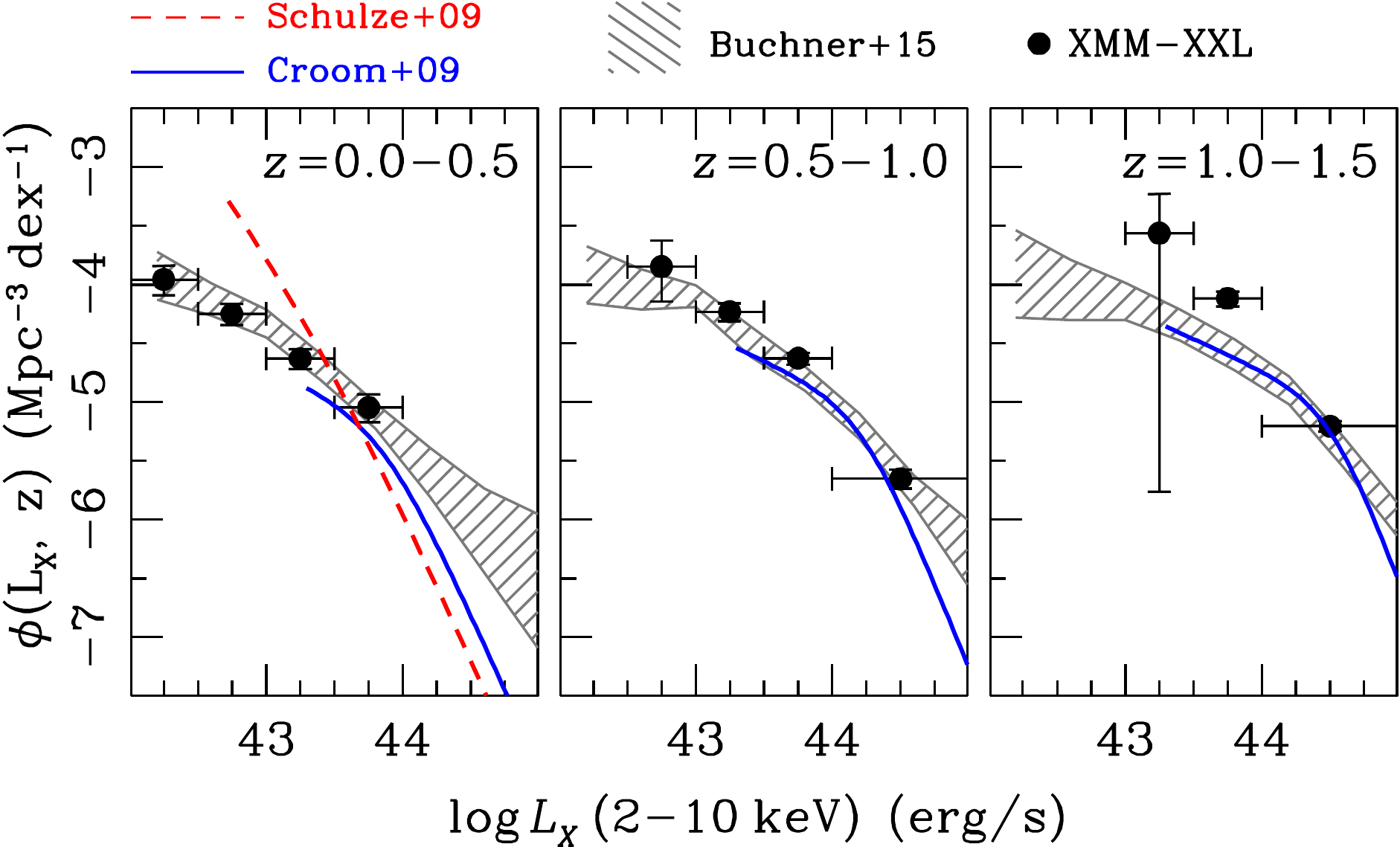}
\end{center}
\caption{Space density per luminosity dex of X-ray selected unobscured
AGN ($\rm  N_H =  10^{20}-10^{21.5}\,cm^{-2}$) in comparison  with the
luminosity   function  of   UV/optical  selected   QSOs.    The  black
data-points are  the constraints from  the XMM-XXL-N, the  shaded area
represents the results  of \protect\cite{Buchner2015}.  The blue curve
is   the   LEDE   (Luminosity   Evolution   and   Density   Evolution)
parametrisation  of UV/optical  QSO luminosity  function  presented by
\protect\cite{Croom2009}.  The  red dashed  curve is the  low redshift
broad-line      AGN     luminosity     function      determined     by
\protect\cite{Schulze2009}.  All  the optically type-1  AGN luminosity
functions  are  estimated  at  the  mean  redshift  of  each  redshift
interval.  }\label{fig_uxlf}
\end{figure*}

\begin{figure*}
\begin{center}
\includegraphics[height=1\columnwidth]{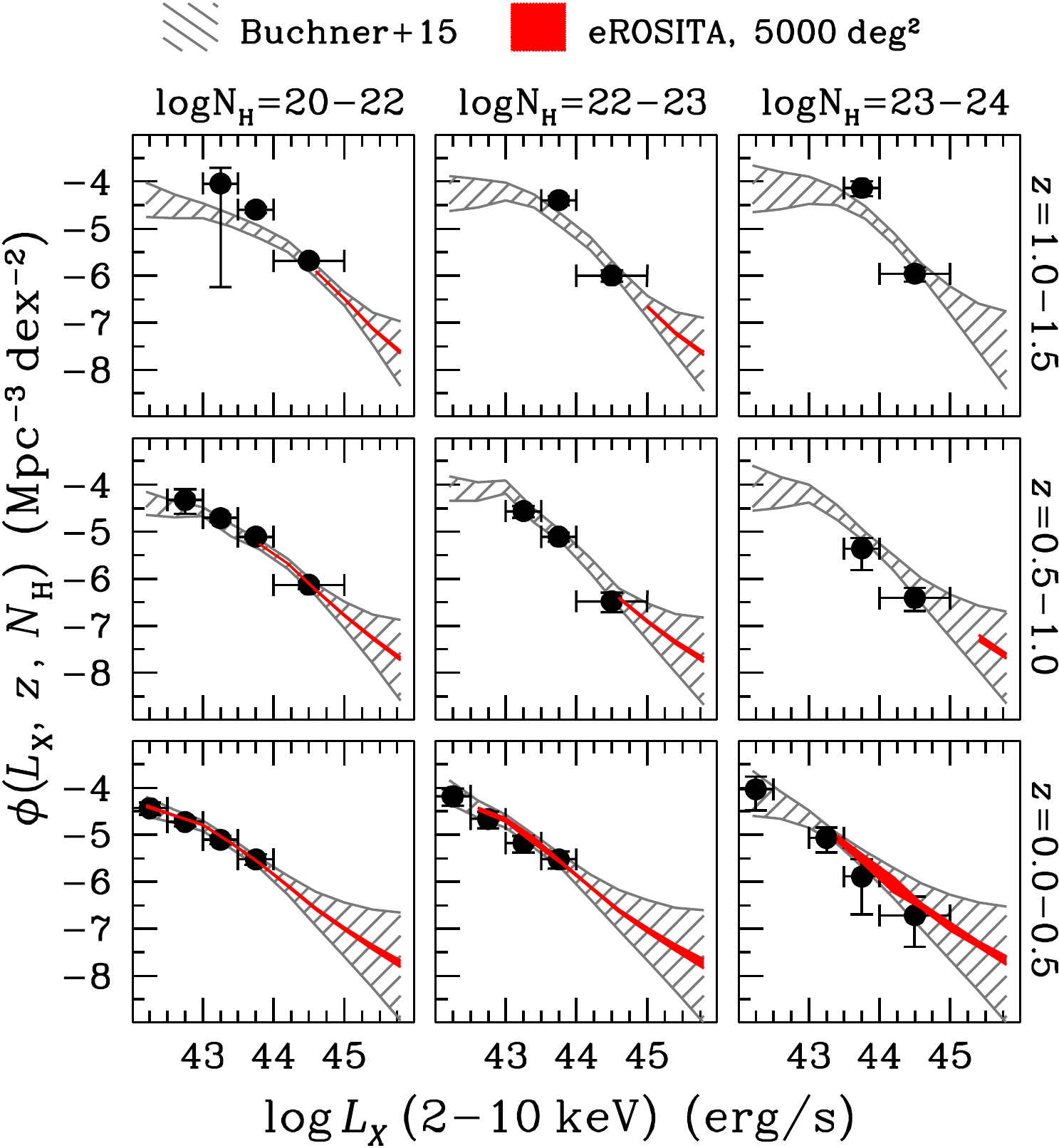}
\end{center}
\caption{The red  shaded regions show the expected  constraints on the
space  density  of  AGN  at  different  redshift  and  column  density
intervals from the eROSITA 4-year observations over a sky area of $\rm
5000\,deg^2$. The red regions are are plotted only for luminosity bins
where the expected number of eROSITA detected sources is greater than
unity. The  black data-points  are  the  constraints from  the 
XMM-XXL-N,   the  grey   hatched  area   represents  the   results  of
\protect\cite{Buchner2015}.  }\label{fig_xlfero}
\end{figure*}

It is further  shown that selecting AGN using  mid-IR colour cuts only
\citep[e.g.][]{Donley2012,  Stern2012}   leads  to  redshift-dependent
incompleteness.  Nevertheless, at  $z\ga1$ these methods are efficient
and relatively complete ($\approx  50-70$\%) in compiling luminous AGN
samples including heavily obscured systems.  \cite{Stern2012} used AGN
and galaxy  template Spectral Energy Distributions  to investigate the
redshift-dependent efficiency  of the AGN selection based  on the WISE
$W1-W2$ colour.  They showed  that suppression of the AGN mid-infrared
emission  relative  to  the   host  galaxy,  either  because  of  dust
extinction or dilution by stellar  light, is an issue for redshifts $z
\la 1.5$.  \cite{Stern2012} showed  that the exact redshift dependence
of the  $W1-W2$ colour-selection efficiency  depends on the  levels of
extinction and dilution of  the AGN radiation.  For example, templates
with  AGN emission  fraction  of  50\% in  the  wavelength range  $\rm
0.1-30\,\mu m$ lie below the  color cut $W1-W2=0.8$ used to select AGN
at all  redshifts below  $z=1.5$.  An AGN  fraction in  the wavelength
range $\rm 0.1-30\,\mu m$ of  75\% reduces the efficiency of selecting
AGN  via the  colour cut  $W1-W2=0.8$  only in  the relatively  narrow
redshift   range    $z\approx   0.5-1$.    The    redshift   dependent
incompleteness in Figure \ref{fig_wise} is likely related to the above
effects,  with  lower redshift  AGN  being  more  diluted than  higher
redshift  sources.   Interestingly  the  level  of  the  line-of-sight
obscuration measured by $\log N_H$ appears to be a second order effect
on the  incompleteness fractions  of Figure \ref{fig_wise}.   This may
because at  fixed redshift more  obscured sources in the  flux limited
XMM-XXL-N sample  are also expected to have  higher X-ray luminosities
and  therefore  a  larger  contrast  of the  integrated  AGN  emission
relative to the host galaxy.

Applying a hydrogen column density cut $\log \rm N_H/cm^{2} = 21.5$ to
the X-ray sample we find  fair agreement with the luminosity functions
of  UV/optically-selected  QSO, particularly  at  $z\ga0.5$.  This  is
consistent with  independent claims that the  column density threshold
that  maximises the  agreement between  the X-ray  unobscured  and the
UV/optically-selected broad-line QSO classes is $\log \rm N_H/cm^{2} =
21.5$ \citep[e.g.][]{Merloni2014}.  The XMM-XXL-N sample also includes
a  small fraction  \citep[10\%][]{Liu2016} of  AGN with  broad optical
emission lines and evidence for X-ray obscuration higher than $\rm N_H
\ga  10^{22} \,  cm^{-2}$ (see  Fig.   \ref{fig_example_type1}).  This
fraction  is   similar  to   previous  estimates  in   the  literature
\cite[e.g.][]{Brusa2003}. These  sources likely include  a fraction of
type-1.8 or earlier Seyferts, which at least in the local Universe are
associated, on  the average, with  hydrogen columns in excess  of $\rm
N_H  = 10^{22}\,cm^{-2}$  \citep{Burtscher2016}.  The  class  of Broad
Absorption  Line (BAL)  QSOs is  known to  be X-ray  underluminous for
their  UV  emission likely  because  of  X-ray  obscuration along  the
line-of-sight but also, in few cases, because of an intrinsically weak
X-ray  continuum  \citep[e.g.][]{Gallagher2006,  Gibson2009,  Luo2013,
Luo2014}.  The fraction of BAL features among optical/UV-selected QSOs
can be  as high as 26\%  \citep{Hewett2003, Reichard2003, Gibson2009}.
\cite{Stalin2011}  studied the  incidence of  BAL troughs  among X-ray
selected  AGN  in  the  XMM-LSS  sample and  reported  a  fraction  of
$7\pm5$\%, i.e. lower than  optical/UV-selected QSOs but comparable to
the fraction  of XMM-XXL-N broad-line QSOs with  X-ray absorption $\rm
N_H  \ga 10^{22} \,  cm^{-2}$ \citep[10\%][]{Liu2016}.   Searching for
BAL features  in the  optical spectra of  the present X-ray  sample is
beyond the scope of this  paper.  Nevertheless, it is interesting that
one   of   the  type-1   and   X-ray   absorbed   sources  in   Figure
\ref{fig_example_type1}  ($\alpha=02:05:41.3$,  $\delta=-0.4:37:07.4$,
$z=0.7220$) shows  evidence for an  absorption trough blueward  of the
Mg\,II broad emission line.

Finally, we make predictions on the AGN space density constraints that
the eROSITA  All Sky Survey \citep{Merloni2012} can  deliver.  We show
that  this mission  will  provide  a nearly  unbiased  census of  AGN,
including heavily  obscured systems approaching  column densities $N_H
\rm \approx  10^{24} \, cm^{-2}$,  at relatively low  redshift $z<0.5$
and for accretion luminosities $L_X\rm  \ga 10^{43} \, erg \, s^{-1}$.
This sample  will be a  unique resource for studying  AGN demographics
and  population properties  at $z<0.5$,  i.e. in  the last  5\,Gyrs of
cosmic time.  At higher redshifts eROSITA will place unique constrains
on the bright-end  of the luminosity function ($L_X\rm  \ga 10^{44} \,
erg \, s^{-1}$) and the  fraction of moderately obscured sources among
such luminous AGN.

\begin{table*}
\begin{centering}
\caption{X-ray AGN space density per luminosity and column density dex
in different redshift and $\log N_H$ intervals. The first column lists
the logarithm (base 10) of the X-ray luminosity range (2-10\,keV band)
of each measurement (units of  erg/s). The subsequent columns list the
logarithm  (base 10)  of the  space  density in  different $\log  N_H$
intervals. For each luminosity bin the median of the logarithmic (base
10) space density is listed. The numbers in the parentheses correspond
to   the  10th   and  90th   percentiles  of   the   $\log_{10}  \phi$
distributions. }\label{table:phi}
\begin{tabular}{c cccc}
\hline
   $\log_{10} L_X$ & \multicolumn{4}{c}{ $\log_{10} \phi \; (\rm Mpc^{-3}\,dex^{-2})$}  \\
\hline
  &  \multicolumn{4}{c}{$z=0.0-0.05$} \\
\hline
  & \multicolumn{1}{c}{$\log N_H = 20-22$}  & \multicolumn{1}{c}{$\log N_H = 22-23$}  & \multicolumn{1}{c}{$\log N_H = 23-24$}  & \multicolumn{1}{c}{$\log N_H = 24-25$}  \\
\hline
42.0--42.5  &  -4.43 (-4.57, -4.32)  &  -4.18 (-4.39, -4.02)  &  -4.03 (-4.48, -3.76)  &  --   \\   
42.5--43.0  &  -4.73 (-4.82, -4.64)  &  -4.67 (-4.86, -4.52)  &  --   &  --   \\   
43.0--43.5  &  -5.11 (-5.20, -5.03)  &  -5.18 (-5.38, -5.02)  &  -5.07 (-5.38, -4.85)  &  --   \\   
43.5--44.0  &  -5.52 (-5.65, -5.41)  &  -5.52 (-5.73, -5.36)  &  -5.89 (-6.69, -5.51)  &  -4.95 (-5.48, -4.65)  \\   
44.0--45.0  &  --   &  --   &  -6.71 (-7.38, -6.31)  &  -5.78 (-6.05, -5.58)  \\   
45.0--46.0  &  --   &  --   &  --   &  --   \\   
\hline
  &  \multicolumn{4}{c}{$z=0.5-1.0$} \\
\hline
  & \multicolumn{1}{c}{$\log N_H = 20-22$}  & \multicolumn{1}{c}{$\log N_H = 22-23$}  & \multicolumn{1}{c}{$\log N_H = 23-24$}  & \multicolumn{1}{c}{$\log N_H = 24-25$}  \\
\hline
42.0--42.5  &  --   &  --   &  --   &  --   \\   
42.5--43.0  &  -4.33 (-4.62, -4.10)  &  --   &  --   &  --   \\   
43.0--43.5  &  -4.71 (-4.79, -4.64)  &  -4.56 (-4.70, -4.45)  &  --   &  --   \\   
43.5--44.0  &  -5.11 (-5.16, -5.06)  &  -5.11 (-5.21, -5.03)  &  -5.36 (-5.81, -5.13)  &  --   \\   
44.0--45.0  &  -6.13 (-6.22, -6.05)  &  -6.48 (-6.72, -6.29)  &  -6.41 (-6.68, -6.20)  &  --   \\   
45.0--46.0  &  --   &  --   &  --   &  -6.84 (-7.34, -6.53)  \\   
\hline
  &  \multicolumn{4}{c}{$z=1.0-1.5$} \\
\hline
  & \multicolumn{1}{c}{$\log N_H = 20-22$}  & \multicolumn{1}{c}{$\log N_H = 22-23$}  & \multicolumn{1}{c}{$\log N_H = 23-24$}  & \multicolumn{1}{c}{$\log N_H = 24-25$}  \\
\hline
42.0--42.5  &  --   &  --   &  --   &  --   \\   
42.5--43.0  &  --   &  --   &  --   &  --   \\   
43.0--43.5  &  -4.04 (-6.24, -3.71)  &  --   &  --   &  --   \\   
43.5--44.0  &  -4.59 (-4.66, -4.53)  &  -4.39 (-4.50, -4.30)  &  -4.13 (-4.31, -4.00)  &  --   \\   
44.0--45.0  &  -5.68 (-5.73, -5.64)  &  -6.00 (-6.13, -5.89)  &  -5.96 (-6.13, -5.82)  &  --   \\   
45.0--46.0  &  --   &  --   &  --   &  -6.85 (-7.34, -6.55)  \\   
\hline

\end{tabular}
\end{centering}
\end{table*}

\begin{table}
\begin{center}
\caption{Fraction of obscured (Compton-thin) AGN at different redshift
and luminosity  intervals. The first column lists  the logarithm (base
10) of the X-ray luminosity range (2-10\,keV band) of each measurement
(units  of erg/s).  The second  column is  the obscured  fraction. The
median      of     the      distribution      of     the      quantity
$f=\frac{\int_{22}^{24}\phi(L_X,  z,   \log  N_H)\,{\rm  d}\log  N_H}{
\int_{20}^{24}   \phi(L_X,  z,   \log  N_H)\,{\rm   d}\log   N_H}$  is
listed. The numbers in the parentheses correspond to the 10th and 90th
percentiles of the distribution. }\label{table:fobscured}
\begin{tabular}{c c}
 \hline
  $\log L_X$  & fraction of obscured  \\
  range            & (Compton-thin) AGN  \\
\hline
    \multicolumn{2}{c}{$z=0.0-0.05$} \\
\hline
42.0--42.5  &   0.69 ( 0.55,  0.79)  \\   
43.0--43.5  &   0.50 ( 0.39,  0.59)  \\
43.5--44.0  &   0.43 ( 0.31,  0.54)  \\ 
\hline
    \multicolumn{2}{c}{$z=0.5-1.0$} \\
\hline
43.5--44.0  &   0.44 ( 0.37,  0.50)  \\   
44.0--45.0  &   0.33 ( 0.25,  0.42)  \\   
\hline
    \multicolumn{2}{c}{$z=1.0-1.5$} \\
\hline
43.5--44.0  &   0.69 ( 0.63,  0.75)  \\   
44.0--45.0  &   0.34 ( 0.28,  0.39)  \\   
\hline
\end{tabular}
\end{center}
\end{table}

\section*{Acknowledgements}

We   thank   the   anonymous   referee   for   useful   comments   and
suggestions. This work benefited  from the {\sc thales} project 383549
that is jointly funded by  the European Union and the Greek Government
in   the  framework   of  the   programme  ``Education   and  lifelong
learning''.  We acknowledge support  from the  FONDECYT Postdoctorados
3160439 (JB)  and the Ministry of Economy,  Development, and Tourism's
Millennium Science  Initiative through grant IC120009,  awarded to The
Millennium Institute  of Astrophysics  MAS (JB). Funding  for SDSS-III
has   been  provided  by   the  Alfred   P.   Sloan   Foundation,  the
Participating Institutions,  the National Science  Foundation, and the
U.S. Department of Energy Office of Science.  The SDSS-III web site is
http://www.sdss3.org/.  SDSS-III  is   managed  by  the  Astrophysical
Research Consortium for the Participating Institutions of the SDSS-III
Collaboration  including  the  University  of Arizona,  the  Brazilian
Participation Group,  Brookhaven National Laboratory,  Carnegie Mellon
University, University of Florida, the French Participation Group, the
German  Participation  Group,  Harvard  University, the  Instituto  de
Astrofisica   de   Canarias,   the  Michigan   State/Notre   Dame/JINA
Participation  Group,  Johns  Hopkins  University,  Lawrence  Berkeley
National Laboratory, Max Planck Institute for Astrophysics, Max Planck
Institute for  Extraterrestrial Physics, New  Mexico State University,
New  York  University,   Ohio  State  University,  Pennsylvania  State
University,  University  of   Portsmouth,  Princeton  University,  the
Spanish Participation Group, University  of Tokyo, University of Utah,
Vanderbilt   University,  University   of   Virginia,  University   of
Washington, and Yale University.

%%%%%%%%%%%%%%%%%%%%%%%%%%%%%%%%%%%%%%%%%%%%%%%%%%

%%%%%%%%%%%%%%%%%%%% REFERENCES %%%%%%%%%%%%%%%%%%

% The best way to enter references is to use BibTeX:

\bibliographystyle{mnras}
\bibliography{/home/age/soft9/BIBTEX/mybib} % if your bibtex file is called example.bib

% Don't change these lines
\bsp	% typesetting comment
\label{lastpage}
\end{document}